\documentclass[twocolumn,amsmath,amssymb,floatfix,longtable]{revtex4-1}

\usepackage{asymptote}
\usepackage{adjustbox}
\usepackage{tikz}
\usepackage[colorlinks,urlcolor=blue,anchorcolor=blue,linkcolor=blue,citecolor=blue,breaklinks=true]{hyperref}
\RequirePackage[normalem]{ulem} %DIF PREAMBLE
\newcommand\soutpar[1]{\let\helpcmd\sout\parhelp#1\par\relax\relax}

\begin{document}
\title{Quantum criticality in the spin-isotropic pseudogap Bose-Fermi Kondo model:\\ entropy, scaling, and the g-theorem}

\author{Zuodong Yu$^{1,2}$}
\thanks{Both authors contributed equally}
\email{richzyu@gmail.com}
\author{Farzaneh Zamani$^{3}$}
\thanks{Both authors contributed equally}
\email{zamani@physik.uni-bonn.de}
\author{Pedro Ribeiro$^{4}$}
\email{pedrojgribeiro@tecnico.ulisboa.pt}
\author{Stefan Kirchner$^{1,2}$}
\email{stefan.kirchner@correlated-matter.com}
\affiliation{$^{1}$ Zhejiang Institute of Modern Physics and Department of Physics, Zhejiang University, Hangzhou, 310027, China \\
$^{2}$ Zhejiang Province Key Laboratory of Quantum Technology and Device, Zhejiang University, Hangzhou 310027, China\\
  $^{3}$ Physikalisches Institut and Bethe Center for Theoretical Physics, Universit\"{a}t Bonn, Nussallee 12, D-53115 Bonn, Germany\\
  $^{4}$ CeFEMA, Instituto Superior T\'{e}nico, Universidade de Lisboa, Av. Rovisco Pais, 1049-001 Lisboa, Portugal}
\date{\today}

\begin{abstract}
We study the behavior of the entropy of the pseudogap Bose-Fermi Kondo model within a dynamical large-$N$ limit, where $N$ is related to the symmetry group of the model. This model is a general quantum impurity model that describes a localized
level coupled to a fermionic bath having a density of states that vanishes
in a power-law fashion near the Fermi energy   and to a bosonic bath possessing a power-law spectral density below a  cutoff energy.
As a function of the couplings to the baths
various quantum phase transitions can occur.
We study how the impurity entropy changes across these zero-temperature transitions and compare our results with predictions based on the g-theorem. This is accomplished by an analysis of the leading and sub-leading scaling behavior.  Our analysis shows that the $g$-theorem does not apply to the pseudogap Bose-Fermi Kondo model at the large-N level. This inapplicability originates from an anomalous contribution to the scaling function in the hydrodynamic regime where $k_B T>\hbar \omega$ which is absent in the quantum coherent regime, {\itshape i.e.}, for $k_B T<\hbar \omega$.
We also compare our results with those obtained for  the Sachdev-Ye-Kitaev model.
\end{abstract}
\maketitle

\section{Introduction}

Quantum phase transitions (QPTs) have been a central topic of condensed matter research \cite{Hertz.76,Sachdev}.
This is due to a number of reasons. There is {\itshape e.g.} mounting evidence of a close link between unconventional superconductivity as  observed in the cuprates, iron selenides, or $4f$-based heavy-electron compounds and the occurrence of the so-called strange metal behavior at elevated temperatures in these superconductors. This strange metal phase is characterized by  a linear-in-temperature relaxation rate and a logarithmically [in temperature ($T$)] increasing specific heat \cite{Hussey.13,Keimer.15}. Moreover, aforementioned unconventional superconductivity  is commonly found at the border of magnetism.  This observation has led to the speculation that the strange metal out of which superconductivity emerges is caused by a quantum critical point (QCP) hidden under the superconducting dome. There is also an increasing amount of evidence, that quantum criticality in strongly correlated metals defies a description in terms of an order parameter functional \cite{Si.01,Coleman.01}. Moreover, at least for rare-earth based intermetallics it has been demonstrated that  the general phase diagram of this materials class can be organized around the different possible QCPs \cite{Si.06,Coleman.10,Si.14,Kirchner.20}.

QPTs are phase transitions that take place at zero $T$ and that can be accessed  through  some control parameter like pressure, chemical doping, or magnetic field.
In contrast to classical phase transitions which occur at non-zero $T$ and are driven by the competition of internal energy and entropy, QPTs are a ground state property. 
The $T=0$ entropy  is  tied to the ground state degeneracy, which is expected to vanish as required by the third law of thermodynamics.
An increase of the entropy at finite $T$ in the vicinity of a continuous QPT  may, however,  still be expected due to the competition of the phases that brings about the QCP. 
This in turn may promote the emergence of novel order, {\itshape e.g.} superconductivity, in order to avoid the accumulation of entropy associated with proximity to the QCP. A situation that appears to be realized in CeRhIn$_5$, a heavy-electron compound, where a QCP with critical  Kondo destruction is hidden beneath the superconducting dome \cite{Park.08}.
CeCu$_{6-x}$Au$_{x}$ is a rare-earth intermetallic compound that undergoes a Kondo-destroying QPT at $x\approx 0.1$ which  separates a magnetic  from a Kondo-screened paramagnetic phase.
The multidimensional entropy landscape of CeCu$_{6-x}$Au$_{x}$ above the QCP and its relation to quantum critical fluctuations  has recently been  mapped out \cite{Grube.17} which demonstrated a direct link between quantum criticality and the  entropy accumulation at finite $T$ near the QCP. 
Such entropy accumulation in the vicinity of quantum criticality may form the basis for dedicated cooling devices in terms of
adiabatic processes across the critical coupling  \cite{Wolf.11}.

In generic bulk systems, one expects the residual ($T=0$) entropy to vanish identically in accordance with the third law of thermodynamics.  
The situation is different in  quantum impurity systems which may possess intermediate coupling fixed points that are characterized by a finite residual entropy. 
Quantum impurity models capture the interplay between a local and discrete quantum mechanical degree of freedom, {\itshape e.g.}, a magnetic moment, that hybridizes with a continuous and gapless bath of fermionic or bosonic modes and thus forms an important testing ground for understanding that interplay. A well-known example is the isotropic two-channel Kondo model which possesses a residual entropy $\sqrt{2}$ which is understood in terms of a Majorana zero mode \cite{Nozieres.80,Tsvelik.85,Kirchner.20b}.
Here, this impurity entropy is defined as the difference between the entropies of the full quantum impurity model and that of the gapless host in which the impurity is embedded, {\itshape i.e.}, the bath. As such, it is per se not bound by thermodynamic requirements and could, {\itshape e.g.}, even increase as $T$ decreases.
The impurity entropy $S$ can in principle be measured in systems  with sufficiently low concentrations of quantum impurities, such that contributions to $S$  beyond the lowest, non-trivial  order in the concentration can be ignored.

In this paper, we address the behavior of the impurity entropy of a class of quantum impurity systems that feature  critical  Kondo destruction. Among the simplest quantum impurity system that can undergo a Kondo-destroying QPT is the pseudogap Kondo model \cite{Withoff.90}. In this model, a QCP separates a Kondo-screened local Fermi liquid phase from a phase where the local moment remains unquenched down to $T=0$.  The critical properties of this model have been studied extensively using numerical and other renormalization group approaches \cite{Gonzalez-Buxton.98,Ingersent.02,Fritz.06}, as well as, {\itshape e.g.}, dynamical large-N \cite{Vojta.01}, local moment \cite{Glossop.03,Glossop.05},   and Monte Carlo methods \cite{Glossop.11,Pixley.12}.
Physical realizations of this model include certain quantum dot structures \cite{Silva.06} and disordered metals containing containing  low concentrations of magnetic moments \cite{Zhuravlev.07}.

\section{Models of critical Kondo destruction}
The Bose-Fermi Kondo model (BFKM) is a quantum impurity model that has been introduced in the context of Kondo-destroying quantum criticality which occurs in certain rare earth-based heavy-electron
intermetallics like, {\itshape e.g.}, CeRhIn$_5$, CeCu$_{6-x}$Au$_{x}$, or YbRh$_2$Si$_2$  \cite{Si.96,Si.99,Si.01} (see also \cite{Sengupta.00}). 
Kondo-destroying quantum criticality has been attracting considerable interest as it defies a description in terms of an order-parameter functional \cite{Si.01,Coleman.01,Senthil.04,Si.14,Kirchner.20}.
This is most clearly reflected in the Fermi volume jump observed in YbRh$_2$Si$_2$ and from thermodynamic and transport properties at finite $T$ above the QCP  and which are indicating a linear-in-$T$ relaxation rate \cite{Paschen.04,Friedemann.10,Friedemann.11}. This linear-in-$T$ relaxation rate has been interpreted in terms of $\omega/T$ scaling of the magnetic response as, {\itshape e.g.}, observed in CeCu$_{6-x}$Au$_{x}$. More recently, $\omega/T$ scaling in the charge response of YbRh$_2$Si$_2$ has been detected in the vicinity of the magnetic QCP \cite{Prochaska.20}.
As non-trivial $\omega/T$ scaling is not expected within the standard Landau-Ginzburg framework for magnetic criticality \cite{Hertz.76,Millis.93}, it can serve as a diagnostic tool for unconventional criticality.
It has been demonstrated that the spin-isotropic BFKM displays $\omega/T$ scaling at its Kondo-destroying QCP \cite{Zhu.04,Cai.20}.
In the BFKM, Kondo screening becomes critical due to the competition with a singular bosonic bath. Its properties have been investigated using a range of methods. 
The large-N limit of the BFKM has been considered in Ref.\ \cite{Zhu.04} while renormalization group (RG) methods have been used in Refs.\ \cite{Zhu.02,Zarand.02}. The model has also been addressed using numerical renormalization group (NRG) generalizations to include bosons  \cite{Bulla.03,Glossop.05a}. The spin-anisotropic BFKM  includes the spin-boson model as a limiting case. 
The BFKM arises within the extended dynamical mean field or EDMFT approach to quantum criticality in rare-earth intermetallics which maps the Kondo lattice model to a BFKM augmented with a self-consistency condition \cite{Si.01,Si.14}. The model also arises in quantum dot structures attached to ferromagnetic leads \cite{Kirchner.05,Zamani.16}.

In  the pseudogap BFKM, gapless bosonic and fermionic baths are coupled to a local moment. 
The dynamics of this model is described by

\begin{eqnarray}
H_{\mbox{\tiny pgBFKM}} &=& H_{\mbox{\tiny bath}}+ H_{\mbox{\tiny b-s}}\\
	H_{\mbox{\tiny bath}} &=& \sum_{k,\sigma} \varepsilon_k c^{\dagger}_{k,\sigma} c^{}_{k,\sigma} 
	+ \sum_{q} \omega_q \boldsymbol{\vec{\phi}}_q^{\dagger}\boldsymbol{\vec{\phi}}_q  \\
		H_{\mbox{\tiny b-s}}					&=& J_K^{\parallel} S^z s_{c}^z + \frac{J_K^{\perp}}{2} \big(S^+ s_{c}^- +S^- s_{c}^+ \big) \nonumber \\
	 &+&g^{\parallel} S^z \big( \phi^{z,\dagger}_{0}+\phi^{z}_{0}\big) 
	+ g^{\perp} \sum_{i=x,y} S^i\big(\phi^{i,\dagger}_0+ \phi^{i}_{0}\big),\nonumber
\end{eqnarray}
where $H_{\mbox{\tiny bath}}$ denotes the bath part and $H_{\mbox{\tiny b-s}}$ describes the coupling between bath and impurity degrees of freedom.
$J_K^{\perp}$ and $J_K^{\parallel}$ are the transversal and longitudinal Kondo exchange coupling constants between the local moment $\boldsymbol{S}$ and the spin density of the fermionic bath at the impurity location, given by $s_{c}^z=\sum_{k,k^\prime}(c^{\dagger}_{k \uparrow}c^{}_{k \uparrow}-c^{\dagger}_{k \downarrow}c^{}_{k \downarrow})$, $s_{c}^+=\sum_k c^\dagger_{k\uparrow}c^{}_{k\downarrow}$, and $s^-=(s^+)^\dagger$. $g^{\perp}$ and $g^{\parallel}$ are the transversal and longitudinal couplings between the local moment and the bosonic bath,
and $\omega_q$ ($\varepsilon_k$) is the bosonic (fermionic) bath dispersion.
The pseudogap density of states (DOS) of the fermionic bath, is characterized by  a power-law dependence as the Fermi energy ($\varepsilon_F=0$) is approached, {\itshape i.e.}, $\sum_k\delta(\omega-\varepsilon_k)\sim |\omega|^r\Theta(D-|\omega|)$, while the bosonic spectral density is characterized by a sub-Ohmic behavior at low energies,  {\itshape i.e.}, $\sum_q [ \delta(\omega-\omega_q)-\delta(\omega+\omega_q)] \sim |\omega|^{1-\epsilon}\mbox{sgn}(\omega)\Theta(\Lambda-|\omega|)$.

The pseudogap BFKM contains as special cases the pure pseudogap Kondo model ($g^{\perp}=g^{\parallel}=0$) and the Bose Kondo model ($r=0$). Each of the two allow to critically destroy  Kondo screening either through the depletion of fermionic screening states or via the coupling to a singular bosonic bath that can compete with spin-flip scattering between fermionic bath and local moment. The combination of the two  possibilities of Kondo screening suppression in the pseudogap BFKM
thus allows to study the interplay of both effects near critical Kondo destruction. As a result, the general phase diagram of this model is correspondingly rich.
So far, it has been studied using perturbative RG in the spin-isotropic case, {\itshape e.g.}, for $g^{\perp}=g^{\parallel}$ and $J_K^{\perp}=J_K^{\parallel}$  \cite{Vojta.03,Kircan.04} and in the easy-axis case (($g^{\perp}=0$ and $J_K^{\perp}=J_K^{\parallel}$) using NRG \cite{Glossop.08,Pixley.13}, and continuous-time quantum Monte Carlo (CT-QMC) methods \cite{Pixley.11,Pixley.13}. Here, we will study the SU(2) symmetric pseudogap BFKM in a dynamical large-N limit, where the SU(2) symmetry group is  enlarged to SU(N).

The dynamical large-N method is not capable of capturing the local Fermi liquid fixed point and instead results in an intermediate coupling fixed point that corresponds to an overscreened multichannel Kondo ground state possessing non-Fermi-liquid properties \cite{Parcollet.98,Cox.93}. 
This short-coming not withstanding, the large-N method yields controlled results for the critical properties of the pseudogap Kondo model and sub-Ohmic BFKM  that are in line with those obtained using NRG and CT-QMC \cite{Glossop.11,Zamani.13,Cai.20}.

In quantum impurity models with a bulk component that is
conformally invariant,  a conformal mapping can
be found to obtain boundary correlators at temperatures $T>0$ from their $T=0$ counterparts \cite{Affleck.91,Cardy.84}.
A two-point correlator of a
primary conformal field $\Phi$ with scaling dimension $\Delta$ exhibits at $T=0$ a
power-law decay $\langle\Phi(\tau,T=0)\:\Phi(0,T=0)\rangle\sim\tau^{-2\Delta}$. This gives rise to a scaling form \cite{Ginsparg.89}
\begin{equation}
\label{eq:scaling_form}
\chi_{\Phi\Phi}(\tau,T) \equiv \langle \Phi(\tau,T)\:\Phi(0,T) \rangle
  \sim \left(\frac{\pi T}{\sin(\pi \tau T)}
\right)^{2\Delta}
\end{equation}
at $T\neq 0$.
The Fourier transform of Eq.~\eqref{eq:scaling_form} implies an $\omega/T$ scaling form of $\chi_{\Phi\Phi}(\omega,T)$, provided $0<2\Delta<1$ (see Appendix \ref{app:FT}).
For the boundary entropy of conformally invariant systems, a relation known as the $g$-theorem exists, linking the boundary contribution to the fixed-point entropy with the renormalization group (RG) flow \cite{Affleck.91}. According to this theorem, the boundary entropy decreases along RG trajectories.
A proof of the $g$-theorem has been provided  in Ref.\ \cite{Friedan.04}.
%%%%%%%%%%%%%%%%%%%%%%%%%%%%%%%%%%%%%%%%%%%%%%

The pseudogap BFKM does not possess conformal invariance. Both
the pseudogap DOS of the fermionic bath and the sub-Ohmic spectral density of the bosonic bath break conformal invariance. Interestingly, a scaling behavior as in Eq.\ \ref{eq:scaling_form} for $\tau \rightarrow \beta/2$ at sufficiently large $\beta$ has been reported to emerge near quantum criticality in a range of models which can be viewed as special cases of the pseudgap BFKM \cite{Kirchner.08,Glossop.11,Pixley.12,Cai.20}.

The emergence of the scaling form (\ref{eq:scaling_form})  is confined to the scaling regime of the associated QCP
and does in itself not imply the applicability of the $g$-theorem.
It is thus not {\it a priori} clear if it is at all possible to relate the boundary entropy to the RG flow. In this paper, we are primarily concerned with the fate of the $g$-theorem of the pseudogap BFKM in the so-called large-$N$ limit specified below. In contrast to earlier results \cite{Kircan.04}, we find that the $g$-theorem is not fulfilled in the pseudogap BFKM. We benchmark our approach against the pure Kondo model which is conformally invariant as well as against the so-called Sachdev-Ye-Kitaev (SYK) model in the large-N limit.
 
In the following, we introduce the Hamiltonian of the pseudogap BFKM in Secs.~\ref{sec:models} and \ref{sec:DLN} discusses the dynamical large-N limit and the evaluation of the impurity entropy. In Sec.~\ref{sec:results}, we present our results for the scaling function in $\omega$ and $\tau$ and the boundary entropy at the various fixed points. This leads us to conclude that in the pseudogap BFKM the boundary entropy does not always decrease along the RG flow. We trace back this breakdown of the $g$-theorem to an anomalous contribution to the $\tau T$-scaling function present at all intermediate fixed points. 
A summary recapitulates our findings and puts them in perspective. Appendices \ref{app:s1}-\ref{app:SYK}
contain supplementary results and details.

%%%%%%%%%%%%%%%%%%%%%%%%%%%%%%%%%%%%%%%%%%%%%%%%%%%%%%%%%%%%%
\section{$SU(N)\times SU(M)$ Model}
\label{sec:models}

\begin{figure}[ht]
\centering
\includegraphics[width=0.45\columnwidth]{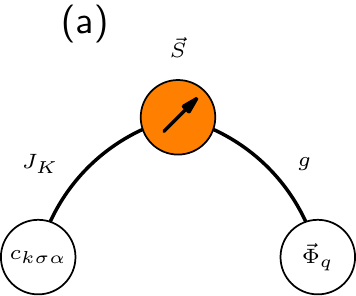}\\
\includegraphics[width=0.3\columnwidth]{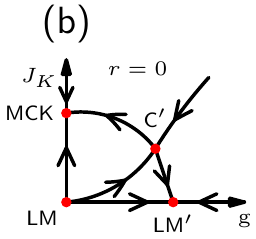}
\includegraphics[width=0.3\columnwidth]{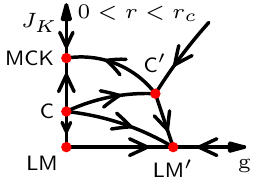}
\includegraphics[width=0.3\columnwidth]{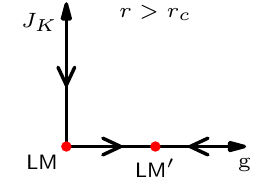}
\caption{(a) Sketch of the BFKM. (b) Flow diagram for different values of $r$ and $0\leq \epsilon \le 1$. For the case considered in this paper ($\kappa =M/N=1/2$) one finds $r_c\approx 0.3115$. For details, see main text.}
\label{fig:fig_1}
\end{figure}

We study the large-$N$ version of the multichannel BFKM, featuring a quantum spin (S) coupled to gapless fermionic ($c$) and bosonic ($\Phi$) excitations, as illustrated in Fig.\ref{fig:fig_1}-(a). 
In the large-$N$ version of the model the $SU(2)$ degree of freedom is generalized to $SU(N)$ and the 
fermionic fields transform under the fundamental representation of $SU(N)\times SU(M)$, where $M$ represents  the number of degenerate charge channels of the fermionic bath. Likewise, the $N$-component bosonic vector fields transform under $SU(N)$.  
The system is thus described by the Hamiltonian
\begin{align}
\label{eq:Ham}
	H=&\sum_{k \sigma \alpha}\varepsilon_{k}c_{k\sigma\alpha}^{\dagger}c_{k\sigma\alpha} +\sum_{q}\omega_{q}\Phi_{q}^{\dagger}\Phi_{q} \nonumber\\
	 &+\frac{J_{\text{K}}}{N}S\cdot s_{c}+\frac{g}{\sqrt{N}}\sum_{q}S\cdot(\Phi_{q}^{\dagger}+\Phi_{q}),
\end{align}
where $\sigma$ and $\alpha$ are, respectively, the SU($N$)-spin and SU($M$)-channel indices and $p$, $q$ are momentum indices.
The total $c$-electron spin-density at the impurity site is
\begin{align}
\label{eq:spinDen}
s_c^i = \sum_{\alpha,\sigma,\sigma'}  \sum_{p,p'} c^{\dagger}_{p\sigma\alpha} t^{i}_{\sigma,\sigma'} c_{p'\sigma'\alpha}.
\end{align}
In this equation,  the generators of su($N$) in the fundamental representation are referred to as $t^i$  ($i=1,\dots,N^2 - 1$). 
The large-$N$ limit is taken in such a way that the ratio $\kappa=M/N$ is kept fixed while $N\rightarrow \infty$ and $M\rightarrow \infty$. 
Note that the fermionic and bosonic baths are fully characterized by their local spectral properties. For the fermions we consider a density of states (DOS) of the form 
\begin{equation}
\label{eq:conductelec}
	A_{c}(\omega)=A_{0}\theta(D-|\omega|)|\omega|^{r},
\end{equation} 
where $D$ is a high-energy cutoff. $A_{0}$ is fixed through $\int d\omega A_{c}(\omega) = 1$.
In what follows, we focus on $r\in [0,1[$.  The pseudogap Kondo model with negative $r$ has {\itshape e.g.} been studied in Refs.\ \cite{Mitchell.13,Cheng.17}.
 The bosonic spectral density is taken to be of the form 
\begin{equation}
	A_{\Phi}(\omega)=A_{\Phi_{0}}\mathrm{sign}(\omega)\theta(\Lambda-|\omega|)|\omega|^{1-\epsilon},
\end{equation}
where $\Lambda$ is a high-energy cutoff, $\epsilon\in[0,1[$ characterizes the sub-Ohmic DOS, and  $A_{\Phi_{0}}$ is chosen such that $\int_{0}^{\Lambda} d\omega A_{\Phi_{0}}(\omega) = 1$.

At $T=0$, the phase space of the pseudogap BFKM encompasses a number of fixed points. The nature of the fixed points and the structure of the flow diagram depend on the values of $r$  and $\epsilon$. For $0<r<1$ and $0\leq \epsilon <1$ there are three cases that need to be distinguished [see Fig.\ref{fig:fig_1}-(b)].  

For $r=0$, the standard Bose-Fermi Kondo model is obtained \cite{Si.01,Zhu.02}. Its large-N version possesses  an over-screened multichannel Kondo phase (MCK) at small $g /J$ which is separated from a critical local moment phase, controlled by (LM'),  by an unstable critical point (C') that features critical Kondo destruction \cite{Zhu.04}. The unstable trivial fixed point (LM) corresponds to the fully decoupled impurity.

In the presence of a pseudogap, {\itshape i.e.}, $r>0$, the suppression of the electronic DOS at the Fermi energy makes the MCK fixed point less stable and leads to the appearance of a further critical point (C).  
Beyond a critical value of $r$, {\itshape i.e.}, $r>r_c$, the suppression of the electronic DOS is so effective that the MCK fixed point disappears leading to a phase diagram  where  LM'  is the only stable fixed point. 
Note that the standard Kondo problem corresponds to $M=1$ and $N=2$. In this exactly screened case, the MCK gives way to single channel Kondo physics which is governed by a strong coupling fixed point ($J_{K}\rightarrow \infty$) and displays Fermi liquid signatures. Within the large-$N$ description adopted in this paper, this strong coupling fixed point is out of reach.

The characterization of each phase and the scaling laws for the different quantities were obtained in \cite{Zhu.02,Zhu.04,Cai.20} both by perturbative RG and large-$N$ methods. In particular, it was found that the $T = 0$ scaling properties in the characteristic quantum critical fan are in line with dynamical $\omega/T$ scaling (see below) \cite{Zhu.02,Zhu.04,Sachdev}.

%%%%%%%%%%%%%%%%%%%%%%%%%%%%%%%%%%%%%%%%%%%%%%%%%%%%%%%%%%%%%%%
%%%%%%%%%%%%%%%%%%%%%%%%%%%%%%%%%%%%%%%%%%%%%%%%%%%%%%%%%%%%%%%

\section{Methods}
\label{sec:DLN}

\subsection{Dynamical large-$N$}

We resort to a pseudofermion representation of the local spin, i.e., $S^i= \sum_{\sigma\sigma'} f^\dagger_\sigma \tau^i_{\sigma,\sigma'} f_{\sigma'}$. Here, $\tau^{i}~(i=1,\dots,N^{2}-1)$ forms an antisymmetric representation of $su(N)$ fixed by imposing the constraint $Q= \sum_\sigma f^\dagger_\sigma f_{\sigma} =q N$. A dynamical Lagrange multiplier $\lambda$ enforces the constraint within the functional integral formalism. We have chosen $q=1/2$ in this work. 

The imaginary time action for the pseudogap BFKM, Eq.\eqref{eq:Ham} is given by
\begin{eqnarray} \label{eq:action}
	S^{\text{DLN}}&=&-\int_{\tau}( c_{k\alpha \sigma}^{\dagger}g_{c}^{-1}c_{k'\alpha \sigma}+\Phi_{q}^{\dagger}g_{\Phi}^{-1}\Phi_{q}+f_{\sigma}^{\dagger}g_{f}^{-1}f_{\sigma}) \nonumber\\
				  &+&\int_{\tau}[\frac{J_{K}}{N}f_{\sigma}^{\dagger}f_{\sigma'}c_{k\sigma'\alpha}^{\dagger}c_{k'\sigma\alpha}-\lambda q_{0}N \\
		&+&\frac{g}{\sqrt{N}}f_{\sigma}^{\dagger}f_{\sigma'}\tau_{\sigma\sigma'}^{i}(\Phi_{q}^{i\dagger}+\Phi_{q}^{i})], \nonumber
\end{eqnarray}
with $g_c(i \omega_n)= (i \omega_n -\epsilon_k)^{-1} $ and $g_\Phi(i \nu_n)= (i \nu_n -\omega_q)^{-1} $ are, respectively, the fermionic and bosonic degrees of freedom of the bath modes where $i\omega_n = 2\pi i(n+1/2) /\beta $ and $i\nu_n = 2\pi i n/\beta $, $n\in\mathbb{ Z}$ are the fermionic and bosonic Matsubara frequencies. The bare pseudofermion Green's function is defined as $g_{f}(i \omega_n)=(i \omega_n -\lambda)^{-1}$.

In the following, we employ a dynamical large-$N$ procedure \cite{Parcollet.98,Vojta.01,Zhu.04,Ribeiro.15} briefly described in Appendix \ref{app:s1} and summarized below for convenience.
The procedure consists of introducing a bosonic Hubbard-Stratonovich field $B$ to decouple the fermionic interacting term. 
This allows us to integrate out the bath degrees of freedom and to recover an action solely in terms of local fields. 
The interacting terms in the action are subsequently decoupled with the help of two sets of bi-local fields $W$ and $Q$. 

For the impurity contribution to the free energy $N f^{\text{DLN}}$ associated with Eq.~(\ref{eq:action}), we write $f_\text{imp}=f^{\text{DLN}}-f_\text{\text{bulk}}$, where $f_\text{\text{bulk}}$ denotes the bath contribution defined as the free energy associated with $H_{\mbox{\tiny bath}}$, Eq.~(\ref{eq:Ham}), for $J_K=0=g$.
As shown in Appendix \ref{app:s1}, the impurity contribution of Eq.~(\ref{eq:action})  can be written as
\begin{eqnarray}
\label{eq:largeNaction}
	f_{\text{imp}}^{}&=&\int_{\tau}[\bar{Q}(\tau) Q(-\tau) +\bar{W}(-\tau)W(\tau)-q_0 \lambda(\tau)]\\
	 &+& T \kappa \mathrm{Tr}\ln(-G_{B}^{-1})-T\mathrm{Tr}\ln(-G_{f}^{-1}), \nonumber
\end{eqnarray}
where $G_B^{-1}=g_B^{-1}-\Sigma_B$ and $G_f^{-1}=g_f^{-1}-\Sigma_f$ and
\begin{equation}\begin{split}
\label{eq:self}
	 \Sigma_B (\tau)=& - g_c(-\tau) Q (\tau) \\
	 \Sigma_f (\tau)=&  \bar{Q}(\tau) + g~\bar{W}(\tau) + g~g_{\Phi}(\tau) W(\tau). 
\end{split}\end{equation}
Note that, the definition of the Hubbard-Stratonovich  fields is slightly different to that  used in Refs.~\cite{Parcollet.98,Zhu.04,Glossop.11,Ribeiro.15}. As shown below, this change in the Hubbard-Stratonovich fields  does not affect the self-energy values at the saddle point but is more convenient in the present context. 
The decoupling scheme chosen here avoids the generation of a temperature-dependent Jacobian, related to $g_c(\tau)$ and $g_{\Phi}(\tau)$, which would need to be properly taken into account when evaluating the impurity entropy. Formally, both definitions of the Hubbard-Stratonovich fields are equivalent and lead to identical values for the self-energies and impurity entropy but the choice adopted here is more convenient when explicitly evaluating the temperature dependence of the impurity entropy.\\
The partition function in Eq.(\ref{eq:largeNaction}) is  suitable for a saddle-point approximation, 
\begin{align}
\label{eq:saddlePoint}
    \dfrac{\delta f_{\text{imp}}}{\delta X}=0,
\end{align}
where $X$ represents $\bar{Q},Q,\bar{W},W,\lambda$.  
The propagators for $f$ and $B$ are determined by the saddle-point equations which relate proper self-energy contributions to the bi-local fields.
In the limit of $N,M\rightarrow \infty$ (while $\kappa$ is kept constant), the saddle-point approximation
\begin{eqnarray}
	\label{eq:sc1}
	&&Q(\tau)=-G_{f}(\tau) \\
	\label{eq:sc2}
	&&\bar{Q}(\tau)=-\kappa G_{B}(\tau)g_{c}(\tau) \\
	\label{eq:sc3}
	&&W(\tau)=-gG_{f}(\tau) \\
	\label{eq:sc4}
	&&\bar{W}(\tau)=-gG_{f}(\tau)g_{\Phi}(-\tau)
\end{eqnarray}
becomes exact \cite{Parcollet.98,Vojta.01,Zhu.04,Ribeiro.15}.

This procedure yields the
following self-consistency equations (for details, see Appendix \ref{app:s1})
\begin{eqnarray} \label{eq:Sigma_f}
	\Sigma_{f}(\tau)&=&\Sigma_{f}^{1}(\tau)+\Sigma_{f}^{2}(\tau)\\
	&=&-\kappa G_{B}(\tau)g_{c}(\tau)-g^{2}G_{f}(\tau) \tilde g_{\Phi}(\tau) \nonumber \\
	\Sigma_{B}(\tau)&=&G_{f}(\tau)g_{c}(-\tau) \label{eq:Sigma_B}
\end{eqnarray}
with $\tilde g_{\Phi}(\tau) = g_{\Phi}(\tau)+g_{\Phi}(-\tau)$. This redefinition is equivalent to extending the sum over $q$ from positive $\omega_q$ to negative values with $\omega_{-q}=-\omega_q$.
The local spin susceptibility at the saddle point is given in terms of the pseudoparticle propagator $G_f$ as
\begin{align}
    \chi(\tau)=G_f(\tau)G_f(-\tau)
\end{align}
while the local $t$-matrix is obtained as the convolution of $G_f(\omega,T)$ and $G_B(\omega,T)$ from 
\begin{align}
    {\mathcal{T}}(\tau)=G_f(\tau)G_B(-\tau).
\end{align}
At the saddle-point, the free energy is  given by  
\begin{align}
\label{eq:free_action_sp}
	f_{\text{imp}}=&\int_{\tau}[ 
	\kappa  G_B(\tau) g_c(\tau) G_f(\tau)
	+  g^2 G_f(- \tau ) g_\Phi(\tau ) G_f(\tau)]\nonumber \\
	 &+T \kappa \mathrm{Tr}\ln(-G_{B}^{-1})-T\mathrm{Tr}\ln(-G_{f}^{-1})-\lambda q_{0}
\end{align}
and  $f_{\text{imp}}^{}$   naturally assumes the form  
\begin{align}
\label{eq:LGT}
    f_{\text{imp}}^{}&=\Phi[G_f,G_B,\lambda]
    -T\text{Tr}{G_f \Sigma_f}+T\kappa \text{Tr}{G_B \Sigma_B} \nonumber \\
    &+T\kappa \text{Tr}{\ln\big ({- G_B^{-1}}\big)}-T\text{Tr}{\ln\big({- G_f}^{-1}\big)}-\lambda q_{0}.
\end{align}

\begin{figure}[t!]
\centering
\includegraphics[width=0.73\columnwidth]{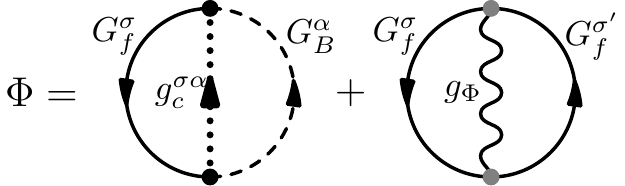}\\
\caption{Leading order contribution. Each vertex contains factor of $\sqrt{\frac{1}{N}}$. Summation is assumed for all $\sigma, \alpha$ which makes the diagram order $N$.}
\label{fig:fig_fey}
\end{figure}

Recognizing $\Phi$ as the Legendre transform of $f_{\text{imp}}$, the stationary condition in Eq.\eqref{eq:saddlePoint} translates to 
\begin{align}
    \label{eq:LWf}
     \dfrac{\delta \Phi}{\delta G_{a}}=\eta_a\Sigma_{a},~(a=f,B),
\end{align}
where $\eta_f=1$ and $\eta_B=-1$. Equation (\ref{eq:LWf})  identifies $\Phi$, depicted in Fig.~\ref{fig:fig_fey},  as the corresponding Luttinger-Ward functional. Our derivation shows that a rigorous  large-$N$ limit is equivalent to a conserving approximation in the Kadanoff-Baym sense\cite{Baym.61,Coleman.05,Rech.06}, see also App.~\ref{app:KBA}. Identifying Luttinger-Ward functionals through associated saddle point limits is one way of constructing such functionals in a non-perturbative manner \cite{Potthoff.06}. 
%The variation of Eq.\eqref{eq:LGT} with respect to $G_i$ yields the Dyson equation, so that

The relation between $G_a$ and $\Sigma_a $  is given by the Dyson  equation $G_{a}^{-1} = g_{a}^{-1} - \Sigma_a$, with $a=B,f$ and where the bare propagators are given by  $g_{B}=-J_{K}$ and $g_{f}=\frac{1}{i\omega-\lambda}$.  
As $B$ is an Hubbard-Stratonovich decoupling field,  its bare part only depends on the coupling constant and any dynamics has to be acquired through interaction effects. 
In addition to Eqs. (\ref{eq:Sigma_f}, \ref{eq:Sigma_B}), $G_{f}$ has to respect the constraint $G_{f}(\tau\rightarrow0^{-})=q_{0}$, which is enforced through the Lagrange multiplier $\lambda$.
For the relation between the free energy and the saddle point action $S$ one finds $f=S/\beta$.
%%%%%%%%%%%%%%%%%%%%%%%%%%%%%%%%%%%%%%%%%
%%%%%%%%%%%%%%%%%%%%%%%%%%%%%%%%%%%%%%%%%%%
From the entropy
\begin{equation}\begin{split}
	s_{\text{full}}=&-N\frac{d f^{DLN}}{d T}
\end{split}\end{equation}
we can split off the impurity contribution, defined by $s_\text{imp}= s_{\text{full}} - s_0$, where $s_{\text{full}}$ is the total entropy of the system  and $s_0$ is the contribution of the bath in the absence of the impurity. 
Note that, defined as a difference, $s_{\text{imp}}$ does not need to obey the second law of thermodynamics.
In the remainder, we will omit the subscript on $s_\text{imp}$, {\itshape i.e.}, $s=s_\text{imp}$.
As detailed in Appendix. \ref{app:Details}, we obtain 
\begin{equation}\begin{split}
	s=&-\int \frac{d\omega}{\pi} \left\{  \partial_T n_{b}(\omega) \kappa\mathrm{Im}[\ln(-G_{B}^{-1}(\omega))]\right.\\
	 &+n_{b}(\omega)\kappa\mathrm{Im}[-G_{B}(\omega)\partial_{T}A(\omega)]\\
   &+\partial_T n_{f}(\omega) \mathrm{Im}[\ln(-G_{f}^{-1}(\omega)) ]\\
 &+n_{f}(\omega)\mathrm{Im}[G_{f}(\omega)\partial_{T}B(\omega)]\\
  &\left.+ \partial_T n_{f}(\omega)\mathrm{Im}[\Sigma_{f}^{1}(\omega)G_{f}(\omega)+\frac{1}{2}\Sigma_{f}^{2}(\omega)G_{f}(\omega)] \right\},
		\label{eq:entropy}
\end{split}\end{equation}
where the imaginary parts of $\partial_{T}A(\omega)$ and $\partial_{T}B(\omega)$ are 
\begin{align}
\label{eq:entropy-formula}
	\mathrm{Im}\,\partial_{T}A(\omega)=&\int \frac{d\omega'}{\pi} G_{f}''(\omega')\\ \times & g_{c}''(\omega'-\omega)\partial_T [n_{f}(\omega')-n_{f}(\omega'-\omega)]\nonumber \\
\mathrm{Im}\,	\partial_{T}B(\omega)=&\int\frac{d\nu}{\pi}g^{2}g_{\Phi}''(\nu)\\ \times & G_{f}''(\omega-\nu) \partial_T[n_{f}(\nu-\omega)+n_{b}(\nu)] \nonumber
\end{align}
and the reals parts are obtained through Kramers-Kronig relations.
%%%%%%%%%%%%%%%%%%%%%%%%%%%%%%%%%%%%%%%%%%%%%%%%%%%%%%%%%%%%%
%%%%%%%%%%%%%%%%%%%%%%%%%%%%%%%%%%%%%%%%%%%%%%%%%%%%%%%%%%%%%
\subsection{Scaling}
\label{sec:scaling}
Asymptotically exact results for the frequency behavior of $G_f$ and  $G_B$ in the $T=0$ limit can be obtained through a scaling ansatz. Following Parcollet {\it et al.} \cite{Parcollet.98}, we set 
\begin{align} \label{eq:scaling_ansatz-1}
	G_f(\tau)&=-A_1\Big( \dfrac{ \tau_0}{\tau}\Big)^{\alpha_{f}}-A_2\Big( \dfrac{ \tau_0}{\tau}\Big)^{\alpha_{f}'}+\ldots,\\
	G_B(\tau)&=-B_1\Big( \dfrac{ \tau_0}{\tau}\Big)^{\alpha_{B}}-B_2\Big( \dfrac{ \tau_0}{\tau}\Big)^{\alpha_{B}'}+\ldots,
	\label{eq:scaling_ansatz-2}
\end{align}
%to next-to-leading order in $1/\tau$, where we consider $\alpha_1<\alpha_2<...$ and $\beta_1<\beta_2<...$ 
valid at $T=0$ and for $\tau \gg \tau_0$, where $\tau_0$ is a short-time cutoff. 
The real-frequency counterpart of these expressions is obtained via analytic continuation of the Fourier transform as outlined in  Appendix \ref{app:FT}. 
For  $G_a(\tau)=-(\frac{1}{\tau})^{\alpha_{a}}$, ($a=f$ or $a=B$), one finds in the $T=0$ limit, 
\begin{align}
	G_B(\omega+i0^+,\beta\rightarrow \infty)= -\dfrac{\pi \tau_0^{\alpha_{B}} }{\Gamma (\alpha_{B})} X^B_{\alpha_{B}}  |\omega|^{\alpha_{B}-1},
\end{align}
\begin{align}
	G_f(\omega+i0^+,\beta\rightarrow \infty) =  -\dfrac{\pi \tau_0^{\alpha_{f}} }{\Gamma(\alpha_{f})}  X^f_{\alpha_{f}}  |\omega|^{\alpha_{f}-1},
\end{align}
where $X^B_{\alpha_{B}}=\tan\big(\frac{\pi \alpha_{B}}{2}\big)+i\, \text{sgn}(\omega)$ and $X^f_{\alpha_{f}}=-\cot\big(\frac{\pi \alpha_{f}}{2}\big)\text{sgn}(\omega)+i$
[see \eqref{eq:appFTboson} and \eqref{eq:appFTfermion}].

Inserting these expressions into the saddle point equations, Eqs.~(\ref{eq:sc1}-\ref{eq:sc4}),  one obtains $\alpha_{f}$ and $\alpha_{B}$ (see Appendix \ref{App:SA}). For the special case $r=0$, similar results have been reported in \cite{Zhu.04,Kirchner.05}. Our results for $\alpha_{f}$ and $\alpha_{B}$ reduce to those of Ref.~\cite{Vojta.01} for the pure pseudogap Kondo model at large N. Moreover, the large-N exponent for the $t$-matrix of the pseudogap Kondo model at the multichannel Kondo fixed point agrees with that of the SU(2)-symmetric pseudogap Kondo model at its strong-coupling fixed point \cite{Bulla.97,Gonzalez-Buxton.98}. Our results for the leading behavior, {\itshape i.e.}, $\alpha_{f}$ and $\alpha_{B}$, are summarized in Table \ref{tab:exponents}.
\begin{table}[h]
	\begin{tabular}{|c|c|c|}
		\hline   
		C & $r+\alpha_{f}=1-\alpha_{B}$\\
		\cline{1-1}
		MCK &$\kappa=\frac{(1-\alpha_{f})\tan(\pi \alpha_{f}/2)}{(r+\alpha_{f})\tan(\pi (r+\alpha_{f})/2)}$ \\
		\hline   
		C'&$\alpha_{f}=\epsilon/2$, $\alpha_{B}=1-(r+\epsilon/2)$\\
		\hline   
		LM'&$\alpha_{f}=\epsilon/2$, $\alpha_{B}=1+(r+\epsilon/2)$\\
		\hline  
	\end{tabular}
	\caption{Scaling exponents of $G_B$ and $G_f$ at the various fixed points.}
	\label{tab:exponents}
\end{table}
From the scaling ansatz for $G_f$ and $G_B$ one also obtains
$\chi(\omega) \sim |\omega|^{2\alpha_f-1}$ and $\mathcal{T}(\omega) \sim |\omega|^{\alpha_f+\alpha_B-1}$ to leading order.
%%%

In the pseudogap BFKM,  both fermionic and bosonic baths break conformal symmetry but for $r=0$ and in the absence of a bosonic bath ($g=0$), the Hamiltonian possesses  conformal symmetry. The resulting invariance  can be used to extend the leading $\omega$ behavior scaling ansatz to the leading $T$ dependence. One obtains
\begin{align}
\label{eq:scaling_ansatz_2}
	G_f(\tau,\beta)&= \Big(  \dfrac{\tau_0 T }{\sin{ \pi \tau T}}\Big )^{\alpha_{f}}+\ldots
\end{align}
and likewise for $G_B$.
The  $\dots$ in Eq.(\ref{eq:scaling_ansatz_2}) stand for subleading corrections.% that vanish for $\omega/\Lambda,T/\Lambda \to 0$. 
This form for $G_f$ and $G_B$, fully determines low energy properties, in particular, the fixed point (or zero-temperature) entropy \cite{Parcollet.98}. 
%\begin{align} \label{eq:scaling_ansatz_2}
%    G_f(\tau)&= - A_1 \Big( \dfrac{\pi \tau_0}{ \beta \sin{ \pi \tau /\beta } }\Big)^{\alpha_1} + ... \\
%     G_B(\tau)&= - B_1 \Big( \dfrac{\pi \tau_0}{ \beta \sin{ \pi \tau /\beta } }\Big)^{\beta_1} + ...
%\end{align}
Equation.(\ref{eq:scaling_ansatz_2}) implies $\omega/T$ scaling, {\itshape i.e.},
\begin{align}
\label{eq:w/T}
& G_f(\omega,T)=T^{\alpha_{f}-1}\Phi_f(\omega/T) + \dots
%\\
%& G_B(\omega,T)=T^{\beta_1-1}\Phi_B(\omega/T),
\end{align}
in the scaling regime, 
provided $\alpha_{f}<1$ which is necessary for the Fourier integral to converge [corresponding statements apply to $G_B(\omega,T)$]. It follows from Eqs.~\eqref{eq:scaling_ansatz_2} and \eqref{eq:w/T} that $G_f(\omega,T=0)\sim \omega^{\alpha_{f}-1}$ and $G_f(\omega=0,T)\sim T^{\alpha_{f}-1}$. The marginal case $\alpha_{f}=1$, which is, {\itshape e.g.}, relevant for the standard Kondo problem,   where the strong-coupling fixed point is described by a boundary conformal field theory, requires an extra energy scale in order to regularize the  Fourier transform (see Appendix \ref{app:FT}). This energy scale can be identified with the Kondo temperature $T_K$. As a result, a somewhat trivial $\omega/T$ scaling ensues in this case  in the limit $T\ll T_K$ and $\omega \ll T_K$.
%\textcolor{blue}{come back to this issue}%However, away from boundary conformal  fixed points, the knowledge of the critical exponents $\alpha_1$ and $\beta_1$ is not enough to predict all low-energy properties. In addition the functional form of the scaling function, $\Phi_f(\omega/T)$, needs to be provided.  

In what follows, we will pay particular attention to the terms represented by the ellipses in Eq.\eqref{eq:scaling_ansatz_2}. The $T=0$ form of these subleading terms can be determined from the saddle-point equations in a fashion analogous to the leading behavior (see Appendix \ref{App:SA}).  As far as the extension to  $T\neq0$ is concerned, a form reminiscent of Eq.~\eqref{eq:w/T} may apply to the subleading terms as well, albeit with a different scaling exponent $\alpha_{f}'>\alpha_{f}$. Even in that case will the sum of
leading and subleading terms together not be of the form of Eq.~\eqref{eq:w/T}.
In other words, the subleading terms necessarily break the $\omega/T$ scaling form of Eq.~\eqref{eq:w/T}.
%Note, also, that the subleading terms represented by $\ldots$ in Eq. \eqref{eq:w/T} may by itself display $\omega/T$ scaling. However, the sum of the two, {\itshape i.e.}, the leading and subleading term together, will not be of that form.
%conformal scaling and this scaling form implies w/T scaling. 
%Show the scaling form\\
%In the present case, both fermionic and bosonic bath break conformal symmetry. As will be shown below, we find w/T scaling at all non-trivial fixed points.
%-subleading may have w/T scaling but the sum of leading and subleading necessarily breaks w/T scaling.
%-perhaps refer to appendix and subleading results.

%%%%%%%%%%%%%%%%%%%%%%%%%%%%%%%%%%%%%%%%%%%%%%%%%%%%%%%%%%%%%
%%%%%%%%%%%%%%%%%%%%%%%%%%%%%%%%%%%%%%%%%%%%%%%%%%%%%%%%%%%%%
\subsection{Numerical solutions}

A numerical solution of the
large-N equations (\ref{eq:sc1}-\ref{eq:Sigma_B}) for given set  \{$r$, $\alpha_\phi$, $J_K$, $g$,$D$, $\Lambda$\} of parameters can be obtained iteratively at $T\neq 0$ and all $\omega$. This is accomplished by Fourier transforming the saddle-point equations to Matsubara space followed by analytic continuation to real frequencies. The self-consistent equation for the self-energies  follow as
\begin{align}
	\Sigma_{f}^{''}(\omega)=&-\kappa\int_{-\infty}^{+\infty}\frac{dx}{\pi}G_{B}^{''}(x)g_{c}^{''}(\omega-x) \nonumber \\ 
	&~~~~~~~~\times [n_{b}(x)+n_{f}(x-\omega)] \nonumber \\
		-&2g^{2}\int_{-\infty}^{+\infty}\frac{dx}{\pi}g_{\Phi}^{''}(x)G_{f}^{''}(\omega-x)\nonumber \\ 
	&~~~~~~~~\times	[n_{b}(x)+n_{f}(x-\omega)], \\
	\Sigma_{B}^{''}(\nu)&=\int_{-\infty}^{+\infty}\frac{dx}{\pi}G_{f}^{''}(x)g_{c}^{''}(x-\nu) \nonumber \\ 
	&~~~~~~~~\times [n_{f}(x)-n_{f}(x-\nu)].
\end{align}

%\begin{equation}\begin{split}
	%X=\int_{-\infty}^{+\infty}\frac{dx}{\pi}\{G_{a}^{''}(x)G_{b}''(\sigma_{b}\nu-\sigma_{a}\sigma_{b}x)[\eta_{a}\sigma_{b}f_{\eta_{a}}(x)-\sigma_{b}\eta_{b}f_{\eta_{b}}(x-\sigma_{a}\nu)]
%\end{split}\end{equation}

In order to resolve the $T=0$ power-law divergences  of  $G_f(\omega,T)$ and $G_B(\omega,T)$, a logarithmically dense energy mesh is used. To improve convergence of the self consistency problem,  a modified Broyden's scheme  is employed\cite{Zitko.09}.
In this work, the criterion used for convergence is that the frequency integral over the absolute value of the difference of two solutions of two consecutive iterations has to be less than $10^{-5}$.
Once convergence has been reached, the impurity entropy $s(T)$, $t$-matrix $\mathcal{T}(\omega,T)$, and local spin susceptibility $\chi(\omega,T)$
can be obtained from $G_f(\omega,T)$ and $G_B(\omega,T)$. 

%%%%%%%%%%%%%%%%%%%%%%%%%%%%%%%%%%%%%%%%%%%%%%%%%%%%%%%%%%%%%%%%%
%%%%%%%%%%%%%%%%%%%%%%%%%%%%%%%%%%%%%%%%%%%%%%%%%%%%%%%%%%%%%%%%%
\subsection{Zero-temperature entropy}
\label{subsec:ZTE}

As  our goal is to test the validity of the $g$-theorem for the pseudogap BFKM, the fixed point value of the entropy is required at all fixed points across the phase diagram. 
The expression in Eq. (\ref{eq:entropy}) simplifies in the  $T=0$ limit provided the local  Green functions, {\itshape i.e.}, $G_f$ and  $G_B$,  display $\omega/T$ scaling, in which case  the only contribution to the impurity entropy comes from the logarithmic terms in Eq. (\ref{eq:entropy}).
As will be demonstrated below, the local Green functions indeed obey $G_a(\omega,T)=T^{\alpha_{a}-1}\Phi_a(\omega/T)$, ($a=B,f$ in this equation distinguishes between the bosonic and fermionic Green function) at all fixed points except the weak-coupling fixed point (LM).
%We can also obtain another formula for zero temperature entropy if all Green's function has $\omega/T$ scaling.

In this case, the leading part of the free energy, for small $T$, is given by 
\begin{align}
	f_{\text{imp}}=&\kappa \int_{-\infty}^{0} \dfrac{d\omega}{\pi} \Bigg[a_{B}\left(\dfrac{\omega}{T} \right)- a_{B}(-\infty) \nonumber \\
	&+2\kappa n_{b}(-\omega)a_{B}\left(\dfrac{\omega}{T} \right)\Bigg] \nonumber\\
	 &+\int_{-\infty}^{0} \dfrac{d\omega}{\pi} \Bigg[a_{f}\left(\dfrac{\omega}{T} \right)-a_{f}(-\infty)\nonumber \\ &-2n_{f}(-\omega)a_{f}\left(\dfrac{\omega}{T} \right)\Bigg],
\end{align}
with
\begin{align*}
a_{f}(\omega/T)&=\arctan\dfrac{G_{f}'(\omega,T)}{G_{f}''(\omega,T)},\\
a_{B}(\omega/T)&=\arctan\dfrac{G_{B}''(\omega,T)}{G_{B}'(\omega,T)}.
\end{align*}
In terms of these functions, the $T=0$ limit of the entropy follows as
\begin{align}
	\label{eq:Sa}
	s=&-\int_{-\infty}^{0} \dfrac{d\tilde{\omega}}{\pi}[\kappa a_{B}(\tilde{\omega})-\kappa a_{B}(-\infty)+2\kappa n_{b}(-\tilde{\omega})a_{B}(\tilde{\omega}) \nonumber \\
			   &+a_{f}(\tilde{\omega})-a_{f}(-\infty)-2n_{f}(-\tilde{\omega})a_{f}(\tilde{\omega})],
\end{align}
where $n_B$ and $n_f$ are the bosonic and fermionic distribution functions.

%%%%%%%%%%%%%%%%%%%%%%%%%%%%%%%%%%%%%%%%%%%%%%%%%%%%%%%%%%%
%%%%%%%%%%%%%%%%%%%%%%%%%%%%%%%%%%%%%%%%%%%%%%%%%%%%%%%%%%%
\section{Results}
\label{sec:results}
We now turn to a discussion of the numerical results, obtained from a self-consistent solution of the saddle-point equations (\ref{eq:Sigma_f}) and (\ref{eq:Sigma_B}), for $T\neq 0$.

\subsection{Green functions}
The local Green functions $G_B$ and $G_f$ display power-law behavior at $T=0$ and low frequency, {\itshape i.e.}, $G_{f,B}(\omega,T=0)\sim\omega^{\alpha_{f,B}-1}$ near the fixed points (C, C', LM', MCK).
$\alpha_B$ and $\alpha_f$ can be obtained from the scaling ansatz, Eqs.~(\ref{eq:scaling_ansatz-1}),(\ref{eq:scaling_ansatz-2}) and are
listed in table \ref{tab:exponents}. 
In the following we focus on a representative set of $\kappa$, $r$ and $\epsilon$ to discuss  our results. We choose $\kappa=1/2$, $r=1/4$ and $\epsilon=2/5$  which corresponds to exponents  $\alpha_{f}=(0.052,0.2,0.2,0.338)$ and $\alpha_{B}=(0.698,0.55,1.45,0.412)$ for fixed points (C, C', LM', MCK), respectively.
%%%%%%%%%%%%%%%% Figure 3 %%%%%%%%%%%%%%%%%%%%%%%%%%%%%%%
\begin{figure}[ht]
	\begin{picture}(0,0)
		\put(-90,-23){\textsf{(a)}}
		\put(34,-23){\textsf{(b)}}
	\end{picture}
	\center
	\includegraphics[width=0.49\columnwidth]{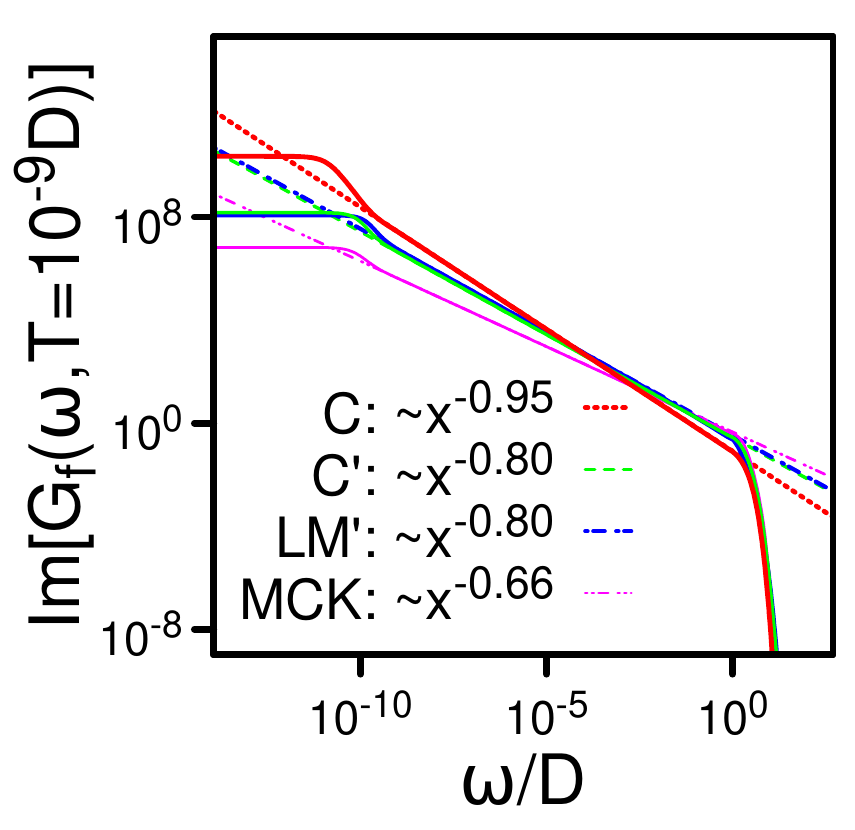}
	\includegraphics[width=0.49\columnwidth]{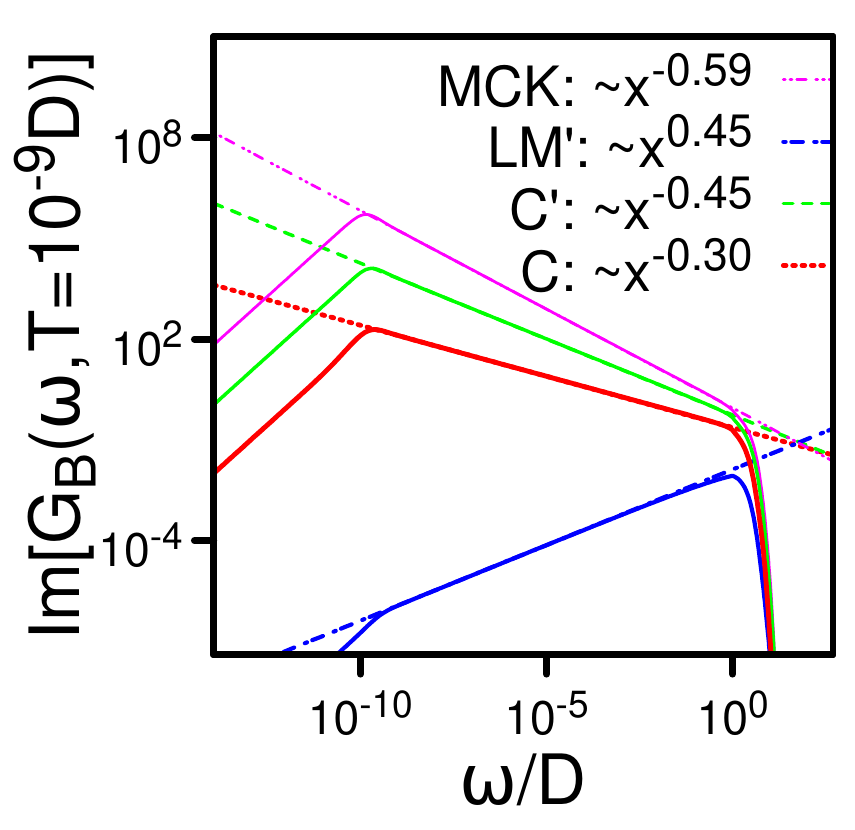}\\
	\caption{Numerical results of Green's function with leading power-law fitting at different fixed points. (a) $\mathrm{Im}[G_{f}(\omega,T=10^{-9}D)]$, (b) $\mathrm{Im}[G_{B}(\omega,T=10^{-9}D)]$.}
	\label{fig:omega-behavior}
\end{figure}
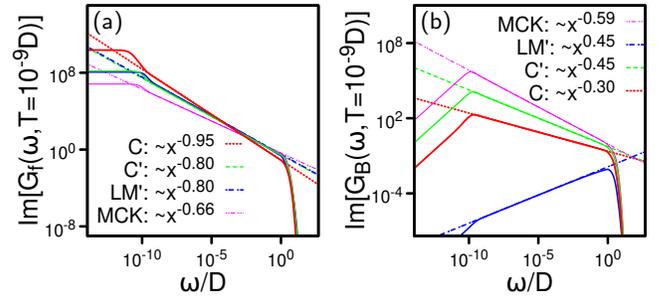
%%%%%%%%%%%%%%%%%%%%%%%%%%%%%%%%%%%%%%%%%%%%%%%%%%%%%%%%%
At $T\neq 0$, by transforming Eq.~\eqref{eq:scaling_form} into real frequencies (see Appendix \ref{app:FT}), one obtains
\begin{eqnarray}
\label{eq:Gf-FT}
G_{f}(\omega)&=&-\left(\frac{2\pi}{\beta}\right)^{\alpha_{f}-1}\tau_{0}^{\alpha_{f}}\mathrm{B}[\frac{\alpha_{f}}{2}-\frac{i\beta\omega}{2\pi},\frac{\alpha_{f}}{2}+\frac{i\beta\omega}{2\pi}]\nonumber\\
			 &&(-\cot{\frac{\pi\alpha_{f}}{2}}\sinh{\frac{\beta\omega}{2}}+i\cosh{\frac{\beta\omega}{2}})\\
G_{B}(\omega)&=&-\left(\frac{2\pi}{\beta}\right)^{\alpha_{B}-1}\tau_{0}^{\alpha_{B}}\mathrm{B}[\frac{\alpha_{B}}{2}-\frac{i\beta\omega}{2\pi},\frac{\alpha_{B}}{2}+\frac{i\beta\omega}{2\pi}]\nonumber\\
			 &&(\tan{\frac{\pi\alpha_{B}}{2}}\cosh{\frac{\beta\omega}{2}}+i\sinh{\frac{\beta\omega}{2}}),
	\label{eq:GB-FT}
\end{eqnarray}
%%%%%%%%%%%%%%%% Figure 4 %%%%%%%%%%%%%%%%%%%%%%%%%%%%%%%
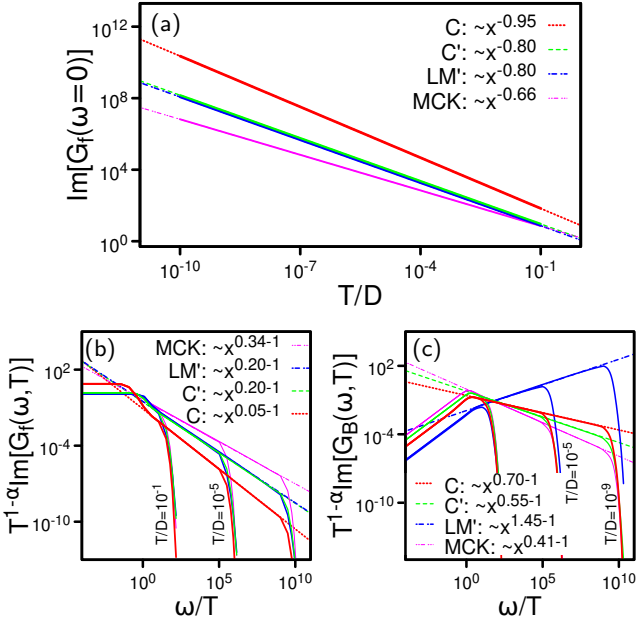
\begin{figure}[h!]
	\begin{picture}(0,0)
		\put(-65,-25){\textsf{(a)}}
		\put(-90,-148){\textsf{(b)}}
		\put(34,-148){\textsf{(c)}}
	\end{picture}
	\center
	\includegraphics[width=0.82\columnwidth]{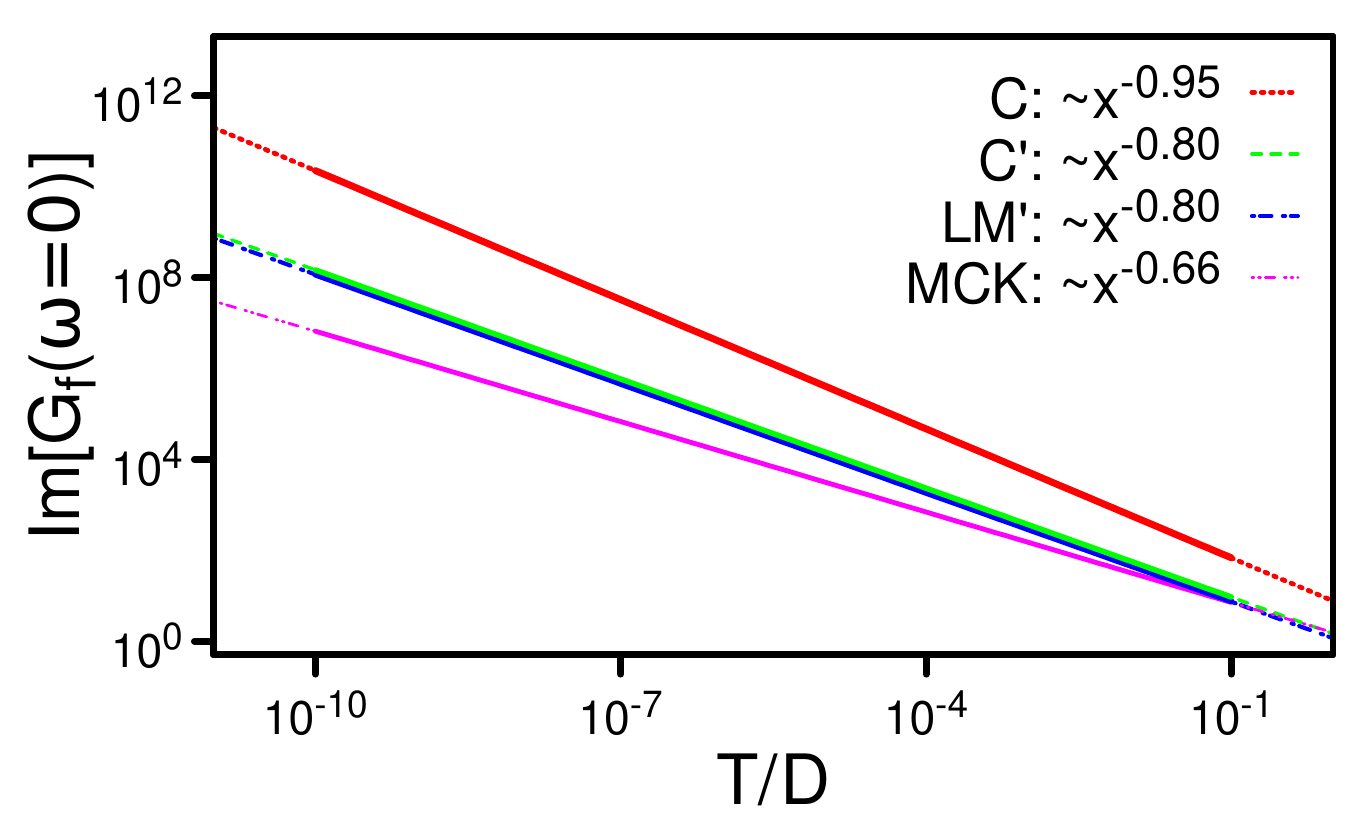}\\
	\includegraphics[width=0.485\columnwidth]{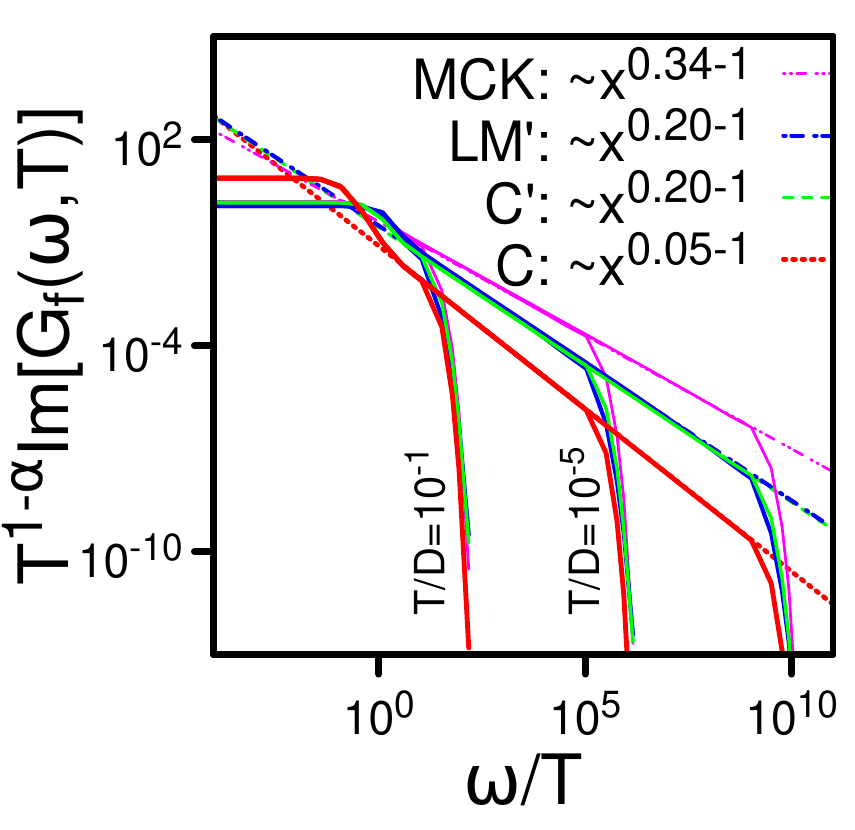}
	\includegraphics[width=0.485\columnwidth]{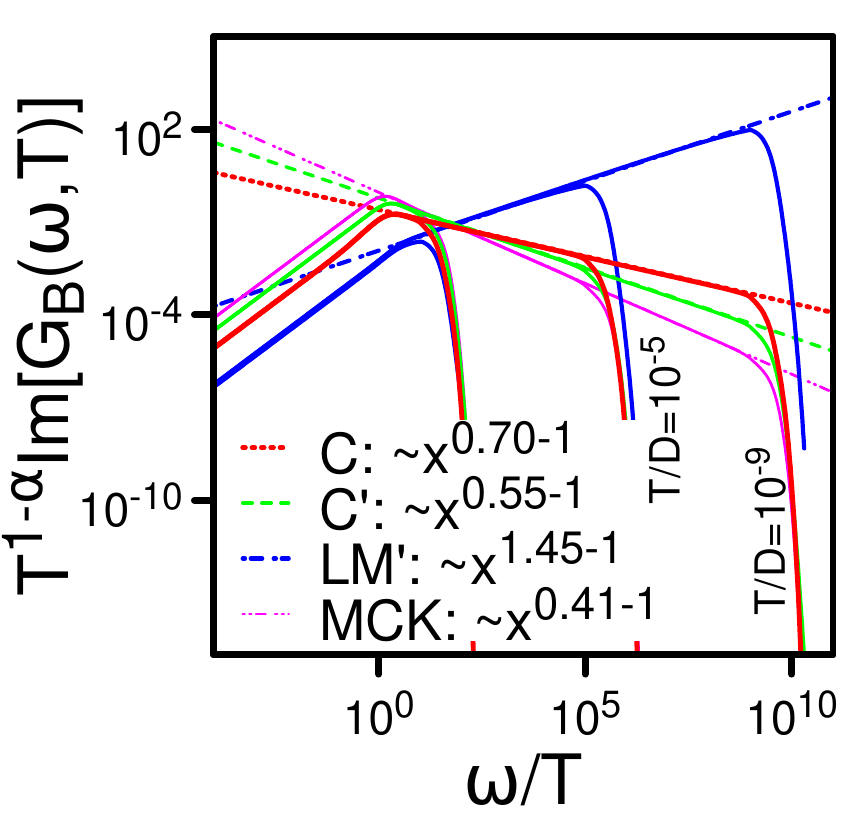}
	\caption{(a) Temperature dependency of $\mathrm{Im}[G_{f}(\omega=0)]$ and the power law fitting at different fixed points. (b), (c) $\omega/T$ scaling behavior of $G_{f}(\omega)$ and $G_{B}(\omega)$ with the exponent obtained from scaling ansatz.}
	\label{fig:T-behavior_and_omegaT}
\end{figure}
%%%%%%%%%%%%%%%%%%%%%%%%%%%%%%%%%%%%%%%%%%%%%%%%%%%%%%%%%
where $B(x,y)$ is the Euler Beta function and a numerical prefactor, equivalent to $A_1$ of Eq.(\ref{eq:app-scal-ansatz1}) and $B_1$ of Eq.(\ref{eq:app-scal-ansatz2}) has been set to one.
Equations (\ref{eq:Gf-FT}) and (\ref{eq:GB-FT}) can be compared with the numerical solution of the saddle-point equations, Eqs.(\ref{eq:Sigma_f}) and (\ref{eq:Sigma_B}), for $T\neq 0$.
First, we establish that the low-$T$ behavior of $G_f(\omega,T)$ and $G_B(\omega,T)$ is in line with the results obtained from the scaling ansatz for $T=0$ and  $\omega\rightarrow 0$.

Figure \ref{fig:omega-behavior} displays Im$[G_f(\omega,T=10^{-9}D)]$ and Im$[G_B(\omega,T=10^{-9}D)]$ ($D$ is defined in Eq.~(\ref{eq:conductelec})) near the intermediate coupling fixed points. Evidently, both Green functions display power-law behavior for $\omega>T$ and below some high-energy cutoff $T_K^*$ which can be identified with min$[\Lambda,T_K^0]$, where $T_K^0$ is the Kondo temperature associated with the Kondo model with $g=0$, $r=0$ and all other parameters left unchanged. It follows from the Kramers-Kronig relation that the real parts of $G_f$ and $G_B$ feature corresponding power-law behavior. Clearly, the numerical results confirm the conclusions drawn from the scaling ansatz. 
%%%%%%%%%%%%%%%% Figure 5 %%%%%%%%%%%%%%%%%%%%%%%%%%%%%%
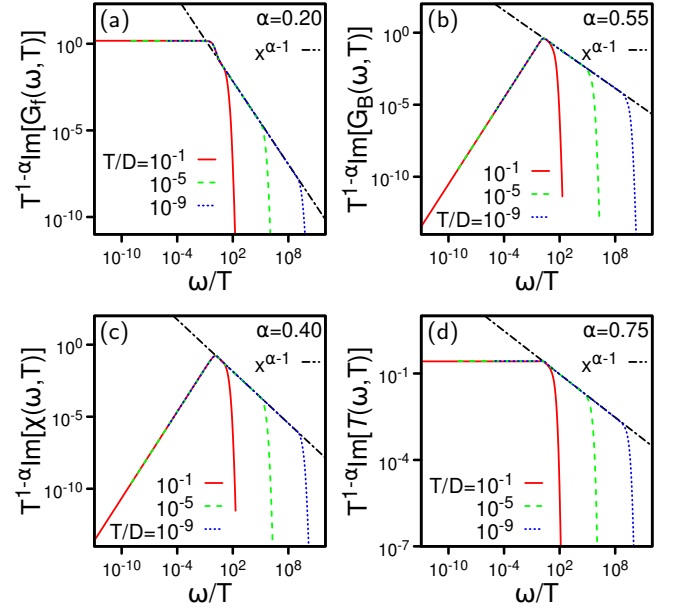
\begin{figure}[ht]
	\begin{picture}(0,0)
		\put(-90,-23){\textsf{(a)}}
		\put(34,-23){\textsf{(b)}}
		\put(-90,-141){\textsf{(c)}}
		\put(34,-141){\textsf{(d)}}
	\end{picture}
	\center
	\includegraphics[width=0.49\columnwidth]{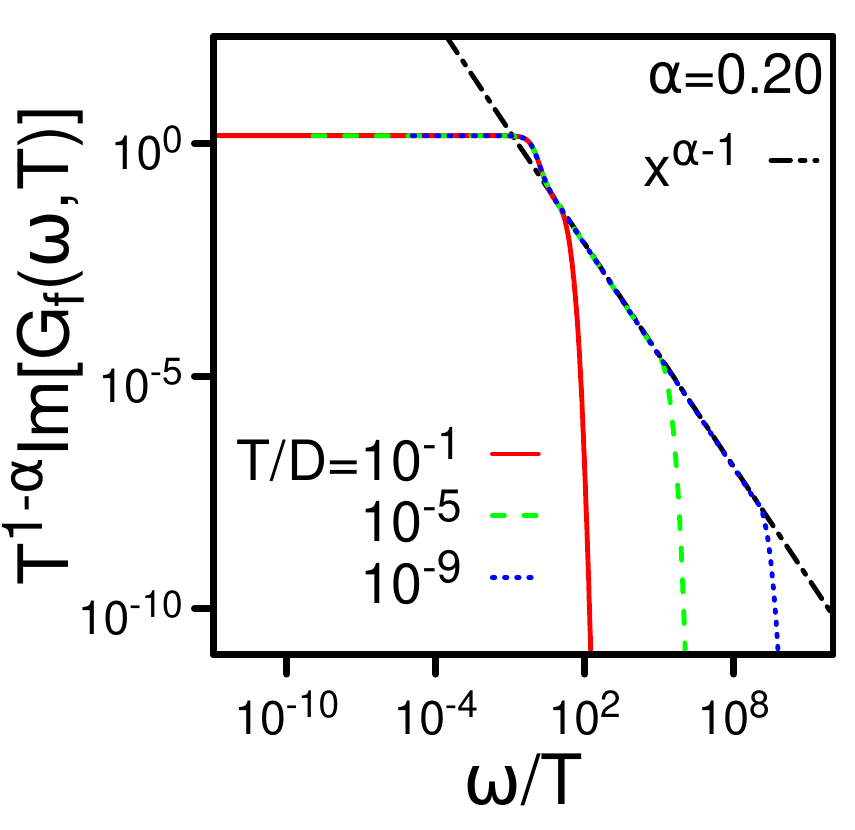}
	\includegraphics[width=0.49\columnwidth]{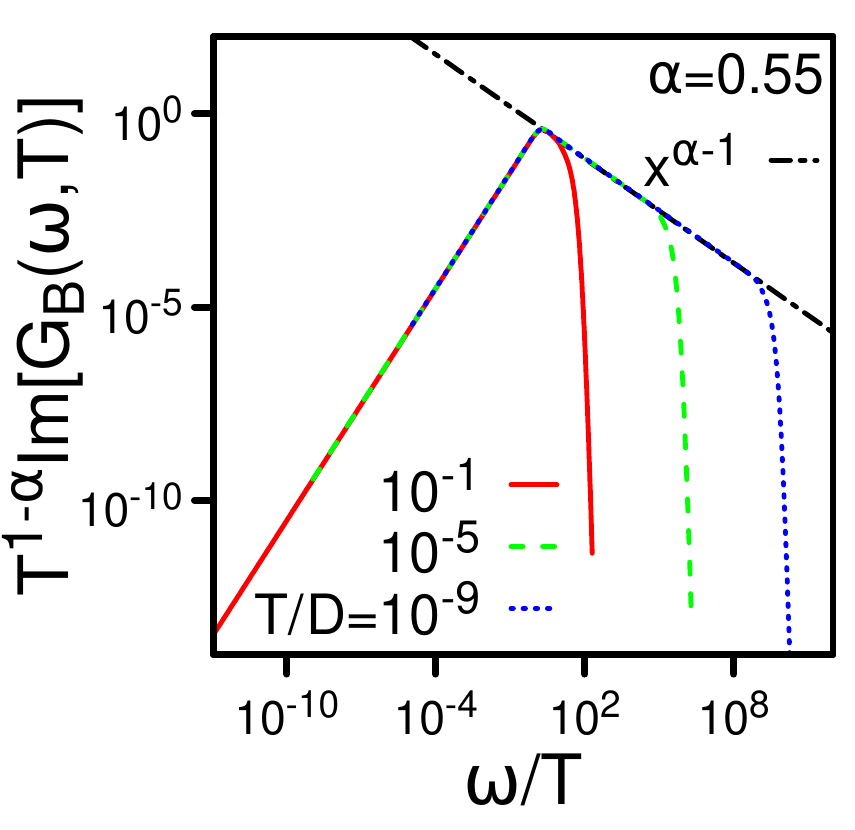}\\
	\includegraphics[width=0.49\columnwidth]{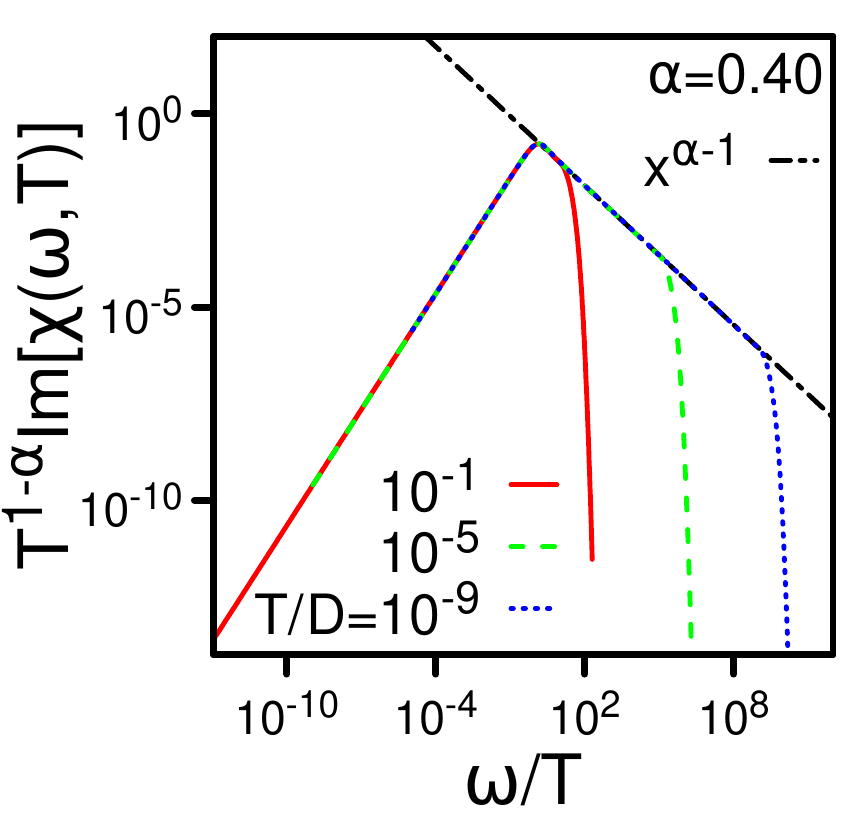}
	\includegraphics[width=0.49\columnwidth]{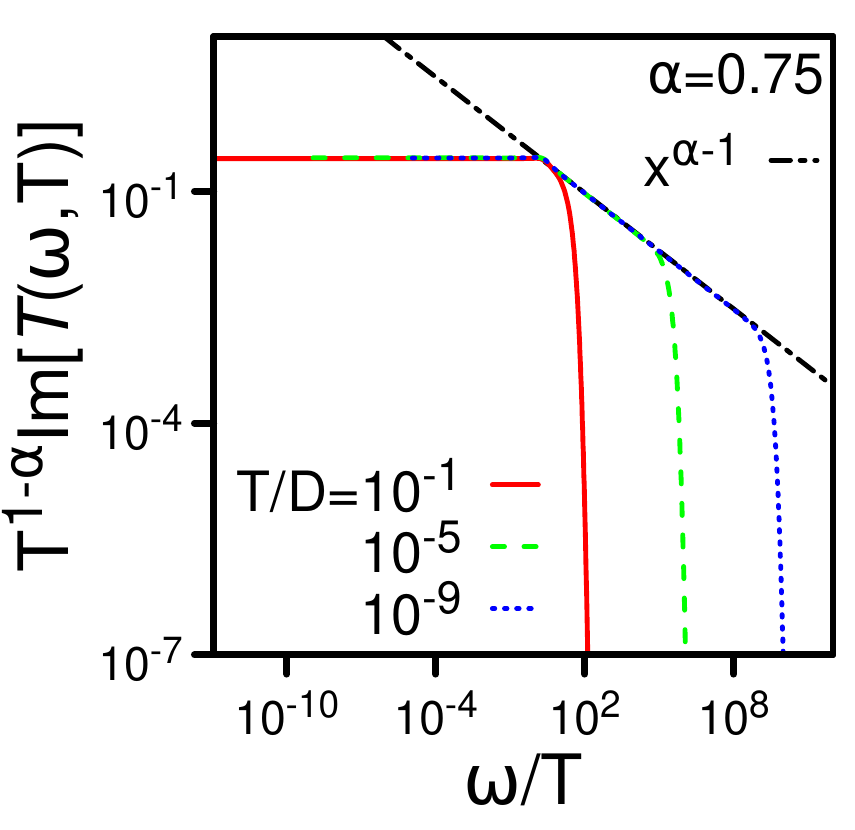}
	\caption{Numerical results of $\omega/T$ scaling at C' fixed point for (a) $G_{f}(\omega)$. (b) $G_{B}(\omega)$, (c) $\chi(\omega)$, (d) $\mathcal{G}(\omega)$. The parameters are $r=1/4, \epsilon=2/5, \kappa=1/2$}
	\label{fig:scaling-C'}
\end{figure}
%%%%%%%%%%%%%%%%%%%%%%%%%%%%%%%%%%%%%%%%%%%%%%%%%%%%%%%%%

From the numerical results we can obtain the $T$ behavior of the Green functions $G_f$ and $G_B$ and consequently obtain the static susceptibility from $\chi_{\mbox{\tiny stat}}(T)=\mbox{Re}[\chi(\omega=0,T)]$. Figure \ref{fig:T-behavior_and_omegaT}(a) shows the $T$ dependence of $G_f(\omega=0,T)$. Power-law behavior is found in the scaling regime associated with each of the intermediate coupling fixed points. The observed power law in $T$ is compatible with the $\omega$-behavior of $G_f(\omega,T=0)$ and points towards $\omega/T$ scaling. This is indeed observed as shown in Fig. \ref{fig:T-behavior_and_omegaT}(b) for $G_f$ and Fig. \ref{fig:T-behavior_and_omegaT}(c) for $G_B$ near the fixed points C, C', LM' and MCK. In each case, we find that the Green function $G_a(\omega,T)$ ($a=f$ or $B$) obeys
\begin{align}
\label{eq:omega-ov-T}
     G_a(\omega,T)=T^{\alpha_{a}-1}\Phi_f(\omega/T) 
\end{align} 
to leading order and with a scaling exponent that agrees within numerical uncertainty with the corresponding exponent from Table \ref{tab:exponents}. From the scaling behavior of  $G_f$ and $G_B$ one can infer a related  scaling  for $\mathcal{T}$ and $\chi$. This is explicitly demonstrated  in Fig.\ref{fig:scaling-C'} for the critical point C'.

\begin{figure}[ht]
	\begin{picture}(0,0)
		\put(-90,-38){\textsf{(a)}}
		\put(34,-38){\textsf{(b)}}
	\end{picture}
\center
\includegraphics[width=0.99\columnwidth]{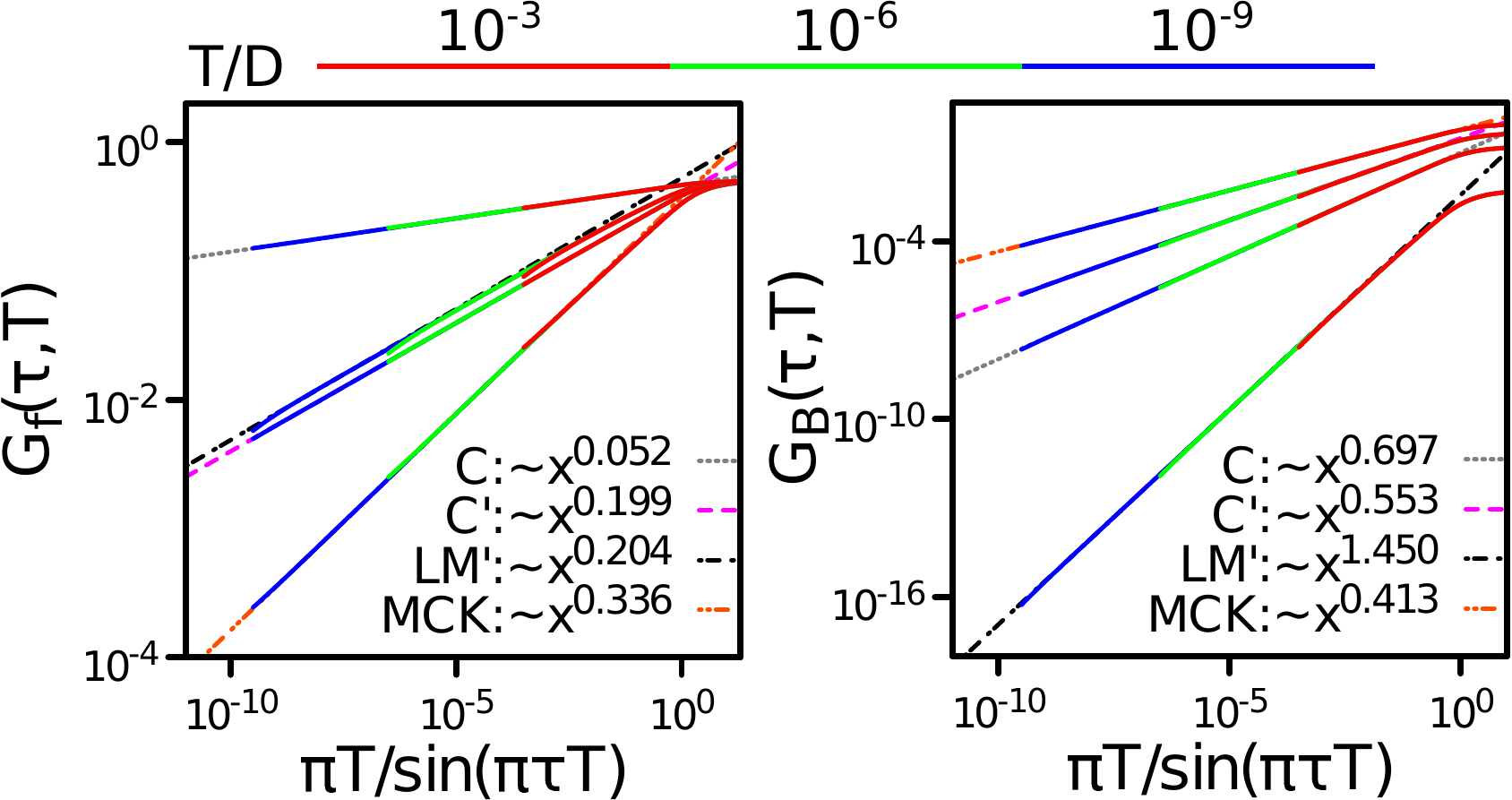}
\caption{Numerical results of Green's function with leading power law fitting at different fixed points. (a) $G_{f}(\tau,T)$. (b) $G_{B}(\tau,T)$ at the intermediate fixed points C, C', LM' and MCK. The dashed lines indicate the leading power-law behavior. The exponent is in line with $\alpha_f$ and $\alpha_B$ from Table \ref{tab:exponents}.}
	\label{fig:tau-behavior}
\end{figure}
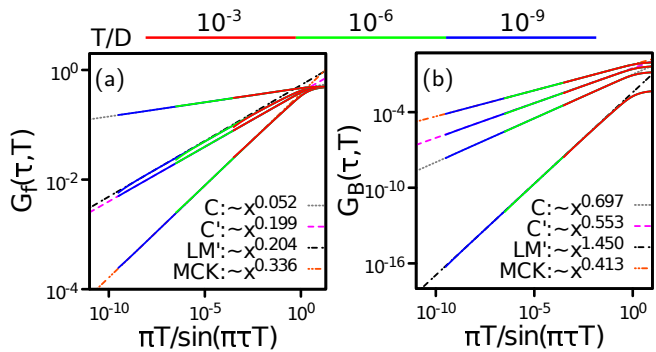

The imaginary-time ($\tau)$ dependence of the various correlation functions is obtained from 
\begin{align}
\label{eq:omega-to-tau}
    \Phi(\tau,T)=-\eta \int d\!\omega \dfrac{e^{-\omega\tau}}{e^{-\omega/T}-\eta} \mbox{Im}[\Phi(\omega+i0^{+},T)],
\end{align}
where $0<\tau\leq 1/T$ and $\eta=-$ ($\eta=+$) for a fermionic (bosonic) correlation function. The $\tau$ dependence of $G_f(\tau,T)$ and $G_B(\tau,T)$ at the intermediate fixed points C, C', LM' and MCK is shown in Fig.\ref{fig:tau-behavior}.
It follows that the Green functions $G_f(\tau,T)$ and $G_B(\tau,T)$  collapse in terms of $\pi T/\sin{(\pi \tau T)}$ at the intermediate fixed points C, C', LM' and MCK.
It follows that local multi-particle correlators display a related scaling due to a Wick-like decomposition of higher correlation functions, valid at the saddle point, in terms of $G_f(\tau,T)$ and $G_B(\tau,T)$.  
The observed scaling to leading order is compatible with  
\begin{align}
\label{eq:tauscaling}
    G_{a}(\tau)=-\left (\frac{\pi \tau_0 T}{ \sin{(\pi \tau T)}} \right)^{\zeta}~(0<\tau<1/T)
\end{align}
for $a=f,B$ and with the scaling exponent $0<\zeta<1$ such that the results of Appendix \ref{app:FT} apply.
Such scaling collapse is reminiscent of the one expected for a  boundary conformal field theory and would suggest that the boundary entropy in the pseudogap BFKM respects the g-theorem.
As will be discussed in the next section, we do, however, observe an extra contribution to Eq.(\ref{eq:tauscaling}) in the dissipative regime, {\itshape i.e.}, for $T>\omega$, which is compatible with $\omega/T$ scaling but affects $G_a(\tau,T)$ near $\tau\approx 1/(2T)$. As a result, the g-theorem is violated in the pseudogap BFKM. 
%%%%%%%%%%%%%%%% Figure 6 %%%%%%%%%%%%%%%%%%%%%%%%%%%%%%%
%%%%%%%%%%%%%%%%%%%%%%%%%%%%%%%%%%%%%%%%%%%%%%%%%%%%%%%%%
%
%%%%%%%%%%%%%%%%%%%%%%%%%%%%%%%%%%%%%%%%%%%%%%%%%%%%%%%%%%%%%%
%%%%%%%%%%%%%%%%%%%%%%%%%%%%%%%%%%%%%%%%%%%%%%%%%%%%%%%%%%%%%%
\subsection{Boundary Entropy}
Having established the behavior of $G_f$ and $G_B$ in the vicinity of the intermediate fixed points C, C', LM' and MCK, we are in a position to obtain the boundary entropy at the various fixed point and assess the applicability of  the $g$-theorem to the pseudogap BFKM at the large-N level. This theorem addresses the behavior of the impurity entropy along  RG trajectories and states that the value of $s$ decreases along the RG flow which has been rigorously proven for boundary conformal models \cite{Affleck.91,Friedan.04}.
An earlier study of the pseudogap BFKM in the limit of large N concluded that the $g$-theorem is obeyed \cite{Kircan.04}. In contrast, our analysis reveals that the $g$-theorem does not apply to this model.\\
The results of Ref. \cite{Kircan.04} are based on Eq.~(\ref{eq:Sa}) and the conformal scaling form, Eq.~(\ref{eq:tauscaling}), which together result in 
\begin{align}
	\label{eq:Saa}
	s^{*}=&\int_{0}^{1} \frac{du}{\pi}\frac{2}{u^{2}-1}\Big[-\frac{\kappa}{u} \arctan(u\cot(\frac{\pi \alpha_{B}}{2}))\nonumber \\
	   +&\kappa \arctan(\cot(\frac{\pi \alpha_{B}}{2}))+u\arctan(u\cot(\frac{\pi \alpha_{f}}{2}))\nonumber \\
    -&\arctan(\cot(\frac{\pi \alpha_{f}}{2}))\Big] 
\end{align}
for the $T=0$ boundary entropy at a fixed point with  exponents $\alpha_f$  and $\alpha_B$ for $G_f$ and $G_B$ respectively.
%%%%%%%%%%%%%%%%%%%%%%%%%%%%%%%%%%%%%%%%%%%%%%%%%%%%%%%%
\begin{figure}[ht]
	\begin{picture}(0,0)
	\end{picture}
\centering
\includegraphics[width=0.65\columnwidth]{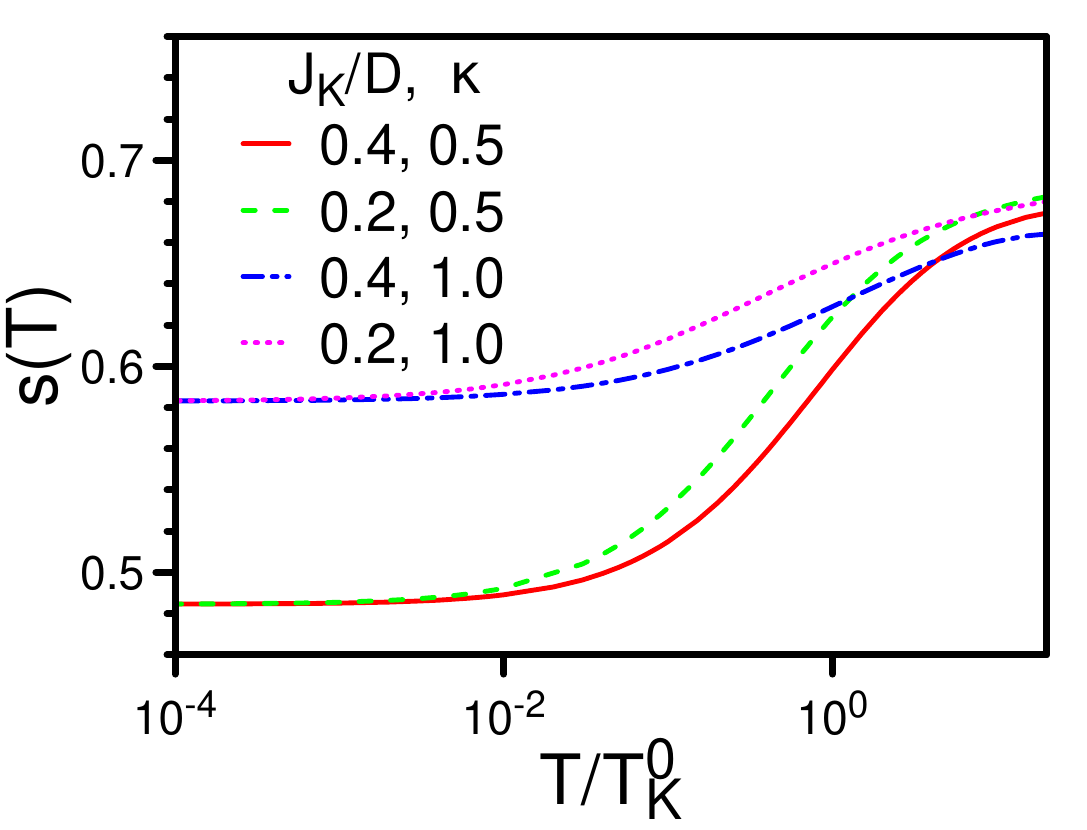}
\caption{The impurity entropy $s(T)$ for $r=0$ and $g=0$ vs. $T/T_K^0$ for $r=0$. The fixed point values at MCK ($s(T=0)$) and LM ($s(T/T_K^0\rightarrow \infty)$) are recovered independently of the value of $J_K$. $s(T=0)$ depends on  $\kappa=M/N$, see Ref.\cite{Parcollet.98}.}
\label{fig:entropy-test}
\end{figure}
%%%%%%%%%%%%%%%%%%%%%%%%%%%%%%%%%%%%%%%%%%%%%%%%%%%%%%%%%%%%
%

Before turning to our results for the fixed-point entropy at the various intermediate fixed points, we will discuss several benchmarks to demonstrate the reliability of our evaluation which is based on Eq.~(\ref{eq:entropy-formula}).\\
The degeneracy of the weak-coupling fixed point LM  at $J_K=0=g$, $N!/(Q!(N-Q)!)$ is tied to the constraint associated with the totally anti-symmetric representation, $\sum_{\sigma=1}^N f^\dagger_\sigma f^{}_\sigma =Q$. The associated boundary entropy is thus $s_{\mbox{\tiny LM}}=\ln 2$ as we have chosen $Q=N/2$.\\
For the special case $r=0$, $g=0$, as discussed by Parcollet {\it et al.}, the strong coupling fixed point of the model  is amenable to a conformal field theory description which, for $Q=N/2$, results in $s_{\mbox{\tiny MCK}}^{r=0}=(1+\kappa)[f(1+\kappa)-2f(2+2\kappa)]/\pi$, where
\begin{align}
    f(x)=\int_0^{\pi/x} du \ln(\sin{u}),
\end{align}
(see Ref. \cite{Parcollet.98} for details).\\
Fig.~\ref{fig:entropy-test} shows that our evaluation of $s(T)$ in the special case with ($r=0$, $g=0$) of the (pseudogap) BFKM indeed reproduces $s_{\mbox{\tiny LM}}$ and $s_{\mbox{\tiny MCK}}^{r=0}$ for given $\kappa$, independently of $J_K$.\\ 
A further test is provided by evaluating the entropy in the large-$N$ version of the Sachdev-Ye-Kitaev (SYK) model, which  is defined by $\Sigma(\tau)=-J^{2}G(-\tau)^{3}$,
together with the Dyson equation, linking $G$ and $\Sigma$ and where $J$ is a coupling constant. As discussed in detail in Appendix \ref{app:SYK}, our calculation of the entropy $s_{\mbox{\tiny SYK}}(T)$ in the SYK model reproduces the analytical prediction for $T=0$ and high $T$ \cite{Maldacena.16}. Appendix \ref{app:SYK} also contains a discussion of the scaling properties of $G$ and $\Sigma$.

%%%%%%%%%%%%%%%%%%%%%%%%%%%%%%%%%%%%%%%%%%%%%%%%%%%%%%%%%%%%
\begin{figure}[ht]
	\begin{picture}(0,0)
		\put(34,102){\textsf{(a)}}
		\put(157,102){\textsf{(b)}}
	\end{picture}
\centering
\includegraphics[width=0.488\columnwidth]{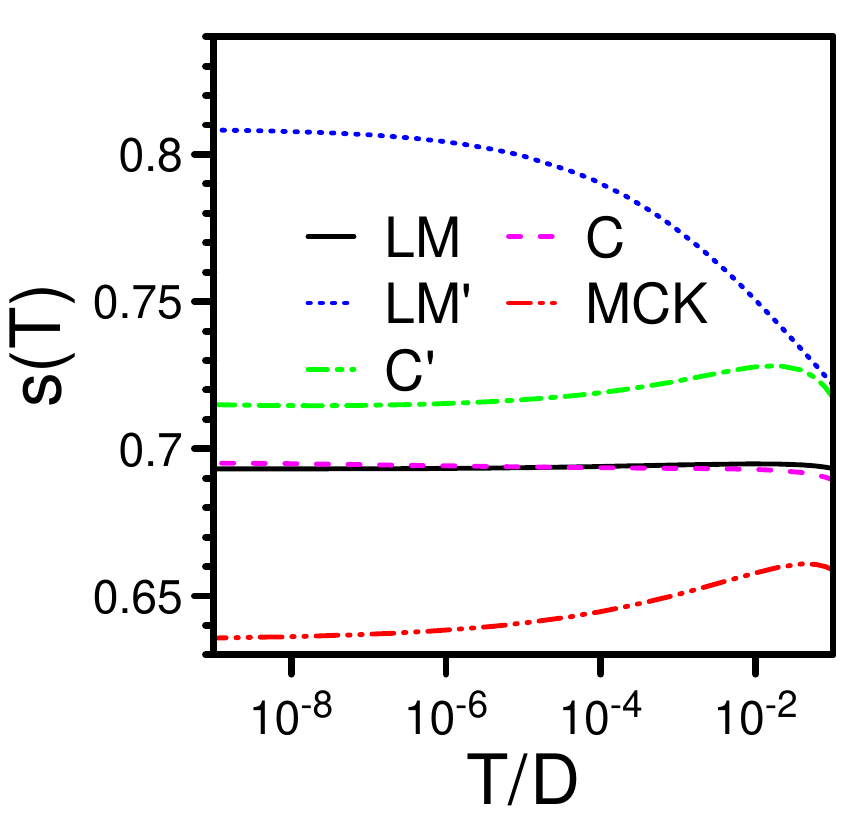}
\includegraphics[width=0.488\columnwidth]{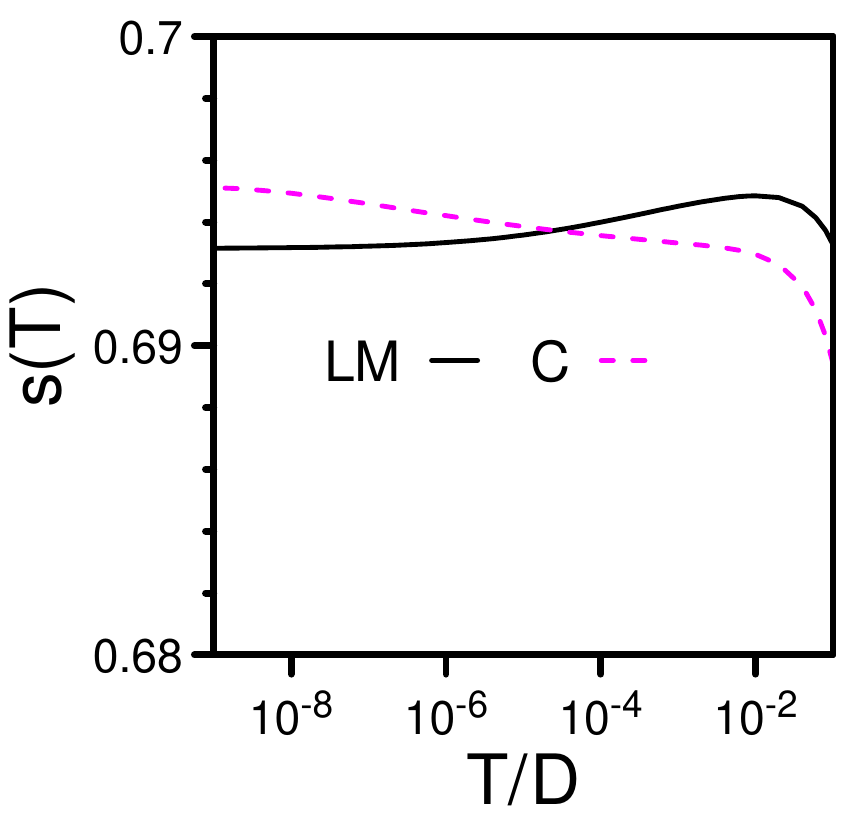}
\caption{(a) Boundary entropy of the pseudogap BFKM for $r=1/4, \epsilon=2/5, \kappa=1/2$ and  coupling constants $J_K$ and $g$ such that $s(T=0)=s_{\mbox{\tiny MCK}}$ for $J_K=0.7D,g=0$,
$s(T=0)=s_{\mbox{\tiny LM}}$ for $J_K=0.325D,g=0$, $s(T=0)=s_{\mbox{\tiny C}}$ for $J_K=0.45D,g=0$,
$s(T=0)=s_{\mbox{\tiny C'}}$ for $J_K=0.8D, g=0.375D$, and
$s(T=0)=s_{\mbox{\tiny LM'}}$ for $J_K=0,g=0.325D$.
(b) Same as (a) for values of $s(T\rightarrow 0)$ between 0.68 and 0.7 to show the differences between $s_{\mbox{\tiny C}}$ and $s_{\mbox{\tiny LM}}$.}
\label{fig:entropy-result}
\end{figure}
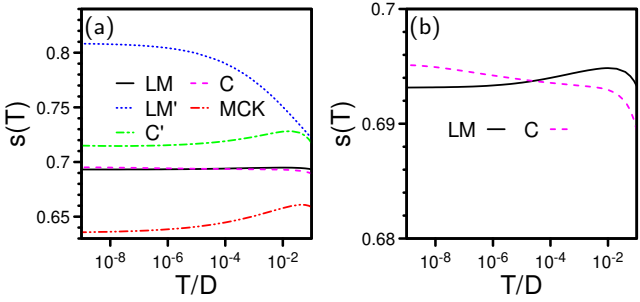
%%%%%%%%%%%%%%%%%%%%%%%%%%%%%%%%%%%%%%%%%%%%%%%%%%%%%%%%

Having established the reliability of Eq.~(\ref{eq:entropy-formula}) in evaluating the boundary entropy, we can apply it to the generic peudogap BFKM with $r\neq0$, $\epsilon \neq 0$ and arbitrary $J_K$ and $g$. In Fig.~\ref{fig:entropy-result}, we show typical results for $r=1/4, \epsilon=2/5, \kappa=1/2$ and a range of coupling constants that lead to flows to different fixed points. One can infer from the results shown in Fig.~\ref{fig:entropy-result} that fixed point LM' is characterized by an impurity entropy that is considerably larger than that associated with the other fixed points. A comparison of the values of $s(T=0)$ with Fig.~\ref{fig:fig_1} implies that the RG flow towards LM' is in contradiction  to expectations based on the $g$-theorem. The same is true for, {\itshape e.g.}, the RG flow from C to C'. Thus, we conclude that the $g$-theorem is not fulfilled in the pseudogap BFKM.
As both the pseudogap DOS and the power spectral density of the bosonic bath (with short-ranged coupling constants $J_K$ and $g$) break conformal invariance of the Hamiltonian, this conclusion may not be completely unexpected.
%%%%%%%%%%%%%%%%%%%%%%%%%%%%%%%%%%%%%%%%%%%%%%%%%%%%%%%%%

\subsection{Scaling function \& entropy flow}

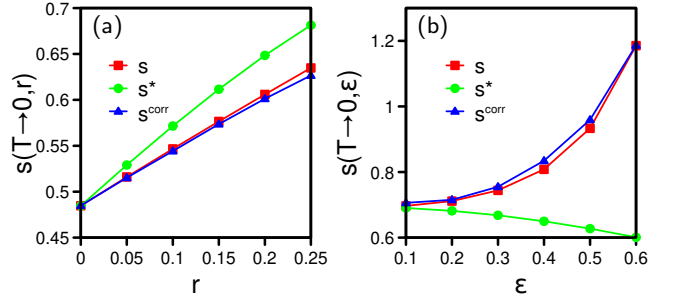
\begin{figure}[ht]
\centering
\begin{picture}(0,0)
	\put(34,102){\textsf{(a)}}
	\put(157,102){\textsf{(b)}}
\end{picture}
\includegraphics[width=0.488\columnwidth]{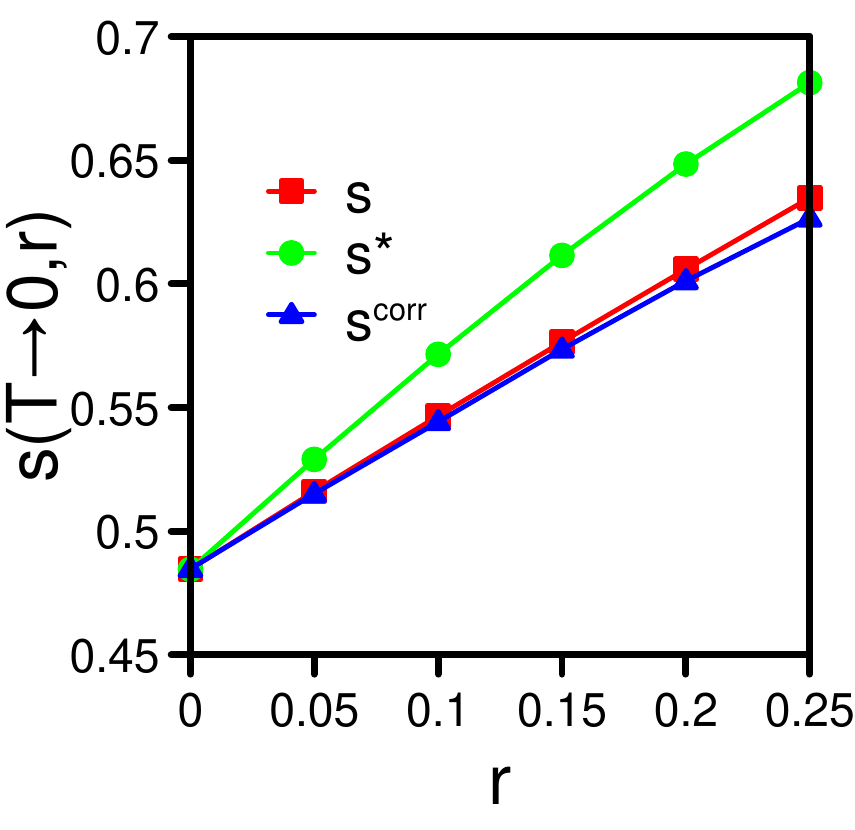}
\includegraphics[width=0.488\columnwidth]{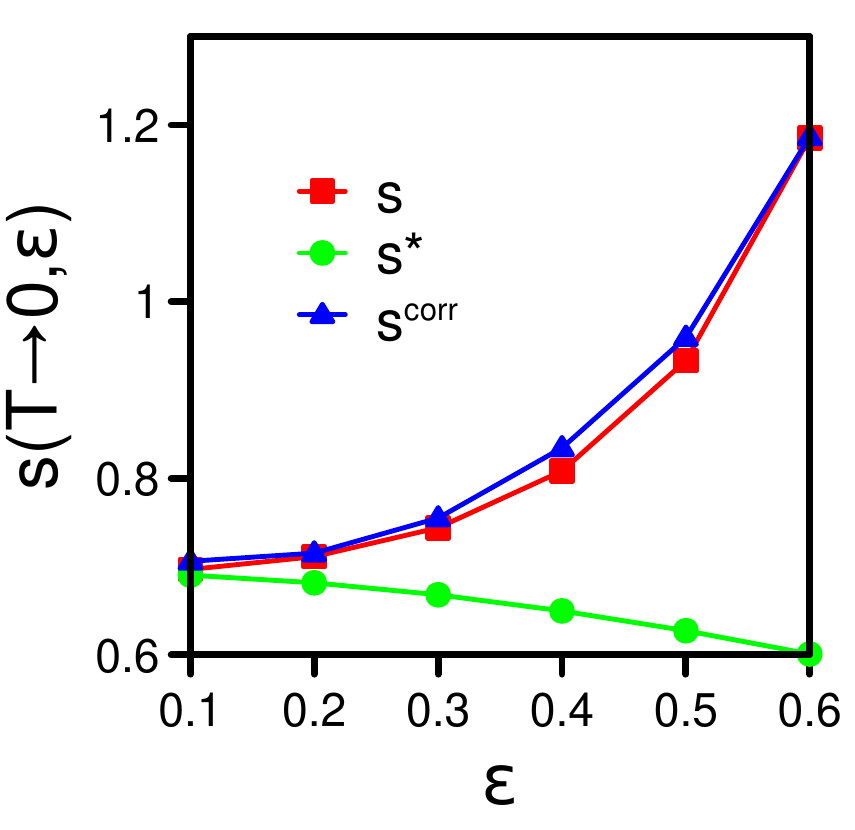}
\caption{ Comparison of different approach of determining the residual boundary entropy, labeled by $s$, $s^{*}$, and $s^{\mbox{\tiny corr}}$. The specifics of each of these approaches are given in the main text. 
(a) $r$ dependence of $s$, $s^{*}$, and $s^{\mbox{\tiny corr}}$ at the MCK fixed point  and (b) $\epsilon$ dependence of $s$, $s^{*}$, and $s^{\mbox{\tiny corr}}$ at the LM' fixed point with difference $\epsilon$.}
\label{fig:entropy_comparison}
\end{figure}

Can we understand why Eq.~(\ref{eq:Saa}) is inappropriate to evaluate the residual boundary entropy of the pseudogap BFKM?
In Fig.~\ref{fig:entropy_comparison}, we show $s^*$, obtained from evaluating Eq.~(\ref{eq:Saa}) together with $s(T\rightarrow 0)$ based on expression Eq.~(\ref{eq:entropy-formula}), and $s^{\mbox{\tiny corr}}$ which is obtained from a correction scheme to be outlined below.
In Fig.~\ref{fig:entropy_comparison}(a), this comparison is shown as a function of $r$ for the pseudogap MCK fixed point
while Fig.~\ref{fig:entropy_comparison}(b) contrast $s(T\rightarrow 0)$ and $s^{\mbox{\tiny corr}}$ with $s^{*}$ as a function of $\epsilon$. Clearly, the difference between $s^{*}$ and $s(T\rightarrow 0)$ grows with $r$ for MCK and with $\epsilon$ for LM'.
%%%%%%%%%%%%%%%%%%%%%%%%%%%%%%%%%%%%%%%%%%%%%%%%%%%%%%%%%%
%%%%%%%%%%%%%%%%%%%%%%%%%%%%%%%%%%%%%%%%%%%%%%%%%%%%%%%%%%

%%%%%%%%%%%%%%%%%%%%%%%%%%%%%%%%%%%%%%%%%%%%%%%%%%%%%%%%%%
\begin{figure}[ht]
\centering
\begin{picture}(0,0)
	\put(34,100){\textsf{(a)}}
	\put(157,100){\textsf{(b)}}
	\put(34,-17){\textsf{(c)}}
	\put(157,-17){\textsf{(d)}}
\end{picture}
\includegraphics[width=0.488\columnwidth]{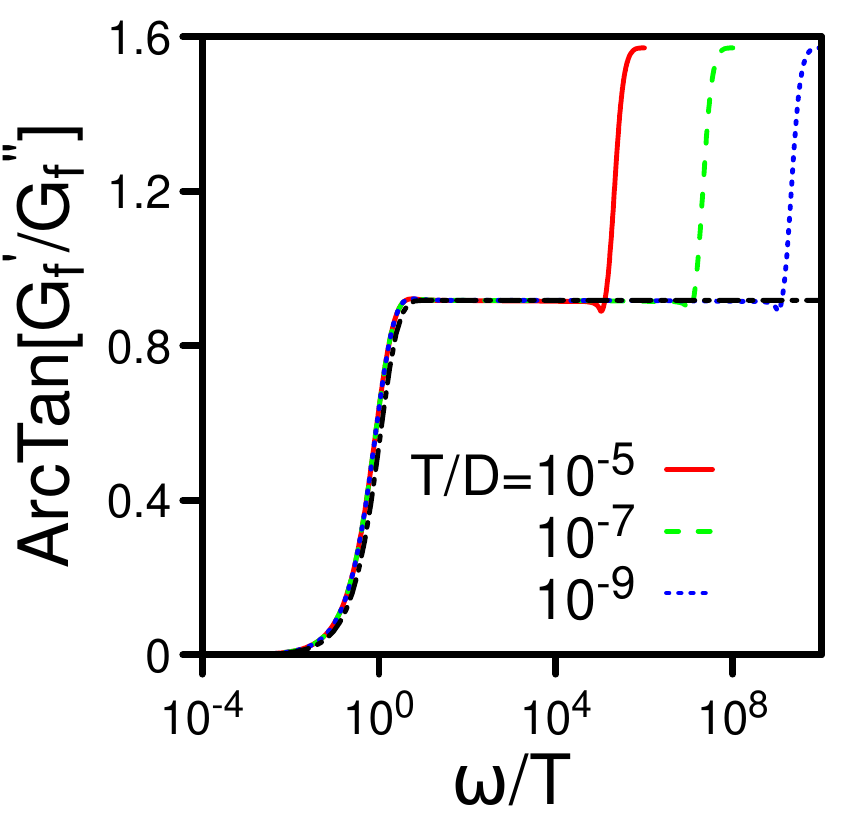}
\includegraphics[width=0.488\columnwidth]{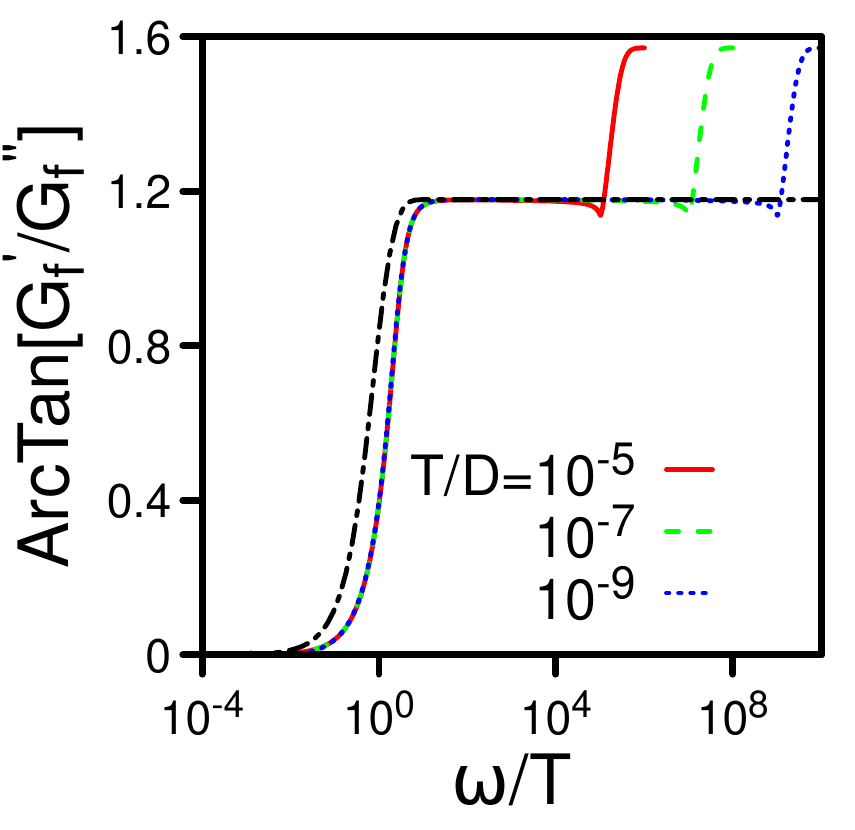}\\
\includegraphics[width=0.488\columnwidth]{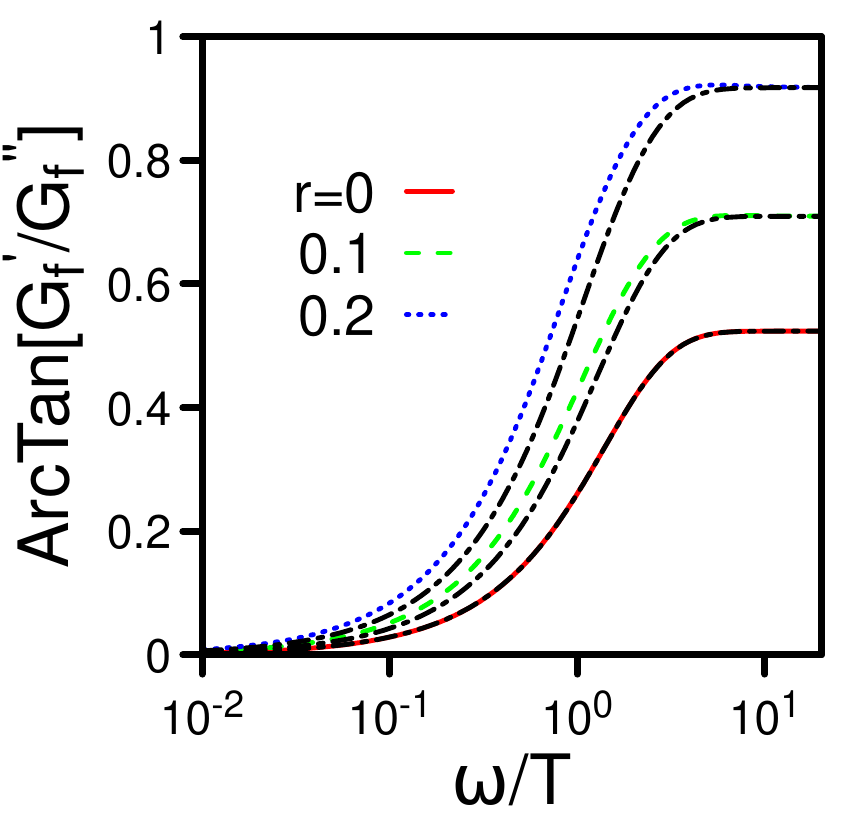}
\includegraphics[width=0.488\columnwidth]{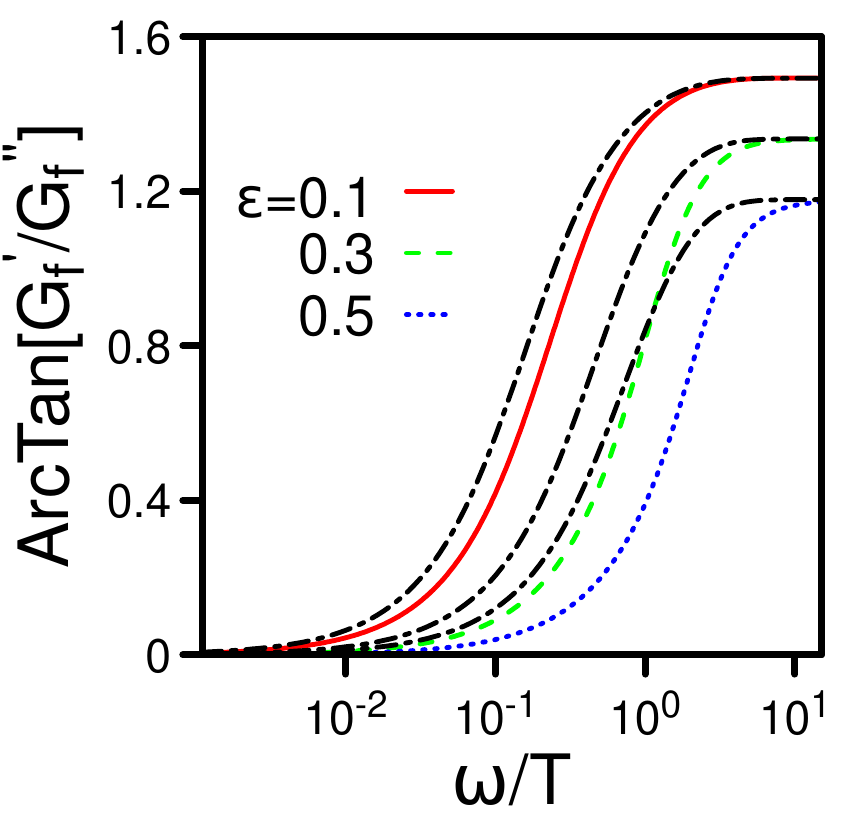}
\caption{Comparison of  $\arctan\big[\mbox{Re}\{G_f\}/\mbox{Im}\{G_f\}\big]$ obtained from  Eq.~(\ref{eq:tauscaling}) and denoted by black dash-dotted lines with results obtained from  self-consistent solutions of the saddle point equations.
(a) Comparison for r=0.2 at the MCK fixed point. 
(b) Comparison for $\epsilon$=0.5 at the LM' fixed point.
(c) $r$-dependent comparison at the MCK fixed point at low $T$ and for $\omega \lesssim T$.  
(d) $\epsilon$-dependent comparison at the LM' fixed point at low $T$ and for $\omega \lesssim T$.}
\label{fig:arcTan}
\end{figure}
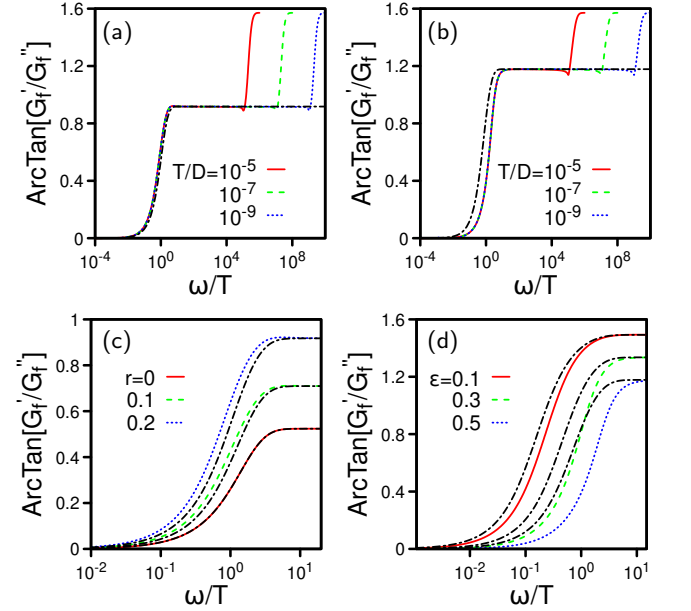
%%%%%%%%%%%%%%%%%%%%%%%%%%%%%%%%%%%%%%%%%%%%%%%%%%%%%%%%%%

In order to trace the origin of this difference, we compare  $a_{f}(\omega/T)=\arctan\frac{G_{f}'(\omega,T)}{G_{f}''(\omega,T)}$, evaluated with the self-consistently determined $G_f$ and with the analytically continued Fourier transform of Eq.~(\ref{eq:tauscaling}), determined in App. \ref{app:FT}, for the pseudogap MCK and the LM' fixed points.  It is worth noting that this comparison is parameter free.
As shown in Fig.~\ref{fig:arcTan}  there is by and large good agreement   for $G_f$ near the pseudogap MCK fixed point for $r=0.2$, shown in Fig.~\ref{fig:arcTan}(a)  and near LM' for  $\epsilon=0.5$, depicted in Fig.~\ref{fig:arcTan}(b). This overall good agreement is also implied by the results shown in Fig.~\ref{fig:tau-behavior}.
The deviations occurring for large argument, {\itshape i.e.}, for $\omega \gg T$, are caused by the high-energy cutoff of the scaling regime and are also visible in Fig.~\ref{fig:T-behavior_and_omegaT}. Outside of the scaling regime, $a_{f}$ no  longer shows $\omega/T$ scaling. 
A further difference becomes visible when zooming into the quantum dissipative regime where $\omega<T$, as shown in Fig.~\ref{fig:arcTan}(c) for MCK and Fig.~\ref{fig:arcTan}(d) for LM'. This deviation grows with $r$, see  Fig.~\ref{fig:arcTan}(c), and with $\epsilon$ as demonstrated in Fig.~\ref{fig:arcTan}(d). Our conclusion that this difference underlies the discrepancy between $s(T\rightarrow 0)$ and $s^{*}$ is further corroborated by using the numerical  $a_{f}(\omega/T)$ and $a_{B}(\omega/T)$ in Eq.~\ref{eq:Sa} and ignoring the part outside of the scaling regime for $\omega\gg T$. This leads to an estimate for the residual boundary entropy which is called $s^{\mbox{\tiny corr}}$ in  Fig.~\ref{fig:entropy_comparison} and which agrees well with $s(T\rightarrow 0)$.\\
The correction to Eq.~(\ref{eq:tauscaling}) shown in Fig.~\ref{fig:arcTan}(c) for MCK and Fig.~\ref{fig:arcTan}(d) for LM' is confined to the quantum dissipative regime where $T\gg \omega$. This suggests that this correction vanishes as $T\rightarrow 0$. At $T\neq 0$, it does however, leads to a linear-in-$T$ contribution to the free energy and thus a contribution to the residual boundary entropy $s$.
For this reason, we will refer to the contribution for $\omega<T$ shown in Fig.~\ref{fig:arcTan} as anomalous.
The results from the scaling ansatz which, strictly speaking, operates at $T=0$ but its results  apply to the quantum coherent regime, {\itshape i.e.}, where $\omega\gg T$ are in line with this conclusion. As discussed in Appendix~\ref{App:SA}, where the scaling ansatz is  extended to the next leading order, no anomalous term appears in the $T=0$, $\omega \rightarrow 0$ solution of the self-consistency equations.

In order to understand the effect of the anomalous contribution on $G(\tau)$, it is useful to notice that the kernel of Eq.~(\ref{eq:omega-to-tau}) for $\tau=1/(2T)$ is equal to $[2\cosh{(\omega/(2T))}]^{-1}$. 

%%%%%%%%%%%%%%%%%%%%%%%%%%%%%%%%%%%%%%%%%%%%%%%%%%%%%%%%%%
\begin{figure}[ht]
	\begin{picture}(0,0)
		\put(34,100){\textsf{(a)}}
		\put(157,100){\textsf{(b)}}
	\end{picture}
\centering
\includegraphics[width=0.488\columnwidth]{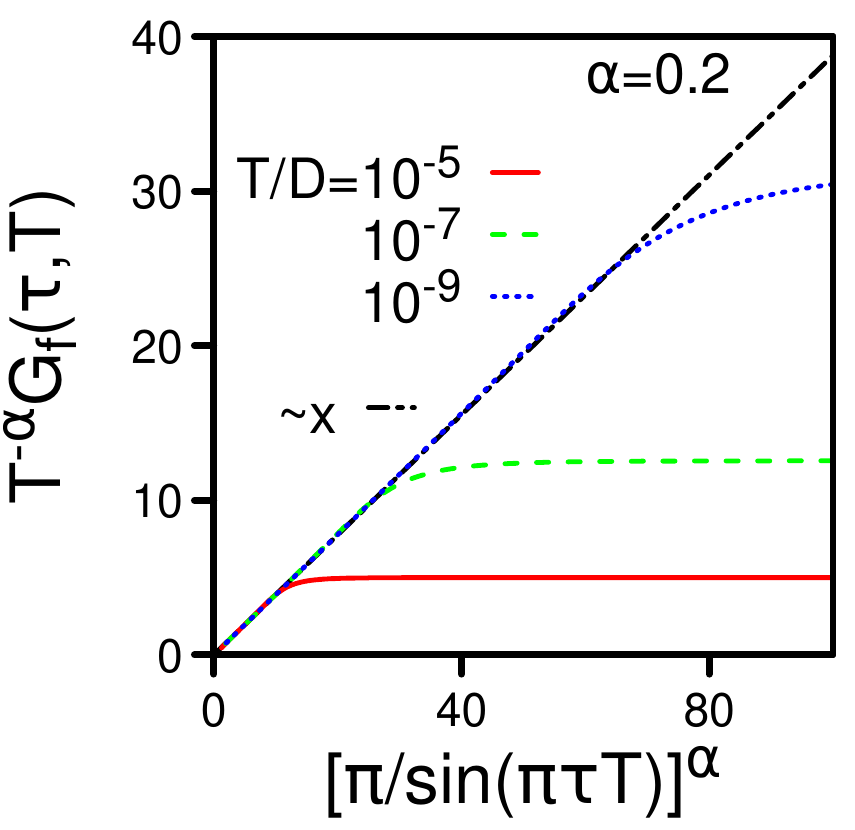}
\includegraphics[width=0.488\columnwidth]{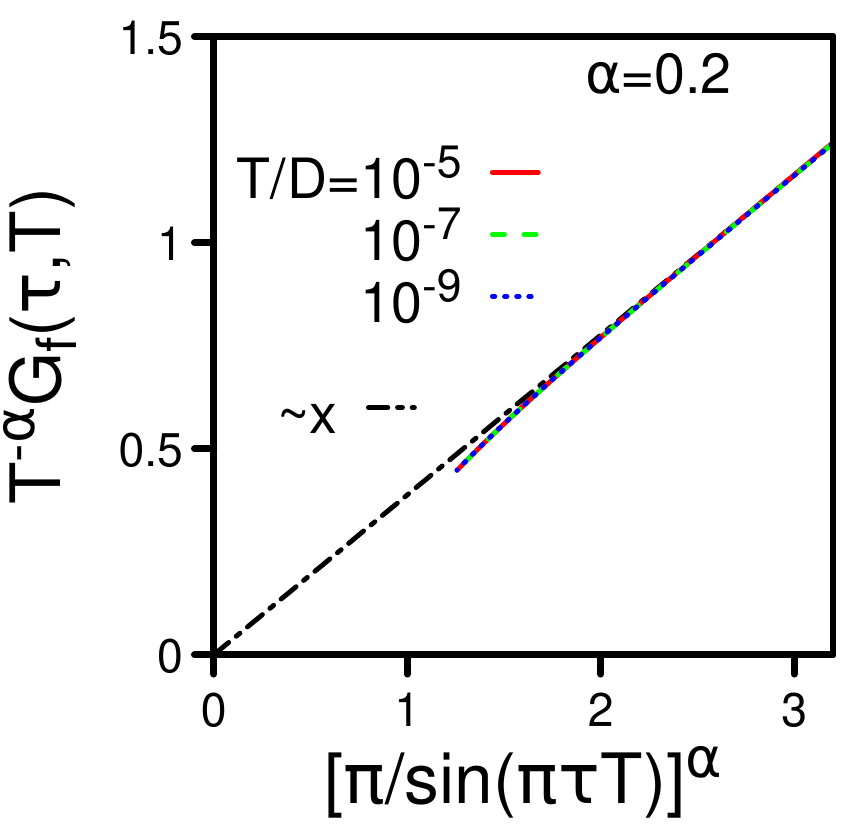}
\caption{The behavior of $G_{f}(\tau)$ of LM' fixed point with $\epsilon=0.4$. (a) The temperature dependency and (b) near $\beta/2$ regime.}
\label{fig:fig_tau}
\end{figure}
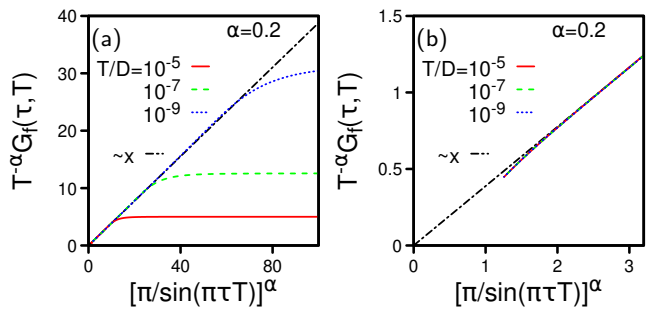
%%%%%%%%%%%%%%%%%%%%%%%%%%%%%%%%%%%%%%%%%%%%%%%%%%%%%%%%%%

Thus, $G(\tau=1/(2T))$ is primarily determined by $G(\omega,T)$ in the hydrodynamic regime. For a function $G$ that displays $\omega/T$ scaling, {\itshape i.e.}, $G(\omega,T)=T^{\alpha-1}\Phi(\omega/T)$, it follows from Eq.~(\ref{eq:omega-to-tau}) that $G(\tau,T)=T^{\alpha}\widetilde{\Phi}(\tau T)$.
In Fig.~\ref{fig:fig_tau}, we plot $\widetilde{\Phi}(\tau T)$ as a function of $[\pi/\sin(\pi\tau T)]^{\alpha}$. Fig.~\ref{fig:fig_tau}(a) shows that in terms of $x=[\pi/\sin(\pi\tau T)]^{\alpha}$, 
\begin{align}
   \widetilde{\Phi}(x)\sim x
\end{align}
in the scaling regime except for the region near $x\approx \pi^\alpha$ which corresponds to $\tau\approx 1/(2T)$, where a tiny deviations from $\widetilde{\Phi}(x)\sim x$ occurs as shown in Fig.~\ref{fig:fig_tau}(b). For comparison, we provide a plot similar to Fig.~\ref{fig:fig_tau} for MCK in the conformally invariant $r=0$ case, see Fig.~\ref{fig:r0_fit}.
It is the deviation from $\widetilde{\Phi}(x)\sim x$ that corresponds to the anomalous contribution shown in Fig.~\ref{fig:arcTan}(b). For $\tau \rightarrow 0^{+}$, corresponding to large $x$, $G_f(\tau)$ is fixed by the constraint so that the $T$ dependence  in Fig.~\ref{fig:fig_tau}(a) for large values of $x$ has to be $\sim T^{\alpha}$. The part in  Fig.~\ref{fig:fig_tau}, on the other hand, that obeys  $\widetilde{\Phi}(x)\sim x$, is compatible with the scaling of Eq.~(\ref{eq:tauscaling}).

%%%%%%%%%%%%%%%%%%%%%%%%%%%%%%%%%%%%%%%%%%%%%%%%%%%%%%%%%%%%
\subsection{$T\neq 0$ behavior of the boundary entropy}

Having analyzed the fate of the $g$-theorem in the pseudogap BFKM and traced back the origin of the inapplicability of the $g$- theorem to the scaling function, we turn to the ramifications for the $T\neq 0$ behavior of the boundary entropy.\\  
Generically, one expects an entropy accumulation near a QCP, {\itshape i.e.},  tuning the system across the critical coupling $g=g_c$ on an isentropic $T_{\mbox{\tiny s=const}}(g)$ at low but non-vanishing $T$ one expects a minimum close to $g_c$. If the $g$-theorem is fulfilled, one expects that $s(T)$ decreases as $T$ is lowered while models that defy it show an increase in $s(T)$ as $T$ decreases \cite{Friedan.04,Florens.04}. 

\begin{figure*}[ht]
\vspace{-0.5cm}
\begin{picture}(0,0)
	\put(32,115){\textsf{(a)}}
	\put(175,115){\textsf{(b)}}
	\put(320,115){\textsf{(c)}}
	\put(32,-11){\textsf{(d)}}
	\put(175,-11){\textsf{(e)}}
	\put(320,-11){\textsf{(f)}}
\end{picture}
\includegraphics[height=0.55\columnwidth]{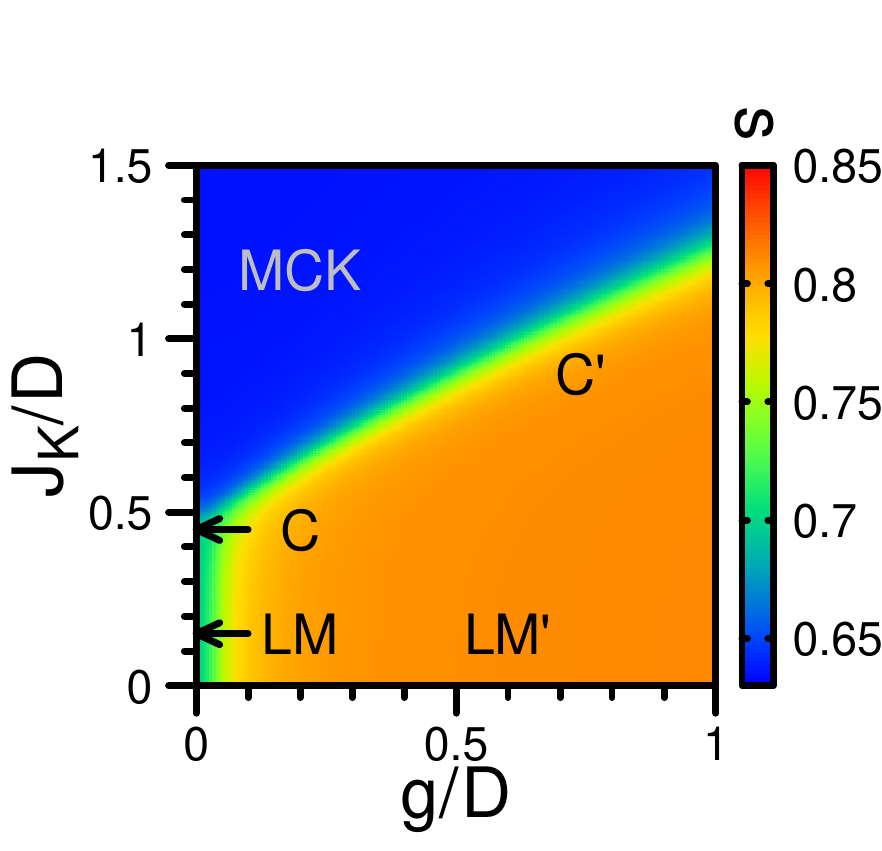}
\includegraphics[height=0.55\columnwidth]{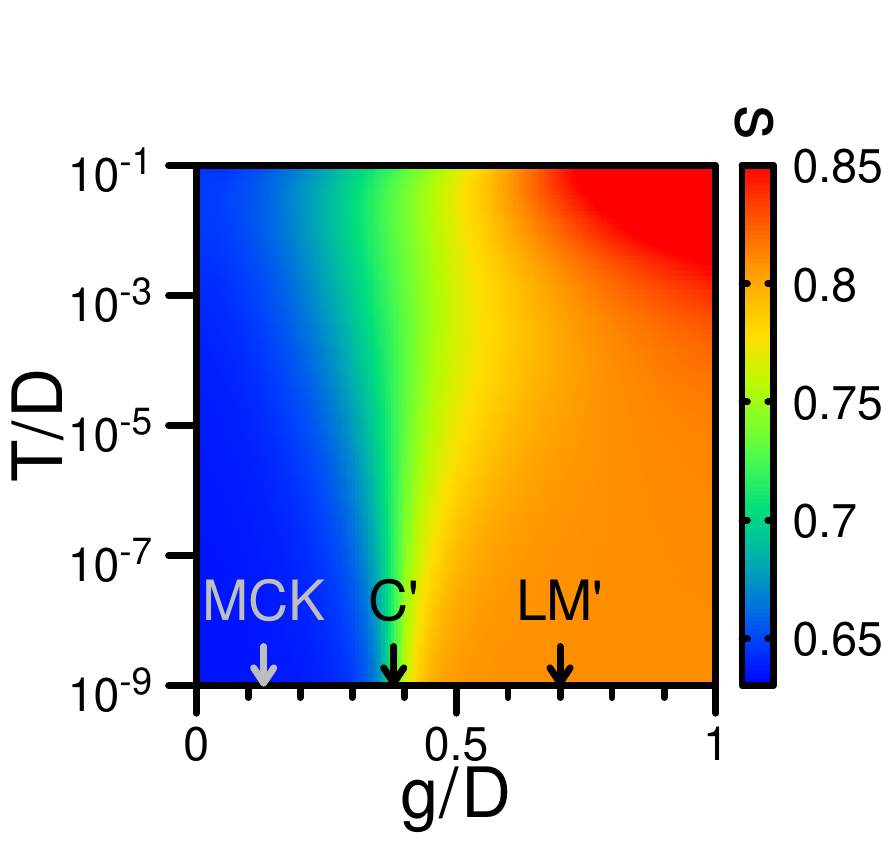}
\includegraphics[height=0.55\columnwidth]{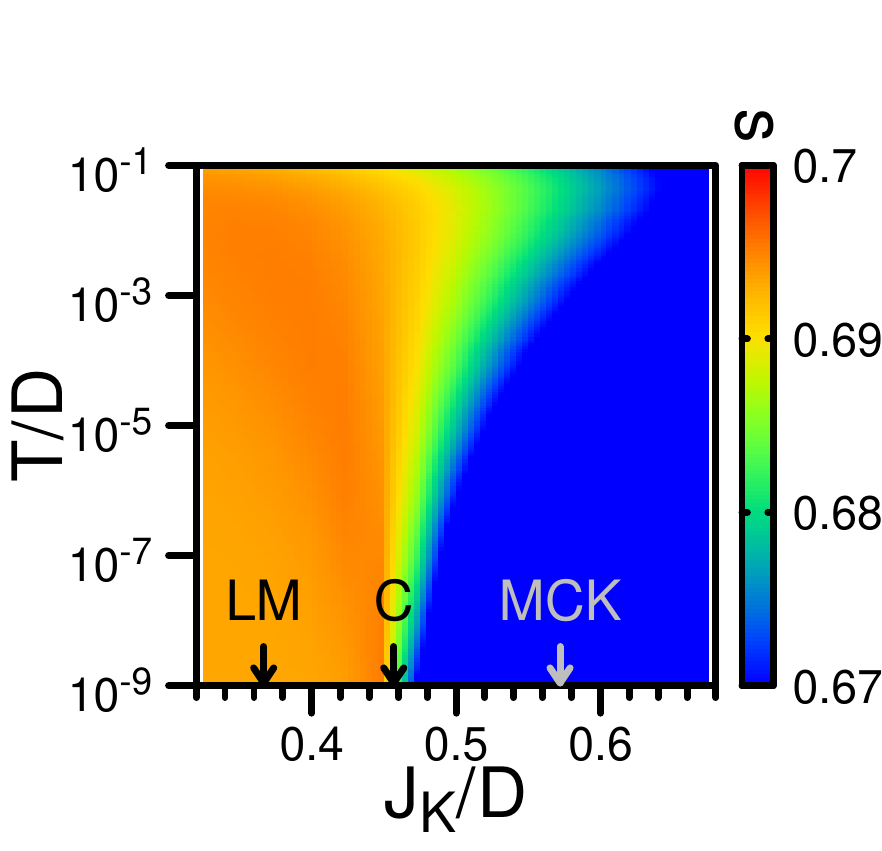}\\
\vspace{-0.4cm}
\includegraphics[height=0.55\columnwidth]{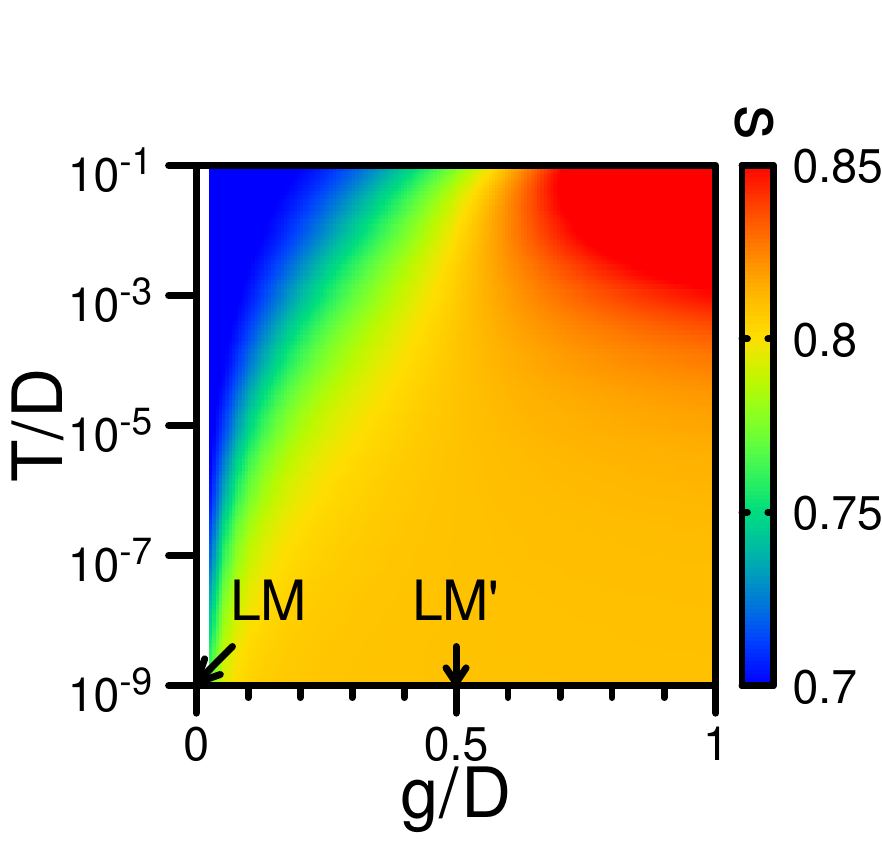}
\includegraphics[height=0.55\columnwidth]{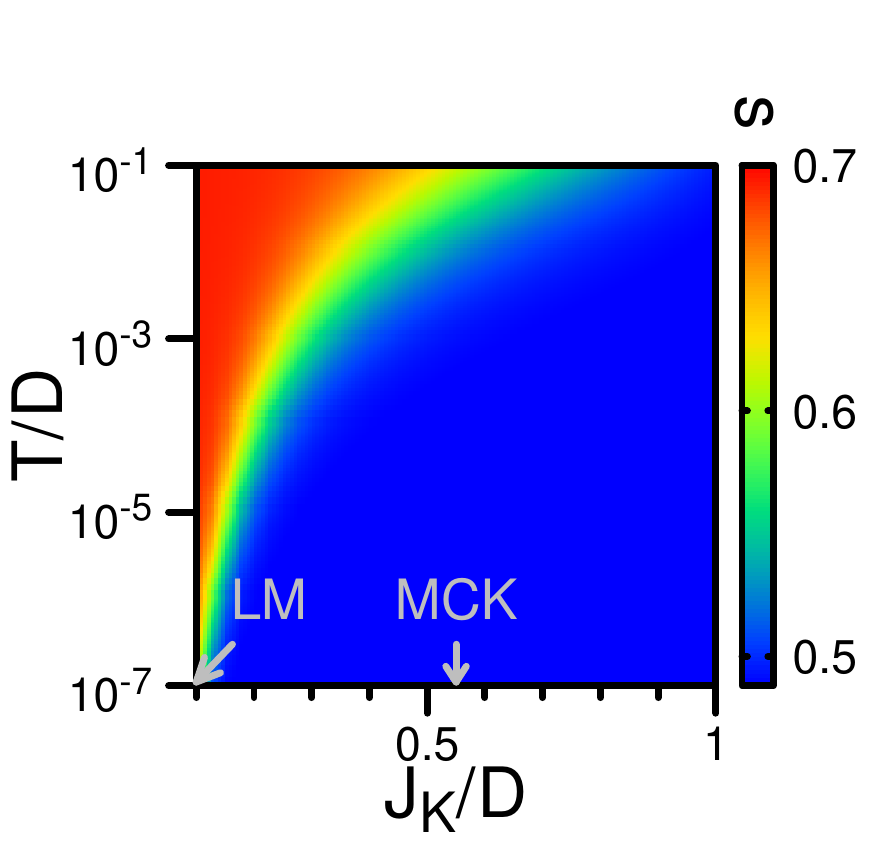}
\includegraphics[height=0.475\columnwidth]{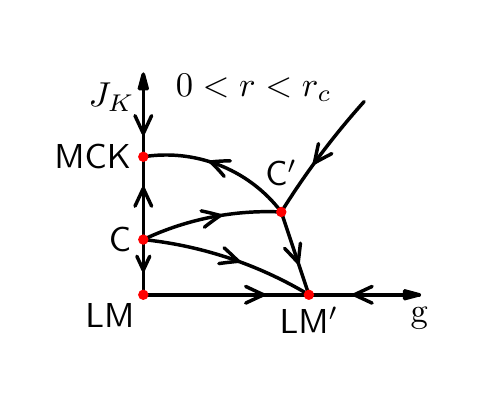}\\
\caption{(a) Boundary entropy $s(J_K,g,T_0)$ at $T_0=10^{-7}D$ as a function of $J_{k}$ and $g$. (b) $s(J_K^{0},g,T)$ across C' at $J_{k}^{0}=0.8D$ vs. $g$ and $T$. (c) $s(J_K,g=0,T)$ across C vs. $J_{k}$ and $T$. (d) $s(J_K=0,g,T)$ near LM' vs. $g$ and $T$. (e)  $s(J_K,g=0,T)$ near the MCK of $r=0$ case vs. $J_{k}$ and $T$. (f) Flow diagram for $0<r<r_{c}$ case.}
\label{fig:s-contour}
\end{figure*}
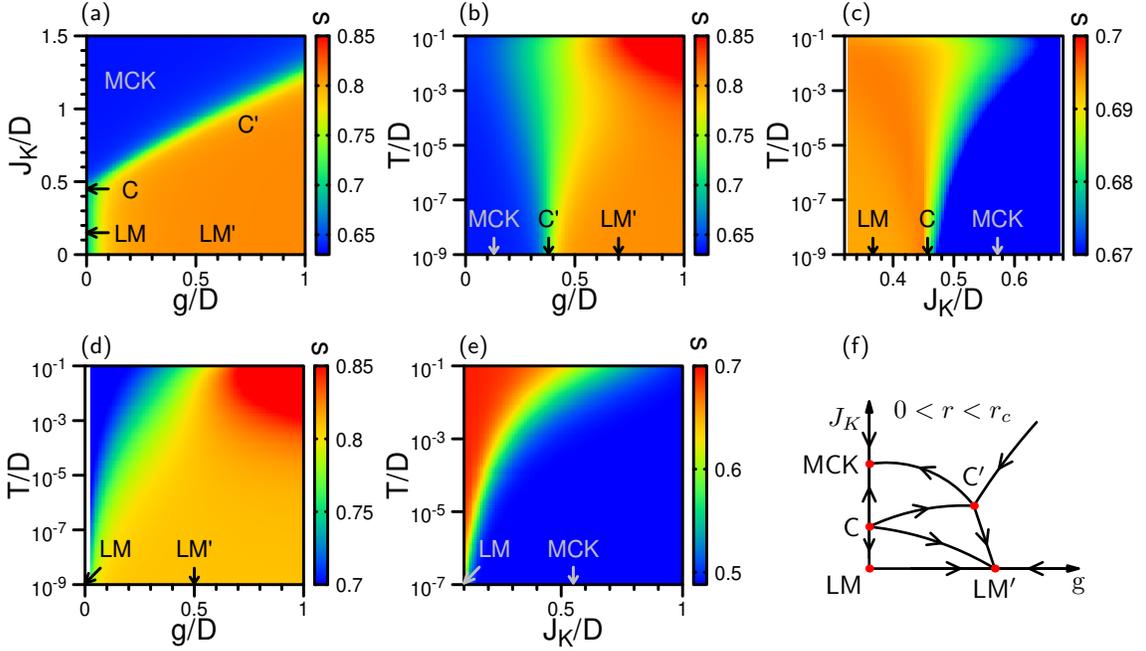

In Fig.~\ref{fig:s-contour}, the boundary entropy $s(T)$ is shown at non-zero $T$ across the phase diagram of the pseudogap BFKM. As shown in Fig.~\ref{fig:s-contour}(a), where $s(T=10^{-7}D)$ is shown as a function of the coupling constants $J_K$ and $g$, LM' gives rise to a phase with an enhanced $s$ compared to the value of $s$ near MCK and the separatrix between MCK and LM', where it is controlled by the flow to C'. 
Consequently, due to the inapplicability of the $g$-theorem for $g\neq 0$, one does not observe an entropy accumulation above C'. Instead, as the system flows from the vicinity of C' at an elevated $T$ to LM' as $T\rightarrow 0$, the boundary entropy $(T)$ increases.
In contrast, for $g=0$, the pseudogap BFKM fulfills the $g$-theorem and this is reflected in the finite-$T$ values of $s$ vs. $J_K$, which is demonstrated in Fig.~\ref{fig:s-contour}(c). For completeness,  Fig.~\ref{fig:s-contour}(d) and Fig.~\ref{fig:s-contour}(e) show the boundary entropy $s(T)$ vs. $T$  and $J_K$ near MCK [in Fig.~\ref{fig:s-contour}(d)] and $T$ and $g$ near LM' [in Fig.~\ref{fig:s-contour}(e)]. 

%%%%%%%%%%%%%%%%%%%%%%%%%%%%%%%%%%%%%%%%%%%%%%%%%%%%
%%%%%%%%%%%%%%%%%%%%%%%%%%%%%%%%%%%%%%%%%%%%%%%%%%%%%%%%%%%%
%%%%%%%%%%%%%%%%%%%%%%%%%%%%%%%%%%%%%%%%%%%%%%%%%%%%%%%%%%%%
\section{Summary}
\label{sec:summary}

We have studied the impurity entropy of  the spin-isotropic  pseudogap Bose-Fermi Kondo model in a dynamical large-N limit. Our primary focus in this study has been the applicability  of the $g$-theorem which relates the residual boundary  entropy to the RG flow to the pseudogap Bose-Fermi Kondo model.  For the $g$-theorem to be valid, the boundary entropy has to decrease along RG trajectories. 
The pseudogap Bose-Fermi Kondo Hamiltonian  lacks conformal invariance due to the pseudogap density of states of the fermionic bath as well as the subOhmic spectral density of the bosonic bath which would otherwise guarantee that the model fulfills the $g$-theorem.  

We addressed the problem by evaluating  the impurity entropy at the large-N level directly from the free energy in that limit and showed that this method is equivalent to a Luttinger-Ward based approach. The correctness of our entropy evaluation is substantiated by applying it to the large-N limits of the standard  $SU(N)\times SU(M)$ symmetric Kondo model and  the Sachdev-Ye-Kitaev or SYK model where exact results for the impurity entropy are available.

In the pseudogap Bose-Fermi Kondo model, energy-over-temperature ($\omega/T$) scaling is found at all intermediate fixed points. 
We also found a scaling form in $\tau T$  for local, {\itshape i.e.}, impurity correlators which implies $\omega/T$ scaling and appears to be compatible with that  obtained from boundary conformal field theory and which is, {\itshape e.g.}, shown in Fig.~\ref{fig:tau-behavior}.

On top of scaling form in $\tau/\beta$ we also  identified an anomalous contribution in the regime where $\hbar \omega<k_B T$, {\itshape i.e.}, in the so-called hydrodynamic regime and which is absent in the quantum coherent regime ($\hbar \omega>k_B T$), where the asymptotically exact scaling behavior is amenable to the scaling ansatz summarized in Appendix~\ref{App:SA}. 
This contribution is present at all non-trivial fixed points except for the multichannel Kondo fixed point for  $r=0$ but is largest for LM'.

Our main conclusion is the finding that the $g$-theorem is not obeyed in the pseudogap Bose-Fermi Kondo model at the large-N level. We  traced this violation back to the anomalous contribution to the large-N scaling functions. As a result, entropy accumulation is generally not observed at the critical fixed point located at intermediate couplings (C' in our notation and located at $J_K^*$, $g^*$) as the residual boundary entropy at LM' is larger than that at C' (see, {\itshape e.g.}, Fig.~\ref{fig:entropy-result}).
Instead, we are able to observe an impurity entropy decrease as temperature rises for certain parameter ranges of the model.

The results reported here are based on the large-N method and anchored around $N\rightarrow \infty$. How our results generalize to finite values of $N$  is not only a relevant but also a largely open question. Where a comparison is possible, the
direct comparison of the leading behavior between the SU($N$)-symmetric large-$N$ limit utilized here and the  SU(2) case indicates that the large-$N$ limit is regular and yields results that are close to those of the SU(2) case, at least for $r=0$  \cite{Cai.20}. On the other hand, no anomalous correction to the scaling function 
of Eq.~(\ref{eq:tauscaling}) of the type we discussed here appears  in essentially exact Monte-Carlo studies of the finite-N counterparts \cite{Glossop.11,Pixley.12,Pixley.13,Cai.20}.

One possibility to reconcile these two observations could be that the singular behavior is subleading. After all, the issue of singularities within the large-$N$ approach is delicate and cases are known where the leading-order behavior in $N$ appears to be regular while subleading corrections turn out to be singular \cite{JeanZinnJustin}.  This possibility is also in line with  the following observation:
The critical point C' for $r=0$ and $\epsilon \longrightarrow 1^{-}$ describes the critical Kondo destruction observed in a class of heavy electron materials within the extended dynamical mean field or EDMFT approach \cite{Si.01,Si.14,Kirchner.20}. 
The residual entropy $s$ in the easy-axis and SU(2) symmetric cases of the $r=0$ Bose-Fermi Kondo model is known to vanish in the limit $\epsilon \longrightarrow 1^{-}$ \cite{Dai.08}. In contrast, at the large-$N$ level, longitudinal fluctuations are sub-leading and $s$ remains finite as  $\epsilon \longrightarrow 1^{-}$.

An interpretation of the $g$-theorem in a quantum information theorem context has recently been provided \cite{Casini.16}. In light of this interpretation, the results of Ref.~\cite{Pixley.15} appear to be consistent with the conclusion that the $g$-theorem is  not fulfilled even in models of critical Kondo destruction  away from the large-$N$ limit investigated here.

%%%%%%%%%%%%%%%%%%%%%%%%%%%%%%%%%%%%%%%%%%%%%%%%%%%%%%%%%%%%%%%%%%%%%%
%%%%%%%%%%%%%%%%%%%%%%%%%%%%%%%%%%%%%%%%%%%%%%%%%%%%%%%%%%%%%%%%%%%%%%
%%%%%%%%%%%%%%%%%%%%%%%%%%%%%%%%%%%%%%%%%%%%%%%%%%%%%%%%%%%%%%%%%%%%%%
\begin{acknowledgments}
We gratefully acknowledge useful discussions with  Qimiao Si, Andreas Wipf, and Jean Zinn-Justin.
This work is  in part supported by the National Key R\&D Program of the MOST of China, Grant No. 2016YFA0300202 and the National Science Foundation of China, Grant No. 11774307.
F.\ Zamani acknowledges financial support by the Deutsche
Forschungsgemeinschaft (DFG) through SFB/TR 185 (277625399) and the Cluster of Excellence ML4Q (390534769). 
 P. Ribeiro acknowledges support by FCT through
the Investigador FCT contract IF/00347/2014 and Grant No. UID/CTM/04540/2019.
S. Kirchner acknowledges support by MOST of Taiwan, Grant No. 108-2811-M-009-500 and the hospitality of the Institute of Physics of NCTU in Hsinchu, Taiwan.
\end{acknowledgments}

\renewcommand{\thefigure}{A\arabic{figure}}
\setcounter{figure}{0}

\appendix
\section{Dynamical Large-$N$}
\label{app:s1}
The overall strategy of the dynamical large-$N$ approach is to cast the action into  a form $S=NS_{\text{eff}}$  so that the saddle-point approximation becomes exact in the limit $N\rightarrow\infty$.
We start from the action of the SU($N$)$\times$SU($M$) symmetric pseudogap BFKM, Eq.(\ref{eq:action}), 
\begin{eqnarray}
	S^{\text{DLN}}&=&-\int_{\tau} c_{k\alpha \sigma}^{\dagger}g_{c}^{-1}c_{k'\alpha \sigma}+\Phi_{q}^{\dagger}g_{\Phi}^{-1}\Phi_{q}+f_{\sigma}^{\dagger}g_{f}^{-1}f_{\sigma}\nonumber\\
				  &+& \int_{\tau} \frac{J_{K}}{N}f_{\sigma}^{\dagger}f_{\sigma'}c_{k\sigma'\alpha}^{\dagger}c_{k'\sigma\alpha} -\lambda q_{0}N\\
	&+&\frac{g}{\sqrt{N}}f_{\sigma}^{\dagger}f_{\sigma'}\tau_{\sigma\sigma'}^{i}(\Phi_{q}^{i\dagger}+\Phi_{q}^{i}).\nonumber
\end{eqnarray}
Introducing  a bosonic Hubbard-Stratonovich field $B$ to decouple the Kondo interaction and integrating out the $c$ and $\Phi$ fields leads to
\begin{eqnarray}
	S^{\text{DLN}}&=&\int_{\tau,\tau'}\sum_{\sigma}f_{\sigma}^{\dagger}(\tau)(-g_{f}^{-1}(\tau,\tau'))f_{\sigma}(\tau')\nonumber\\
	&+&\int_{\tau,\tau'}\sum_{\alpha}B_{\alpha}^{\dagger}(\tau)\frac{1}{J_{K}}\delta(\tau-\tau')B_{\alpha}(\tau')\nonumber\\
	&+&\int_{\tau,\tau'}\sum_{\alpha\sigma}[B_{\alpha}(\tau)f_{\sigma}^{\dagger}(\tau)\sum_{kk'}\frac{g_{c}(\tau,\tau')}{N}f_{\sigma}(\tau')B_{\alpha}^{\dagger}(\tau')]\nonumber\\
	&+&\int_{\tau,\tau'}\frac{g^{2}}{N}\sum_{q\sigma\sigma'}f_{\sigma}^{\dagger}(\tau)f_{\sigma}(\tau')(-g_{\Phi}(\tau,\tau'))f_{\sigma'}^{\dagger}(\tau')f_{\sigma'}(\tau)\nonumber\\
	&+&(N^{2}-1)\mathrm{Tr}\ln(-g_{\Phi}^{-1}) -M\mathrm{Tr}\ln\frac{1}{J_{K}}-q_0 N\int_{\tau}\lambda(\tau)\nonumber\\
	&-&g^{2}q_{0}^{2}\sum_{q}\int_{\tau,\tau'}g_{\Phi}(\tau,\tau')-NM\mathrm{Tr}\ln(-g_{c}^{-1}).
\end{eqnarray}
Additional Hubbard-Stratonovich fields $Q$ and $W$ are introduced to decouple the two quartic terms in the previous expression. Thus,
\begin{eqnarray}
	S^{\text{DLN}}&=&\int_{\tau,\tau'}N\bar{Q}(\tau,\tau')Q(\tau',\tau)
	+\int_{\tau,\tau'}N\bar{W}(\tau',\tau)W(\tau,\tau')\nonumber\\
	&+&\int_{\tau,\tau'}\sum_{\alpha}B_{\alpha}^{\dagger}(\tau)(-G_{B}^{-1}(\tau,\tau'))B_{\alpha}(\tau')\nonumber\\
	&+&\int_{\tau,\tau'}\sum_{\sigma}f_{\sigma}^{\dagger}(\tau)(-G_{f}^{-1}(\tau,\tau'))f_{\sigma}(\tau')-q_0N\int_{\tau}\lambda(\tau) \nonumber\\
	&+&(N^{2}-1)\mathrm{Tr}\ln(-g_{\Phi}^{-1})-M\mathrm{Tr}\ln\frac{1}{J_{K}}\nonumber\\
	&-&g^{2}q_{0}^{2}\sum_{q}\int_{\tau,\tau'}g_{\Phi}(\tau',\tau)-NM\mathrm{Tr}\ln(-g_{c}^{-1})
\end{eqnarray}
where $G^{-1}_f$ and $G^{-1}_B$ are defined as
\begin{eqnarray}
	&&G_{f}^{-1}(\tau,\tau')\nonumber\\
	&=&g_{f}^{-1}(\tau,\tau')-\bar{Q}(\tau,\tau')
	-gg_{\Phi}(\tau,\tau')W(\tau,\tau')-g\bar{W}(\tau,\tau'),\nonumber\\
	&&G_{B}^{-1}(\tau,\tau')=-\frac{1}{J_{K}}\delta(\tau-\tau')+g_{c}(\tau',\tau)Q(\tau,\tau').
\end{eqnarray}
Integrating out $B$ and $f$ fields leads to
\begin{eqnarray}
	S^{\text{DLN}}&=&\int_{\tau,\tau'}N\bar{Q}(\tau,\tau')Q(\tau',\tau)
	+\int_{\tau,\tau'}N\bar{W}(\tau',\tau)W(\tau,\tau')\nonumber\\
	&+&M\mathrm{Tr}\ln(-G_{B}^{-1})-N\mathrm{Tr}\ln(-G_{f}^{-1})      -M\mathrm{Tr}\ln\frac{1}{J_{K}}  \nonumber\\
	&+&(N^{2}-1)\mathrm{Tr}\ln(-g_{\Phi}^{-1})-q_{0}N\int_\tau\lambda(\tau) \nonumber\\
	&-&g^{2}q_{0}^{2}\sum_{q}\int_{\tau,\tau'}g_{\Phi}(\tau',\tau)-NM\mathrm{Tr}\ln(-g_{c}^{-1})\nonumber\\
	&=&NS_{\text{eff}}.
	\label{totalS}
\end{eqnarray}
Upon discarding  terms independent of $B$, $f$ and $\lambda$, one is lead to
\begin{eqnarray}
\label{eq:appendixA.6}
	S_{\text{eff}}&=&\int_{\tau,\tau'}\bar{Q}(\tau,\tau')Q(\tau',\tau)
    +\bar{W}(\tau',\tau)W(\tau,\tau')\nonumber\\
    &+&\frac{M}{N}\mathrm{Tr}\ln(-G_{B}^{-1}(\tau,\tau')) \nonumber\\
	&-&\mathrm{Tr}\ln(-G_{f}^{-1}(\tau,\tau'))-q_{0}\int_\tau\lambda(\tau).
\end{eqnarray}
This form of the action is suitable for taking the  saddle-point limit
$\delta_{Q, \bar{Q}, W, \bar{W}, \lambda}S_{\text{eff}}=0$ which results in a set of  equations, 
\begin{eqnarray}
\label{eq:App-DLNEQ}
	&&Q(\tau)=-G_{f}(\tau)\nonumber\\
	&&\bar{Q}(\tau)=-\kappa G_{B}(\tau)g_{c}(\tau)\nonumber\\
	&&W(\tau)=-gG_{f}(\tau)\nonumber\\
	&&\bar{W}(\tau)=-gG_{f}(\tau)g_{\Phi}(-\tau),
\end{eqnarray}
which become exact in the limit $N\rightarrow \infty$.
This set of equations is augmented by the constraint $G_{f}(\tau\rightarrow0^{-})=q$ that results from the fully antisymmetric representation of the impurity spin algebra.
Note that at the saddle-point level, $\lambda$ is independent of $\tau$.
%
%
%Apply saddle point condition of $\bar{Q},Q,\bar{W},W$ one can write impurity contribution of free-energy density in terms of Green's function as
%\begin{eqnarray}
%	\label{eq:f}
%	f_{\text{imp}}&=&T[\kappa\ln (-G_{B}^{-1}(i\omega))-\ln (-G_{f}^{-1}(i\omega)))]-q\lambda\nonumber\\
%	&+&\int_{\tau}\kappa G_{B}(\tau)G_{f}(-\tau)g_{c}(\tau)+g^{2}g_{\Phi}(\tau) G_{f}(\tau)G_{f}(-\tau)\nonumber
%\end{eqnarray}
%%%%%%%%%%%%%%%%%%%%%%%%%%%%%%%%%%%%%%%%%%%%%%%%%%%%%%%%%%%%%%%%%%%%%%
%%%%%%%%%%%%%%%%%%%%%%%%%%%%%%%%%%%%%%%%%%%%%%%%%%%%%%%%%%%%%%%%%%%%%%
\section{Details of the boundary entropy calculation}
\label{app:Details}

In this appendix, we detail the derivation of Eq.~(\ref{eq:entropy}) for the boundary entropy.
From the definition of the impurity entropy $s$, we have
\begin{align}
    s=-\dfrac{d f_{\text{imp}}}{d T}=-\dfrac{\partial f_{\text{imp}}}{\partial T}-\sum_{X_i}\dfrac{\delta f_{\text{imp}}}{\delta X_i}\dfrac{\partial X_i}{\partial T},
\end{align}
where $X_i$ runs over the set  $\{\bar{Q},Q,\bar{W},W,\lambda\}$. The saddle-point conditions imply that the sum over $X_i$ vanishes. Thus, we can obtain $s$ without considering the  $T$ dependence of the set $\{\bar{Q},Q,\bar{W},W,\lambda\}$. Starting from
\begin{align}
	f_{\text{imp}}=&\int_{\tau}\bar{Q}(\tau)Q(-\tau)
	+\bar{W}(-\tau)W(\tau)-q_{0}\lambda(\tau)\nonumber\\
    +&T\kappa\mathrm{Tr}\ln(-G_{B}^{-1}) -T\mathrm{Tr}\ln(-G_{f}^{-1}),
\end{align}
and transforming to real frequencies, one can take perform the derivative with respect to $T$ on the distribution functions once $G_B$ and $G_f$ have been expressed in terms of $\{\bar{Q},Q,\bar{W},W,\lambda\}$. As a result,
\begin{widetext}
\begin{align}
\label{eq:AppEqB}
 s^{\text{DLN}}=&-\int \frac{d\omega}{\pi}\partial_{T}\{n_{b}(\omega)\kappa\mathrm{Im}[\ln(-G_{B}^{-1}(\omega))]+n_{f}(\omega)\mathrm{Im}[\ln(-G_{f}^{-1}(\omega))]+n_{f}(\omega)\mathrm{Im}[\bar{Q}(\omega)Q(\omega)]-n_{f}(\omega)\mathrm{Im}[\bar{W}(\omega)W(\omega)]\}\nonumber \\
 =&-\int \frac{d\omega}{\pi}\{\frac{dn_{b}(\omega)}{dT}\kappa\mathrm{Im}[\ln(-G_{B}^{-1}(\omega))]+n_{b}(\omega)\kappa\mathrm{Im}[-G_{B}(\omega)\partial_{T}(Q(\tau)g_{c}(-\tau))_{\omega}]\nonumber \\
 +&\frac{dn_{f}(\omega)}{dT}\mathrm{Im}[\ln(-G_{f}^{-1}(\omega))]+n_{f}(\omega)\mathrm{Im}[-G_{f}(\omega)\partial_{T}(gW(\tau)g_{\Phi}(\tau))_{\omega}] \nonumber \\
 +&\frac{dn_{f}(\omega)}{dT}\mathrm{Im}[\bar{Q}(\omega)Q(\omega)]-\frac{dn_{f}(\omega)}{dT}\mathrm{Im}[\bar{W}(\omega)W(\omega)]\}. 
\end{align}
\end{widetext}
The saddle-point condition implies $Q(\tau)=G_{f}(\tau)$, $W(\tau)=-gG_{f}(\tau)$, $\bar{Q}(\tau)=\Sigma_{f}^{1}(\tau)$ and $\bar{W}(\tau)=\Sigma_{f}^{2}(\tau)/2$, in terms of which Eq.~(\ref{eq:AppEqB})  reduces to Eq.~\ref{eq:entropy}.

%%%%%%%%%%%%%%%%%%%%%%%%%%%%%%%%%%%%%%%%%%%%%%%%%%%%%%%%%%%%%%%%%%%
\section{Equivalence of the dynamic large-$N$ limit with an associated conserving Kadanoff-Baym scheme}
\label{app:KBA}
In this appendix we explicitly demonstrate the equivalence of the dynamical large-$N$ approach with that based on a conserving approximation regarding the calculation of the impurity entropy.
Conserving approximations  are invariant with respect to a set of symmetry transformations and respect the related Ward identities which link vertex corrections and self-energies at each order of perturbation   theory.  Within the Kadanoff-Baym approach, conserving approximations are constructed through the  stationary condition of an associated Luttinger-Ward functional $\Phi$. $\Phi$-derivability of an approximation is often taken to be tantamount to  it being conserving \cite{Baym.61}.
It was pointed out in Ref.~\cite{lebanon2006} that a principal difference exists between the Kadanoff-Baym and a true large-$N$ scheme.  It was already shown in Eq.~(\ref{eq:LGT}) that the saddle-point free energy at the dynamical large-$N$ level naturally assumes the form of the Legendre transform of a Luttinger-Ward functional.  Given the importance of a faithful determination of the residual boundary entropy for assessing the validity of the $g$-theorem, we derive explicitly Eq.~(\ref{eq:entropy}) from the large-$N$ equations.

At leading order in $N$,
the Luttinger-Ward functional $\Phi(G)$  is given by 
\begin{eqnarray}
		\label{eq:Y}
		\Phi(G) &=& - \int d\tau \left\{ \boldsymbol{v}^\dagger \left[\boldsymbol{G}_{B}(\tau)\otimes 
		\boldsymbol{G}_{f}(-\tau) \otimes \boldsymbol{g}_{c}(\tau) \right]
		\boldsymbol{v}  \right. \nonumber 
		\\
		& &  \left. +  \boldsymbol{w}^\dagger \left[ \boldsymbol{G}_{f}(\tau) \otimes \boldsymbol{G}_{f}(-\tau) \otimes \boldsymbol{g}_{\Phi}(\tau) \right] \boldsymbol{w}
		\right\}
\end{eqnarray}
with $v_{\sigma_f,\sigma_c}^{\alpha_B,\alpha_c} = 1/\sqrt{N} \delta_{\sigma_f,\sigma_c} \delta_{\alpha_B,\alpha_c}$ and $w_{\sigma_f,s_f}^i= g t^i_{\sigma_f,s_f} /\sqrt{N} $
together with the constraint $\boldsymbol{G}_{f}(\tau \rightarrow 0-) =q \boldsymbol{1}$ [the set $\{t^i\}$ (1,$\dots~N^{2}-1$)  was defined in Eq.(\ref{eq:spinDen})][see also  Eq.~(\ref{eq:LGT})].
The constitutive relation  $\delta \Phi(G)=-T \mathrm{Tr}[\Sigma \delta G]$ together with the assumption that $\boldsymbol{G}_{B}$, $\boldsymbol{G}_{f}$, $\boldsymbol{g}_{c}$ and $\boldsymbol{g}_{\Phi}$ are diagonal  in their respective indices, results in Eqs.(\ref{eq:Sigma_f},\ref{eq:Sigma_B}).
The impurity contribution to the free energy is given by 
\begin{eqnarray} 	\label{eq:f_KB}
	f_{\text{imp}}^{\text{KB}}&=&T \left\{ \mathrm{Tr}\left[\kappa\ln (-G_{B}^{-1})-\ln (-G_{f}^{-1})\right] \right\}-\lambda q_{0} \nonumber\\
	&+& \int d\tau \left\{\kappa \left[ g_{B}^{-1}(-\tau)-G_{B}^{-1}(-\tau) \right] G_{B}(\tau)\right. \nonumber\\
	&-& \left.  \left[g_{f}^{-1}(\tau)-G_{f}^{-1}(\tau) \right] G_{f}(-\tau) \right\} +\Phi(G)
\end{eqnarray}
computed at the extremal points where $\delta f_{\text{imp}}/\delta G_a = 0 $ and $\partial f_{\text{imp}}/\partial \lambda = 0 $. 
When taking the derivative of $f_{\text{imp}}$ with respect to $T$, we can ignore the dependence  of  $G_{f}$, $G_{B}$ and $\lambda$ on $T$ by virtue of these stationary conditions, see App.~\ref{app:Details}. The impurity entropy is thus given by  \cite{Coleman.05,Rech.06}
\begin{eqnarray}
\label{eq:S}
s^{\text{KB}}&=&-\int \frac{d\omega}{\pi}\frac{dn_{b}}{dT} \times  \\  && 
\left\{\kappa \left[\mathrm{Im}\ln(-G_{B}^{-1})+\mathrm{Im}\Sigma_{B}\mathrm{Re}G_{B} \right]
			 -\mathrm{Re}\tilde{\Sigma}_{\Phi}\mathrm{Im}g_{\Phi}\} \right. \nonumber\\
			 &&+\frac{dn_{f}}{dT} \left[\mathrm{Im}\ln(-G_{f}^{-1})+\mathrm{Im}\Sigma_{f}\mathrm{Re}G_{f}-\kappa\mathrm{Re}\tilde{\Sigma}_{c}\mathrm{Im}g_{c} \right]\nonumber
\end{eqnarray}
where the auxiliary quantities $\tilde{\Sigma}_{c}(\tau)=N\Sigma_{c}(\tau)=-G_{B}(-\tau)G_{f}(\tau)$ and $\tilde{\Sigma}_{\Phi}(\tau)=\frac{1}{N}\Sigma_{\Phi}(\tau)=g^{2}G_{f}(\tau)G_{f}(-\tau)$ are used.
By construction, the constitutive relation reproduce the dynamical large-$N$ equations of Eq.(\ref{eq:App-DLNEQ}).\
%Using those, it is easy to show that the saddle point values of the free energy also coincide.
From the identity
\begin{equation}\begin{split}
	n_{b}(\nu)[n_{f}(\omega)-n_{f}(\omega-\nu)]=-n_{f}(\omega)n_{f}(-\omega+\nu)
\end{split}\end{equation}
it follows that
\begin{equation}\begin{split}
	&n_{b}(\nu)\frac{d}{dT}[n_{f}(\omega)-n_{f}(\omega-\nu)]\\
	=&-\frac{dn_{f}(\omega)}{T}n_{f}(-\omega+\nu)-n_{f}(\omega)\frac{dn_{f}(-\omega+\nu)}{dT}\\
	&-\frac{dn_{b}(\nu)}{dT}[n_{f}(\omega)-n_{f}(\omega-\nu)],
\end{split}\end{equation}
by taking the derivative with respect to $T$.
This further implies
%As a result,
%we can transform the second and forth terms of %Eq.~\eqref{eq:entropy} as follows
%
\begin{equation}\begin{split}
	&\int \frac{d\nu}{\pi}n_{b}(\nu)\kappa\mathrm{Im}\left[G_{B}(\nu)\partial_{T}A(\omega)\right] \\
=&\int \frac{d\nu}{\pi}  \left\{  \frac{dn_{f}(\nu)}{dT} \left[ \kappa g_{c}''(\nu)\tilde{\Sigma}_{c}'(\nu) +
G_{f}''(\nu)\Sigma_{f}^{1\prime}(\nu)  \right] \right.\\
 &- \left. \kappa \frac{dn_{b}(\nu)}{dT}G_{B}'(\nu)\Sigma_{B}''(\nu) \right\}
\end{split} \end{equation}
and
\begin{equation}\begin{split}
	&-\int \frac{d\nu}{\pi}n_{f}(\nu)\mathrm{Im}[G_{f}(\nu)\partial_{T}B(\omega)]\\
	=&\int\frac{d\nu}{\pi}\frac{dn_{b}(\nu)}{dT}g_{\Phi}''(\nu)\tilde{\Sigma}_{\Phi}'(\nu)\\
	-&\int \frac{d\nu}{\pi}\frac{dn_{f}(\nu)}{dT}g\bar{W}''(\nu)G_{f}'(\nu)\\
	 &+\int \frac{d\nu}{\pi}\frac{dn_{f}(\nu)}{dT}g\bar{W}'(\nu)G_{f}''(\nu)
\end{split}\end{equation}
These last two equations can be used to establish the equivalence of Eq.~\eqref{eq:S} with Eq.~\eqref{eq:entropy}.
%
%%%%%%%%%%%%%%%%%%%%%%%%%%%%%%%%%%%%%%%%%%%%%%%%%%%%%%%%%%%%%%%%%%
\section{Fourier transformation of $(\frac{\pi \tau_{0}}{\beta \sin(\pi \tau/\beta)})^{\zeta}$}
\label{app:FT}

%for $\alpha>1$, the transformation doesn't converge. the contribution of $c_{4}$, $c_{2}$ and $c_{6}$ will violate the $\omega/T$ scaling.
%%%%%%%%%%%%%%%%%%%%%%%%%%%%%%%%%%%%%%%%%%%%%%%%%%%%%%%%%%%%%%%%
\begin{figure}[ht]
\centering
\includegraphics[width=0.6\columnwidth]{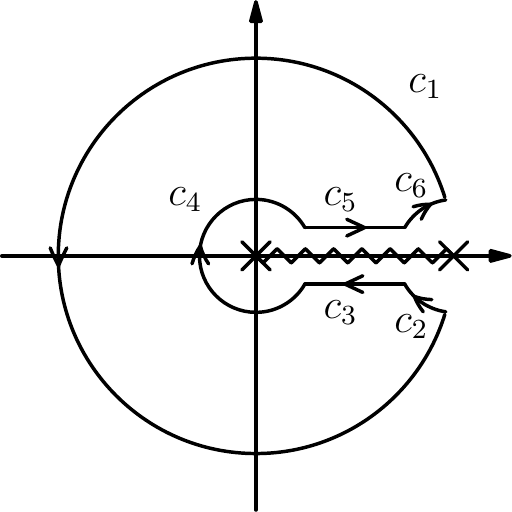}
\caption{The contour $\mathcal{C}=c_1+c_2+c_3+c_4+c_5+c_6$ used to evaluate the integral \eqref{eq:appFT4}. }
\label{fig:FT}
\end{figure}
%%%%%%%%%%%%%%%%%%%%%%%%%%%%%%%%%%%%%%%%%%%%%%%%%%%%%%%%%%%%%%%%
This appendix provides details of performing the Fourier transform of
\begin{align}
\label{eq:appFT1}
    \mathcal{G}_{a}(\tau)=-\left (\frac{\pi \tau_0}{\beta \sin{(\pi \tau/\beta)}} \right)^{\zeta}~(0<\tau<\beta)
\end{align}
which is required in the discussion of the entropy results and the scaling ansatz solution. In Eq.\eqref{eq:appFT1}, 
$\tau_0$ acts as a short-time cutoff and has units of inverse energy. The parameter $a$ distinguishes between bosonic ($a=b$) and fermionic functions ($a=f$), {\itshape i.e.},
\begin{align}
     \mathcal{G}_{a}(\tau)=-\eta  \mathcal{G}_{a}(\tau+\beta)
\end{align}
for $-\beta<\tau<0$ with $\eta=-1$ for $a=b$ and  $\eta=1$ for $a=f$.
We consider $0<\zeta<1$. For fixed $\tau$ it follows that  in the zero-temperature limit ($\beta\rightarrow \infty$) $ \mathcal{G}_{a}(\tau)=-(\tau_0/ \tau)^{-\zeta}.$
The definition of the Fourier transform of Eq.~\eqref{eq:appFT1} is
\begin{align}
    G_{a}(i\omega_n^a)=\int_0^\beta \! d\tau e^{i \omega_n^a \tau}   \mathcal{G}_{a}(\tau),
\end{align}
with the Matsubara frequencies $\omega_n^b=2 n \pi/\beta$ and $\omega_n^f=(2 n+1) \pi/\beta$ for $n=0,\pm 1,\pm 2,\ldots$.  Equation~\eqref{eq:appFT1} can be cast into the form
\begin{align*}
     G_{a}(i\omega_n^a)=-\left (\dfrac{2 \pi i \tau_0}{\beta} \right)^{\zeta}
     \int_0^\beta \! d\tau e^{i \omega_n^a \tau +i \pi \zeta \tau/\beta}  \Big[e^{2\pi i \tau/\beta}-1 \Big]^{-\zeta}
\end{align*}
Performing the substitution $s=e^{2\pi i\tau/\beta}$ maps the integral onto the contour labeled $c_1$ in Fig.~\ref{fig:FT}. The integrand is singular at $s=0$ and $s=1$ and we choose to put the connecting branch cut on the real $s$ axis. Thus,
\begin{align}
\label{eq:appFT4}
I \equiv - \left (\dfrac{2 \pi i}{\beta} \right)^{\zeta-1} \tau_0^{\zeta}\, \oint_{\mathcal{C}}\! dz z^{\frac{m_a+\zeta}{2}-1}\Big (z-1 \Big)^{-\zeta}=0,
\end{align}
where $m_a=\beta \omega_n^a/\pi$ and the contour $\mathcal{C}=c_1+\ldots +c_6$ is depicted in Fig.\ref{fig:FT}. 
As long as $\zeta<1$, the contribution along $c_2$ and $c_6$ vanishes as the radii of these two arcs goes to zero. Similarly, as the radius of the circle $c_4$ shrinks to zero, the contribution to the contour integral along $c_4$ vanishes provided $m>-\zeta$. Thus,
\begin{widetext}
\begin{align}
\label{eq:appFTmatsu}
G_a(i\omega_n^a)=- 2 \left (\dfrac{2 \pi }{\beta} \right)^{\zeta-1}\tau_0^{\zeta}\, \mbox{B}(\dfrac{m+\zeta}{2},1-\zeta) \,
    \begin{cases}
    \sin \big(\dfrac{\pi \zeta}{2}\big), &  a=b, \\
     i \cos \big(\dfrac{\pi \zeta}{2}\big), &  a=f,
    \end{cases}
\end{align}
\begin{align}
\label{eq:appFTfinal}
G_a(\omega+i\delta)=-  \left (\dfrac{2 \pi }{\beta} \right)^{\zeta-1}\tau_0^{\zeta}\, \mbox{B}(\dfrac{\zeta}{2}-\dfrac{i\beta}{2\pi}(\omega+i\delta),\dfrac{\zeta}{2}+\dfrac{i\beta}{2\pi}(\omega+i\delta)) \,
    \begin{cases}
      \tan\big(\dfrac{\pi \zeta}{2}\big) \cosh\big(\dfrac{\beta \omega}{2}\big) + i \sinh \big(\dfrac{\beta \omega}{2}\big), &  a=b, \\
       -\cot \big(\dfrac{\pi \zeta}{2}\big) \sinh\Big(\dfrac{\beta \omega}{2}\big)+i \cosh \big(\dfrac{\beta \omega}{2} \big), &  a=f,
    \end{cases}
\end{align}
\begin{align}
\label{eq:appFTboson}
G_b(\omega+i\delta,\beta\rightarrow \infty)\stackrel{.}{=} -\dfrac{\pi \tau_0^\zeta }{\Gamma (\zeta)}  \,\left[ \tan{\big(\dfrac{\pi \zeta}{2}\big)} |\omega|^{\zeta-1} +i \mbox{sgn}(\omega) |\omega|^{\zeta-1} \right ],  
\end{align}
\begin{align}
\label{eq:appFTfermion}
G_f(\omega+i\delta,,\beta\rightarrow \infty) \stackrel{.}{=}  -\dfrac{\pi \tau_0^\zeta }{\Gamma(\zeta)} \, \left [ -\cot{\big(\dfrac{\pi \zeta}{2}\big)} \mbox{sgn}(\omega)|\omega|^{\zeta-1} +i |\omega|^{\zeta-1}\right ], 
\end{align}
\end{widetext}
where $\mbox{B}(x,y)=\int_0^\infty dt t^{x-1}(1-t)^{y-1}$ is the Euler Beta function. In going from Eq. \eqref{eq:appFTmatsu}  to \eqref{eq:appFTfinal} analytical continuation has been performed.
Equation \eqref{eq:appFTfinal} is a function only of the combination $\beta \omega$. In the zero-temperature limit ($\beta\rightarrow \infty$),  $G_a(\omega)$ displays power-law behavior.
Equations \eqref{eq:appFTboson} and \eqref{eq:appFTfermion} are obtained from Eq. \eqref{eq:appFTfinal} using the relation between the beta and the gamma functions, %$\mbox{B}(x,y)=\Gamma(x)\Gamma(y)/\Gamma(x+y)$, 
as well as  the asymptotic expansion of the gamma function for $\beta \omega \gg 1$,
\begin{align*}
    \Gamma\big(\dfrac{\zeta}{2}+i\dfrac{\beta \omega}{2\pi}\big)\Gamma\big(\dfrac{\zeta}{2}-i\dfrac{\beta \omega}{2\pi}\big)\stackrel{.}{=}2\pi e^{-\beta\omega/2}\big( \dfrac{\beta \omega}{2\pi}\big)^{\zeta-1}\Gamma\big( \zeta\big),
\end{align*}
where $\stackrel{.}{=}$ indicates the leading order term. Therefore, the Fourier transform of Eq. \eqref{eq:appFT1} displays $\omega/T$ scaling of the form $T^{\zeta-1}G_a(\omega)=\Phi(\omega/T)$.\\
If $\zeta\geq 1$, the radius $\tau_0^c$ of  the segments labeled $c_2$ and $c_6$ cannot be contracted to zero. As a result, additional terms, controlled by $\tau_0^c$, contribute to the Fourier transform of Eq. \eqref{eq:appFT1}.

The case $\zeta =1$ is realized for the strong-coupling fixed point of the standard Kondo model where we identify  $\tau_0^c$ with $1/T_K$. Indeed, as $T_K \rightarrow \infty$, 
a {\itshape trivial} $\omega/T$ scaling is found.

%%%%%%%%%%%%%%%%%%%%%%%%%%%%%%%%%%%%%%%%%%%%%%%%%%%%%%%%%%%%%%%%%%%
\section{Scaling ansatz -- leading and subleading behavior}
\label{App:SA}
In the limit of vanishing temperature, $T=0$, the large-$N$ equations allow for an asymptotically exact solution for $\omega \rightarrow 0$. Following Ref.\cite{Parcollet.98}, we make the scaling ansatz
\begin{align}
\label{eq:app-scal-ansatz1}
	G_f(\tau,T=0)&=-A_1\Big(\dfrac{\tau_0}{\tau}\Big)^{\alpha_{f}}-A_2\Big(\dfrac{\tau_0}{\tau}\Big)^{\alpha_{f}'}+\ldots\\
	G_B(\tau,T=0)&=-B_1\Big(\dfrac{\tau_0}{\tau}\Big)^{\alpha_{B}}-B_2\Big(\dfrac{\tau_0}{\tau}\Big)^{\alpha_{B}'}+\ldots
	\label{eq:app-scal-ansatz2}
\end{align}
for the $T=0$ solutions of the saddle point equations at $q=1/2$
with $\alpha_1<\alpha_2$ and $\beta_1<\beta_2$. It follows from Eqs.\eqref{eq:appFTfermion} and \eqref{eq:appFTboson} that
\begin{align}
	&G_f^{-1}(\omega+i0^+,T=0)=-\dfrac{A_1^{-1}}{X_{\alpha_{f}}^f}\dfrac{\Gamma(\alpha_{f})}{\pi \tau_0^{\alpha_{f}}}|\omega|^{1-\alpha_{f}}  \\
	&+\dfrac{A_2}{A_1^2}\dfrac{\Gamma(\alpha_{f})\Gamma(\alpha_{f})}{\Gamma(\alpha_{f}')}\dfrac{\tau_0^{\alpha_{f}'-2\alpha_{f}}}{\pi} \dfrac{X^f_{\alpha_{f}'}}{(X^f_{\alpha_{f}})^2}|\omega|^{1+\alpha_{f}'-2\alpha_{f}}, \nonumber \\
	&G_B^{-1}(\omega+i0^+,T=0)=-\dfrac{B_1^{-1}}{X_{\alpha_{B}}^B}\dfrac{\Gamma(\alpha_{B})}{\pi \tau_0^{\alpha_{B}}}|\omega|^{1-\alpha_{B}}  \\
	&+\dfrac{B_2}{B_1^2}\dfrac{\Gamma(\alpha_{B})\Gamma(\alpha_{B})}{\Gamma(\alpha_{B}')}\dfrac{\tau_0^{\alpha_{B}'-2\alpha_{B}}}{\pi} \dfrac{X^B_{\alpha_{B}'}}{(X^B_{\alpha_{B}})^2}|\omega|^{1+\alpha_{B}'-2\alpha_{B}},
\end{align}
where $X^B_{\alpha}=\tan\big(\frac{\pi \alpha}{2}\big)+i\, \text{sgn}(\omega)$ and $X^f_\alpha=-\cot\big(\frac{\pi \alpha}{2}\big)\text{sgn}(\omega)+i$.
From the saddle point equations we obtain for  $\Sigma_f(\omega+i0^+,T=0)$ and $\Sigma_B(\omega+i0^+,T=0)$ up to and including subleading terms
\begin{widetext}
\begin{align}
	\Sigma_B(\omega)&=\dfrac{\pi A_0 A_1}{\Gamma(\alpha_{f})} B(r+1,\alpha_{f}) X_{r+\alpha_{f}}^f \tau_0^{r+\alpha_{f}-1}|\omega|^{r+\alpha_{f}} \text{sgn}(\omega)  +\dfrac{\pi A_0 A_2}{\Gamma(\alpha_{f}')} B(r+1,\alpha_{f}') X_{r+\alpha_{f}'}^f \tau_0^{r+\alpha_{f}'-1}|\omega|^{r+\alpha_{f}'} \text{sgn}(\omega), \nonumber \\
	\Sigma_{f}^{1}(\omega)&=-\dfrac{\pi \kappa A_0 B_1}{\Gamma(\alpha_{B})} B(r+1,\alpha_{B}) X_{-r-\alpha_{B}}^B \tau_0^{r+\beta_1-1}|\omega|^{r+\alpha_{B}} \text{sgn}(\omega)  -\dfrac{\pi \kappa A_0 B_2}{\Gamma(\alpha_{B}')} B(r+1,\alpha_{B}')
	X_{-r-\alpha_{B}'}^B \tau_0^{r+\alpha_{B}'-1}|\omega|^{r+\alpha_{B}'} \text{sgn}(\omega), \nonumber
\end{align}
and
\begin{align}
	\Sigma_{f}^{2}(\omega)=&-\pi g^2\dfrac{ K_0^2 A_1}{\Gamma(\alpha_{f})} B(2-\epsilon,\alpha_{f}) X_{\epsilon-\alpha_{f}-1}^B \tau_0^{\alpha_{f}-\epsilon}|\omega|^{1+\alpha_{f}-\epsilon} \text{sgn}(\omega)  \nonumber \\
						   &-\pi g^2\dfrac{ K_0^2 A_2}{\Gamma(\alpha_{f}')} B(2-\epsilon,\alpha_{f}') X_{\epsilon-\alpha_{f}'-1}^B  \tau_0^{\alpha_{f}'-\epsilon}|\omega|^{1+\alpha_{f}'-\epsilon} \text{sgn}(\omega). \nonumber
\end{align}
\end{widetext}
which, together with the Dyson equation
\begin{align}
    G_f^{-1}(\omega)& =\omega+\lambda-\Sigma_f(\omega)\\
	G_B^{-1}(\omega)&=J_{K}^{-1}-\Sigma_B(\omega),
\end{align}
results in a set of conditions for $\alpha_{f},\alpha_{B},\alpha_{f}',\alpha_{B}'$ as well as the leading and subleading amplitudes at the various fixed points (except the trivial one at $J_{K}=0$ and $g=0$). $B(x,y)$ in the self-energy expressions denotes the Euler beta function $B(x,y)=\Gamma(x)\Gamma(y)/\Gamma(x+y)$.\\
Specifically, we find the following:
\begin{itemize}
    \item the multichannel Kondo and the pseudogap Kondo fixed point exponents are obtained as solutions of
    \begin{align}
		&\alpha_{f}+\alpha_{B}=1-r \nonumber \\
		& \kappa=\dfrac{(1-\alpha_{f})\tan\big(\pi \alpha_{f}/2 \big)}{(r+\alpha_{f})\tan\big(\pi(r+\alpha_{f})/2 \big)}\nonumber
    \end{align}
    where the first equation results from equating the leading exponents while the second  follows from equating the amplitudes. In a similar fashion, the subleading exponents have to obey
    \begin{align}
		\alpha_{f}'-\alpha_{B}'+2\alpha_{B} &=1-r \nonumber \\
		\Big(\dfrac{A_2}{A_1} \dfrac{\Gamma(\alpha_{f})}{\Gamma(\alpha_{f}')}\Big)^2 &=\kappa \dfrac{\pi^2A_0A_2B_2}{\Gamma(\alpha_{f}')\Gamma(\alpha_{B}')}B(r+1,\alpha_{B}')\nonumber \\ 									 &~~~~\times X_{r+\alpha_{B}'}^B (X_{\alpha_{f}}^f)^2/X^f_{\alpha_{f}'} \nonumber \\
		\Big(\dfrac{B_2}{B_1} \dfrac{\Gamma(\alpha_{B})}{\Gamma(\alpha_{B}')}\Big)^2 &= \dfrac{\pi^2A_0A_2B_2}{\Gamma(\alpha_{f}')\Gamma(\alpha_{B}')}B(r+1,\alpha_{f}')\nonumber \\ 
	 &~~~~\times X_{r+\alpha_{f}'}^f (X_{\alpha_{B}}^B)^2/X^B_{\alpha_{B}'} \nonumber 
    \end{align}
    \item for the critical point C', one finds for the scaling exponents in leading order 
		$\alpha_{f}=\epsilon/2$ and $\alpha_{B}=1-r-\epsilon/2$ while the subleading behavior is characterized by  $\alpha_{f}'=\epsilon$ and $\alpha_{B}'=1-r$.
	\item the leading scaling exponents of the LM' fixed points are $\alpha_{f}=\epsilon/2$ and $\alpha_{B}=1+r+\epsilon/2$. Likewise, we conclude that the subleading exponents are $\alpha_{f}'=\epsilon$ and $\alpha_{B}'=1+r+\epsilon$.
\end{itemize}
%DISCUSS $r^{max}$!

\begin{figure}[h!]
\begin{picture}(0,0)
	\put(35,102){\textsf{(a)}}
	\put(158,102){\textsf{(b)}}
\end{picture}
\centering
\includegraphics[width=0.488\columnwidth]{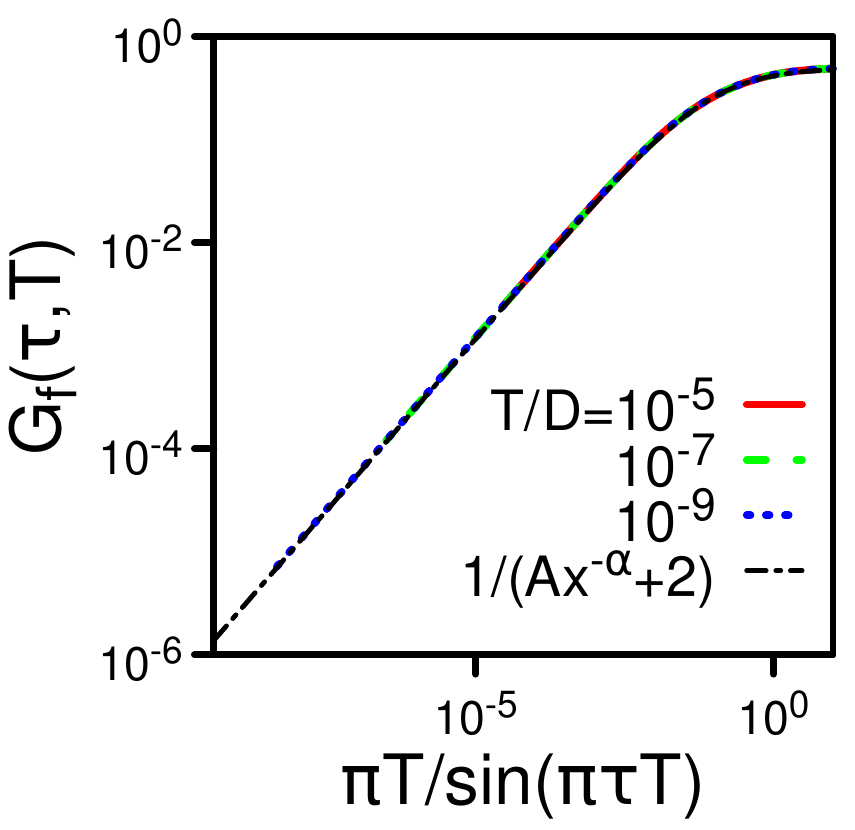}
\includegraphics[width=0.488\columnwidth]{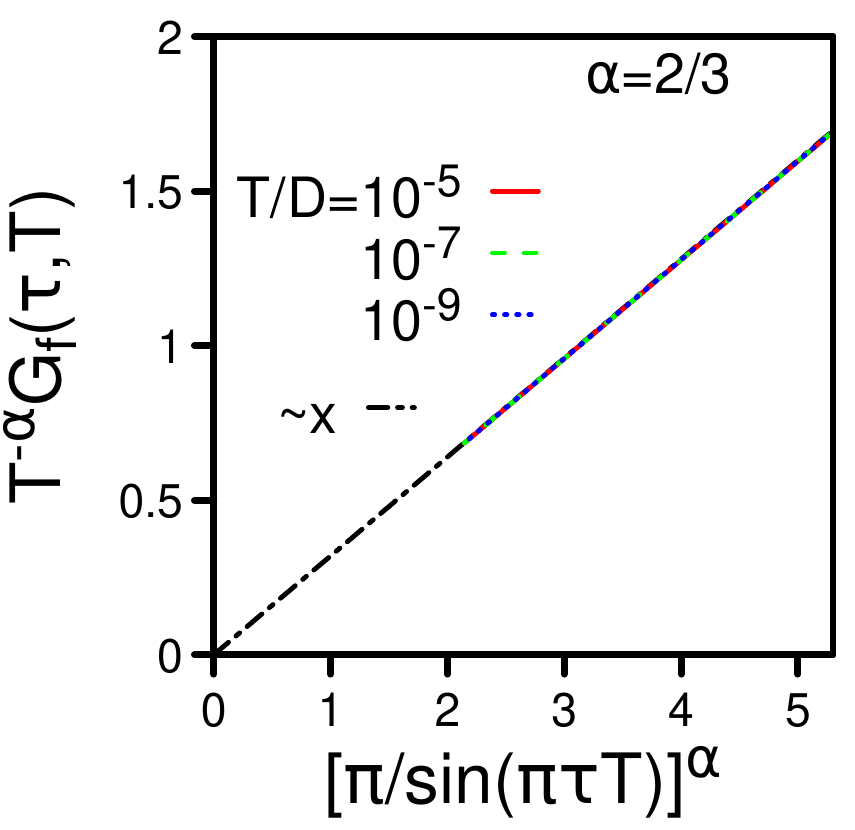}
\caption{Behavior of $G_{f}(\tau)$ for the $r=0$ MCK fixed point case. (a) $G_{f}(\tau)$ at various temperatures. The black dashed line is the fitting of $G_{f}(\tau)$ with $\frac{1}{Ax^{-\alpha_{f}}+2}$ at $r=0$ case. (b) Behavior of $G_{f}(\tau)$ near $\beta/2$ regime.}
\label{fig:r0_fit}
\end{figure}

%%%%%%%%%%%
In the special case $r=0$ and $g=0$, a simple expression for $G_f(\tau)$, valid at any $T$, has been found to describe our numerical results for $G_f(\tau)$
\begin{align}
\label{eq:interpolation}
G_f(\tau)=\Bigg(\dfrac{A_1^{-1}}{\Big(\dfrac{\pi\tau_0/\beta}{\sin{(\pi \tau/\beta)}}\Big)^{\alpha_{f}}}+\dfrac{1}{q_{0}} \Bigg )^{-1},
\end{align}
which is in line with the results of Ref.\cite{Parcollet.98}. Figure \ref{fig:r0_fit} shows a comparison of Eq.~(\ref{eq:interpolation}) with the numerical solution of the large-$N$ equations for the particle-hole symmetric case, {\itshape i.e.}, $q_{0}=1/2$.
%%%
%%%
For $G_B(\tau)$, no equivalent expression valid at all $\tau$ and $T$ has been found.

%%%%%%%%%%%%%%%%%%%%%%%%%%%%%%%%%%%%%%%%%%%%%%%%%%%%%%%%%%%%%%%%%%%
\section{Scaling and  entropy in the SYK model}
\label{app:SYK}

%%%%%%%%%%%%%%%%%% SYK %%%%%%%%%%%%%%%%%%%%%%%%%%%%%%%%%%%%%
\begin{figure}[t!]
	\begin{picture}(0,0)
		\put(34,95){\textsf{(a)}}
		\put(150,95){\textsf{(b)}}
		\put(34,-16){\textsf{(c)}}
		\put(150,-16){\textsf{(d)}}
	\end{picture}
\centering
\includegraphics[width=0.46\columnwidth]{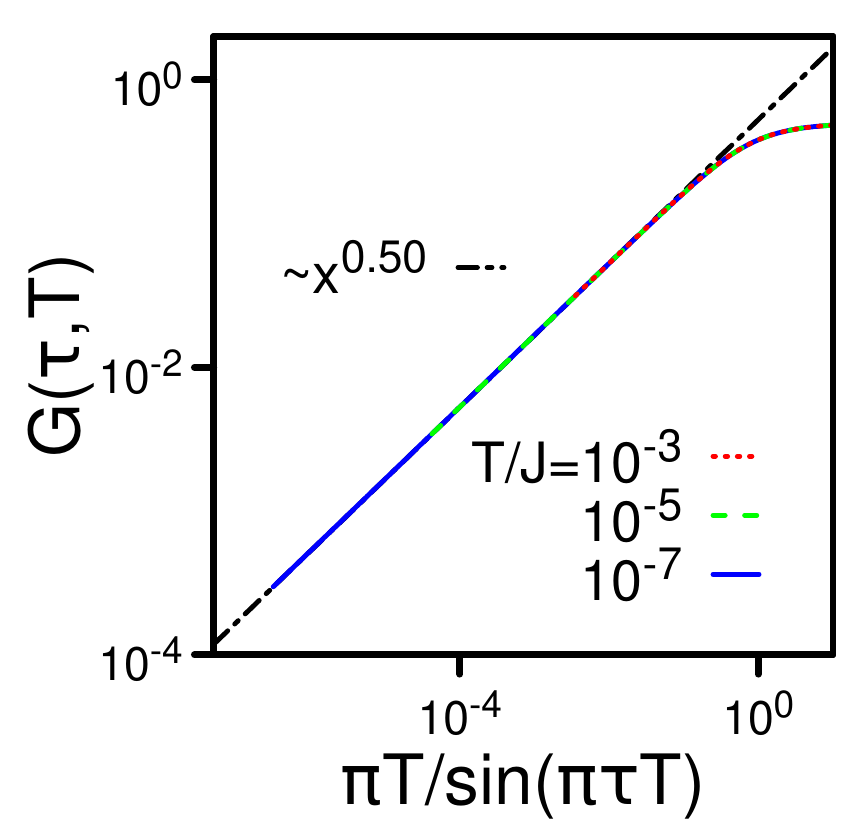}
\includegraphics[width=0.46\columnwidth]{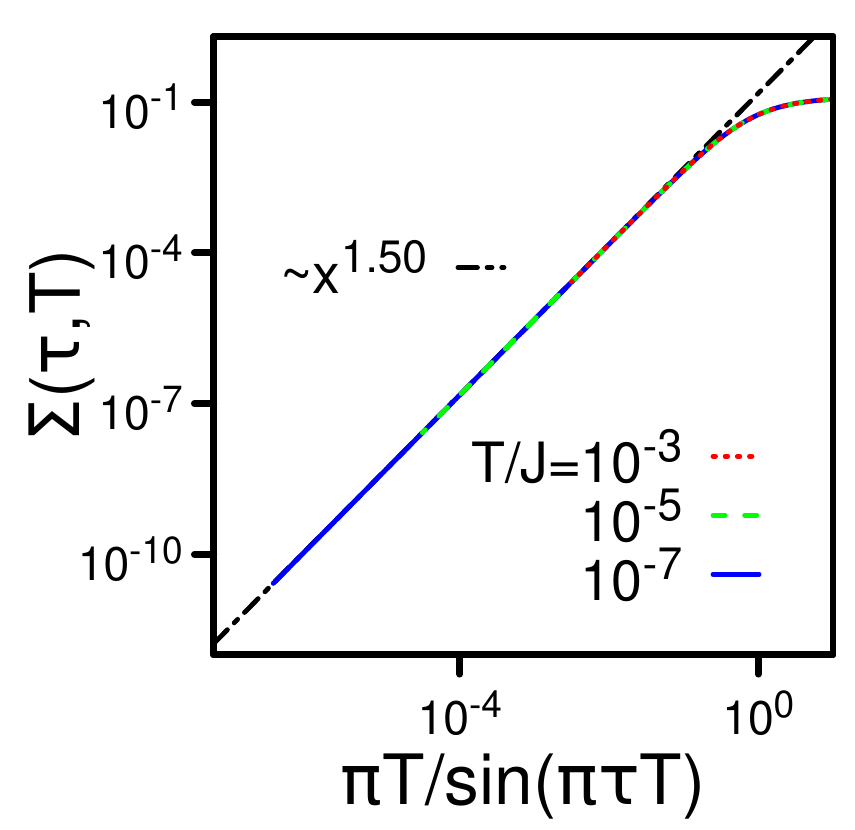}\\
\includegraphics[width=0.46\columnwidth]{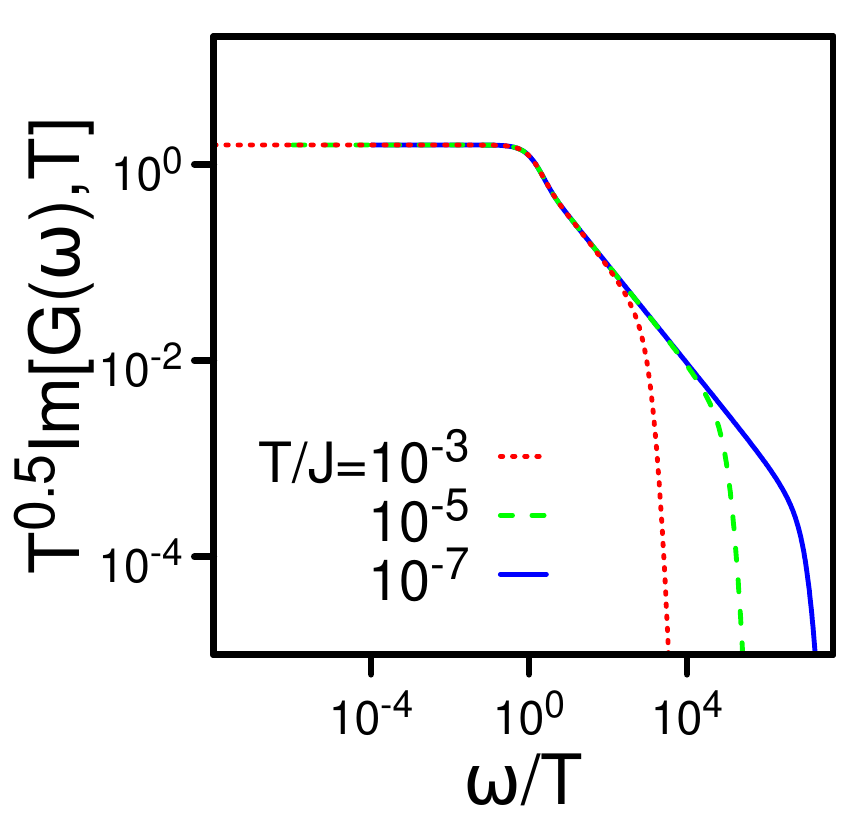}
\includegraphics[width=0.46\columnwidth]{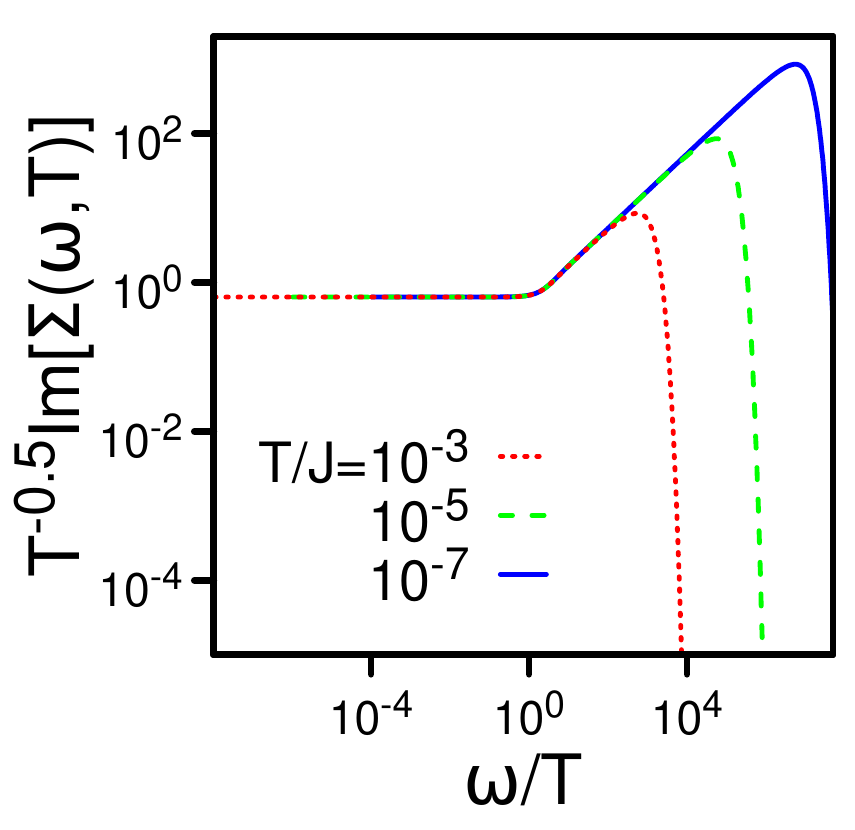}
\caption{Numerical solution of (a) $G(\tau)$ and (b) $\Sigma(\tau)$ at different temperature. $\omega/T$ scaling of $\mathrm{Im}G(\omega)$ and $\mathrm{Im}\Sigma(\omega)$ at different temperature.}
\label{fig:SYK2}
\end{figure}
%%%%%%%%%%%%%%%%%%%%%%%%%%%%%%%%%%%%%%%%%%%%%%%%%%%%%%%%%%

The large-$N$ equation of the SYK model is
\begin{align}
    \Sigma(\tau)=-J^{2}G(-\tau)^{3},
\end{align}

together with the Dyson equation,
\begin{align}
    G(i\omega_{n})=\frac{1}{i\omega_{n}-\Sigma(i\omega_{n})},
\end{align}
where $i\omega_n~(n = \pm 1,\pm 3, \dots )$ are fermionic Matsubara frequencies and $J$ is a coupling constant. 
The strong coupling limit of the SYK model is known to possess an emergent conformal invariance  and the $T=0$ residual entropy has been obtained using analytic methods. One can also show that the entropy approaches $1/2\ln 2$ at high $T$ \cite{Maldacena.16}.
%%%%%%%%%%%%%%%%%%%%%%%%%%%%%%%%%%%%%%%%%%%%%%%%%%%%%%%%%%%%%
\begin{figure}[h!]
\centering
\includegraphics[width=0.7\columnwidth]{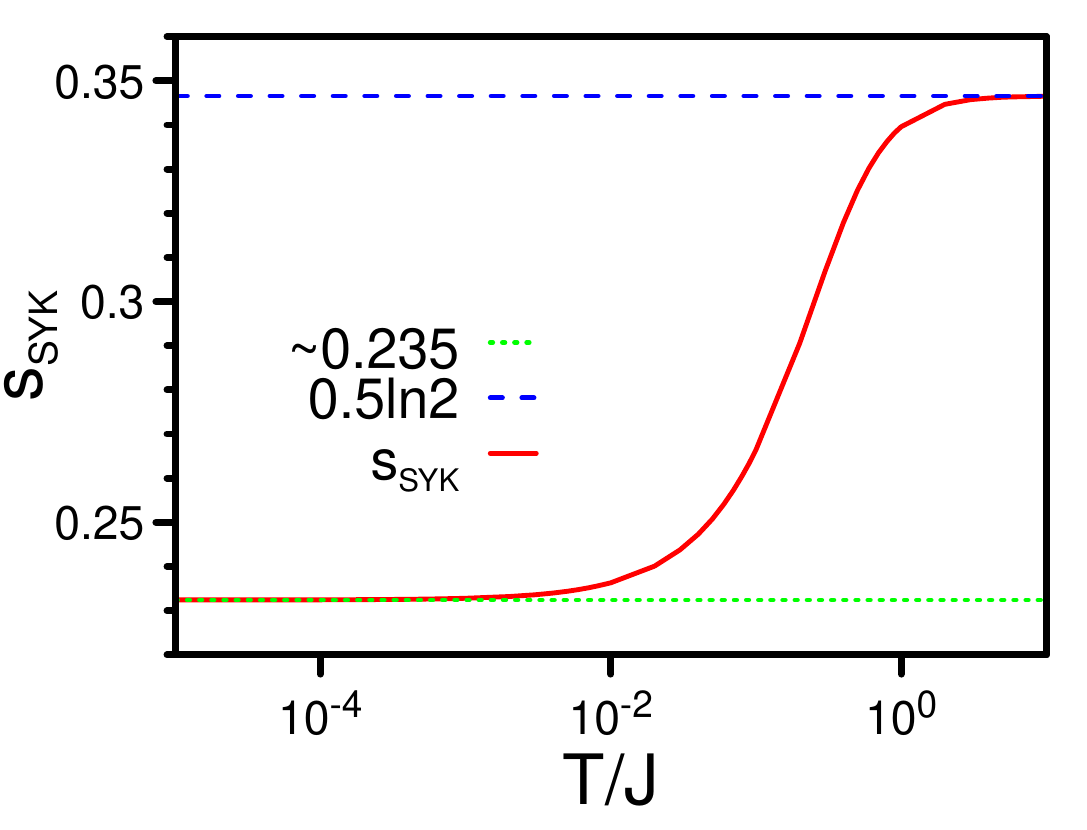}
\caption{Temperature dependence of the entropy of the SYK model. Our result interpolates between the analytical known results for $s_{\mbox{\tiny SYK}}(T=0)$ and $s_{\mbox{\tiny SYK}}(T\gg J)$.}
\label{fig:SYK1}
\end{figure}
%%%%%%%%%%%%%%%%%%%%%%%%%%%%%%%%%%%%%%%%%%%%%%%%%%%%%%%%%%%%%
From the Luttinger-Ward functional,
\begin{align}
	F=&\dfrac{-T}{2}\ln2+\dfrac{1}{2\pi}\int \!d\omega f(\omega)\Bigg\{\mathrm{Im}\Big [ \ln[\dfrac{-G(\omega)^{-1}}{-G_{0}(\omega)^{-1}}\Big]\Big]
	\nonumber \\
	  &+\mathrm{Im}[\Sigma(\omega)G(\omega)]-\dfrac{J^{2}}{4}\int_{\tau}G(\tau)^{4}\Bigg\},
\end{align}
the entropy follows as
\begin{align}
	s_{\mbox{\tiny SYK}}=&\dfrac{1}{2}\ln2-\dfrac{1}{2\pi}\int \!d\omega\dfrac{df(\omega)}{dT}\Bigg(\mathrm{Im}\ln\Big[\dfrac{G_{0}(\omega)}{G_{}(\omega)}\Big] \nonumber \\
	&+\mathrm{Im}\big[\Sigma(\omega)\big]\mathrm{Re}\big[G(\omega)\big]\Bigg) 
\end{align}
Applying the scaling ansatz 
\begin{align}
    G(\tau)=-A\Big(\dfrac{\tau_0}{\tau}\Big)^{\alpha}-B \Big(\dfrac{\tau_0}{\tau}\Big)^{\alpha'}
\end{align}
to the SYK model yields $\alpha=0.5$ and $\alpha'=2 \alpha$.
We solve the self-consistent equation in real frequency space and obtain $G(\omega)$, $G(\tau)$ as well as the entropy $s_{\mbox{\tiny SYK}}(T)$.  The resulting scaling properties of $G(\tau)$, $\Sigma(\tau)$ and the complementary $\omega/T$ scaling of $G(\omega)$ and $\Sigma(\omega)$ are shown in Fig.~\ref{fig:SYK2}. The entropy $s_{\mbox{\tiny SYK}}(T)$, shown in Fig.~\ref{fig:SYK1}, flows from the weak coupling fixed point at high $T$, to the strong-coupling fixed point at $T=0$.
Our results for $s_{\mbox{\tiny SYK}}(T)$ and the exponents ($\alpha=\frac{2}{q}=0.5$) for $G(\omega)$ and $G(\tau)$ agree well with analytical prediction\cite{Maldacena.16}. 
%%%%%%%%%%%%%%%%%%%%%%%%%%%%%%%%%%%%%%%%%%%%%%%%%%%%%%%%%%%%%%%%%%%%
\bibliography{pgBFKM.bib}

%merlin.mbs apsrev4-1.bst 2010-07-25 4.21a (PWD, AO, DPC) hacked
%Control: key (0)
%Control: author (8) initials jnrlst
%Control: editor formatted (1) identically to author
%Control: production of article title (-1) disabled
%Control: page (0) single
%Control: year (1) truncated
%Control: production of eprint (0) enabled
\begin{thebibliography}{73}%
\makeatletter
\providecommand \@ifxundefined [1]{%
 \@ifx{#1\undefined}
}%
\providecommand \@ifnum [1]{%
 \ifnum #1\expandafter \@firstoftwo
 \else \expandafter \@secondoftwo
 \fi
}%
\providecommand \@ifx [1]{%
 \ifx #1\expandafter \@firstoftwo
 \else \expandafter \@secondoftwo
 \fi
}%
\providecommand \natexlab [1]{#1}%
\providecommand \enquote  [1]{``#1''}%
\providecommand \bibnamefont  [1]{#1}%
\providecommand \bibfnamefont [1]{#1}%
\providecommand \citenamefont [1]{#1}%
\providecommand \href@noop [0]{\@secondoftwo}%
\providecommand \href [0]{\begingroup \@sanitize@url \@href}%
\providecommand \@href[1]{\@@startlink{#1}\@@href}%
\providecommand \@@href[1]{\endgroup#1\@@endlink}%
\providecommand \@sanitize@url [0]{\catcode `\\12\catcode `\$12\catcode
  `\&12\catcode `\#12\catcode `\^12\catcode `\_12\catcode `\%12\relax}%
\providecommand \@@startlink[1]{}%
\providecommand \@@endlink[0]{}%
\providecommand \url  [0]{\begingroup\@sanitize@url \@url }%
\providecommand \@url [1]{\endgroup\@href {#1}{\urlprefix }}%
\providecommand \urlprefix  [0]{URL }%
\providecommand \Eprint [0]{\href }%
\providecommand \doibase [0]{http://dx.doi.org/}%
\providecommand \selectlanguage [0]{\@gobble}%
\providecommand \bibinfo  [0]{\@secondoftwo}%
\providecommand \bibfield  [0]{\@secondoftwo}%
\providecommand \translation [1]{[#1]}%
\providecommand \BibitemOpen [0]{}%
\providecommand \bibitemStop [0]{}%
\providecommand \bibitemNoStop [0]{.\EOS\space}%
\providecommand \EOS [0]{\spacefactor3000\relax}%
\providecommand \BibitemShut  [1]{\csname bibitem#1\endcsname}%
\let\auto@bib@innerbib\@empty
%</preamble>
\bibitem [{\citenamefont {Hertz}(1976)}]{Hertz.76}%
  \BibitemOpen
  \bibfield  {author} {\bibinfo {author} {\bibfnamefont {J.}~\bibnamefont
  {Hertz}},\ }\href@noop {} {\bibfield  {journal} {\bibinfo  {journal}
  {Phys.~Rev.~B}\ }\textbf {\bibinfo {volume} {14}},\ \bibinfo {pages} {1165}
  (\bibinfo {year} {1976})}\BibitemShut {NoStop}%
\bibitem [{\citenamefont {Sachdev}(1999)}]{Sachdev}%
  \BibitemOpen
  \bibfield  {author} {\bibinfo {author} {\bibfnamefont {S.}~\bibnamefont
  {Sachdev}},\ }\href@noop {} {\emph {\bibinfo {title} {Quantum Phase
  Transitions}}}\ (\bibinfo  {publisher} {Cambridge University Press},\
  \bibinfo {address} {Cambridge},\ \bibinfo {year} {1999})\BibitemShut
  {NoStop}%
\bibitem [{\citenamefont {Hussey}\ \emph {et~al.}(2013)\citenamefont {Hussey},
  \citenamefont {Gordon-Moys}, \citenamefont {Kokalj},\ and\ \citenamefont
  {McKenzie}}]{Hussey.13}%
  \BibitemOpen
  \bibfield  {author} {\bibinfo {author} {\bibfnamefont {N.~E.}\ \bibnamefont
  {Hussey}}, \bibinfo {author} {\bibfnamefont {H.}~\bibnamefont {Gordon-Moys}},
  \bibinfo {author} {\bibfnamefont {J.}~\bibnamefont {Kokalj}}, \ and\ \bibinfo
  {author} {\bibfnamefont {R.~H.}\ \bibnamefont {McKenzie}},\ }\href {\doibase
  10.1088/1742-6596/449/1/012004} {\bibfield  {journal} {\bibinfo  {journal}
  {Journal of Physics: Conference Series}\ }\textbf {\bibinfo {volume} {449}},\
  \bibinfo {pages} {012004} (\bibinfo {year} {2013})}\BibitemShut {NoStop}%
\bibitem [{\citenamefont {Keimer}\ \emph {et~al.}(2015)\citenamefont {Keimer},
  \citenamefont {Kivelson}, \citenamefont {Norman}, \citenamefont {Uchida},\
  and\ \citenamefont {Zaanen}}]{Keimer.15}%
  \BibitemOpen
  \bibfield  {author} {\bibinfo {author} {\bibfnamefont {B.}~\bibnamefont
  {Keimer}}, \bibinfo {author} {\bibfnamefont {S.~A.}\ \bibnamefont
  {Kivelson}}, \bibinfo {author} {\bibfnamefont {M.~R.}\ \bibnamefont
  {Norman}}, \bibinfo {author} {\bibfnamefont {S.}~\bibnamefont {Uchida}}, \
  and\ \bibinfo {author} {\bibfnamefont {J.}~\bibnamefont {Zaanen}},\
  }\href@noop {} {\bibfield  {journal} {\bibinfo  {journal} {Nature}\ }\textbf
  {\bibinfo {volume} {518}},\ \bibinfo {pages} {179–186} (\bibinfo {year}
  {2015})}\BibitemShut {NoStop}%
\bibitem [{\citenamefont {Si}\ \emph {et~al.}(2001)\citenamefont {Si},
  \citenamefont {Rabello}, \citenamefont {Ingersent},\ and\ \citenamefont
  {Smith}}]{Si.01}%
  \BibitemOpen
  \bibfield  {author} {\bibinfo {author} {\bibfnamefont {Q.}~\bibnamefont
  {Si}}, \bibinfo {author} {\bibfnamefont {S.}~\bibnamefont {Rabello}},
  \bibinfo {author} {\bibfnamefont {K.}~\bibnamefont {Ingersent}}, \ and\
  \bibinfo {author} {\bibfnamefont {J.~L.}\ \bibnamefont {Smith}},\ }\href
  {\doibase 10.1038/35101507} {\bibfield  {journal} {\bibinfo  {journal}
  {Nature}\ }\textbf {\bibinfo {volume} {413}},\ \bibinfo {pages} {804}
  (\bibinfo {year} {2001})}\BibitemShut {NoStop}%
\bibitem [{\citenamefont {Coleman}\ \emph {et~al.}(2001)\citenamefont
  {Coleman}, \citenamefont {P\'{e}pin}, \citenamefont {Si},\ and\ \citenamefont
  {Ramazashvili}}]{Coleman.01}%
  \BibitemOpen
  \bibfield  {author} {\bibinfo {author} {\bibfnamefont {P.}~\bibnamefont
  {Coleman}}, \bibinfo {author} {\bibfnamefont {C.}~\bibnamefont {P\'{e}pin}},
  \bibinfo {author} {\bibfnamefont {Q.}~\bibnamefont {Si}}, \ and\ \bibinfo
  {author} {\bibfnamefont {R.}~\bibnamefont {Ramazashvili}},\ }\href@noop {}
  {\bibfield  {journal} {\bibinfo  {journal} {J.~Phys.~Cond.~Matt.}\ }\textbf
  {\bibinfo {volume} {13}},\ \bibinfo {pages} {R723} (\bibinfo {year}
  {2001})}\BibitemShut {NoStop}%
\bibitem [{\citenamefont {Si}(2006)}]{Si.06}%
  \BibitemOpen
  \bibfield  {author} {\bibinfo {author} {\bibfnamefont {Q.}~\bibnamefont
  {Si}},\ }\href {\doibase https://doi.org/10.1016/j.physb.2006.01.156}
  {\bibfield  {journal} {\bibinfo  {journal} {Physica B: Condensed Matter}\
  }\textbf {\bibinfo {volume} {378-380}},\ \bibinfo {pages} {23 } (\bibinfo
  {year} {2006})}\BibitemShut {NoStop}%
\bibitem [{\citenamefont {Coleman}\ and\ \citenamefont
  {Nevidomskyy}(2010)}]{Coleman.10}%
  \BibitemOpen
  \bibfield  {author} {\bibinfo {author} {\bibfnamefont {P.}~\bibnamefont
  {Coleman}}\ and\ \bibinfo {author} {\bibfnamefont {A.~H.}\ \bibnamefont
  {Nevidomskyy}},\ }\href {\doibase 10.1007/s10909-010-0213-4} {\bibfield
  {journal} {\bibinfo  {journal} {Journal of Low Temperature Physics}\ }\textbf
  {\bibinfo {volume} {161}},\ \bibinfo {pages} {182} (\bibinfo {year}
  {2010})}\BibitemShut {NoStop}%
\bibitem [{\citenamefont {Si}\ \emph {et~al.}(2014)\citenamefont {Si},
  \citenamefont {Pixley}, \citenamefont {Nica}, \citenamefont {Yamamoto},
  \citenamefont {Goswami}, \citenamefont {Yu},\ and\ \citenamefont
  {Kirchner}}]{Si.14}%
  \BibitemOpen
  \bibfield  {author} {\bibinfo {author} {\bibfnamefont {Q.}~\bibnamefont
  {Si}}, \bibinfo {author} {\bibfnamefont {J.~H.}\ \bibnamefont {Pixley}},
  \bibinfo {author} {\bibfnamefont {E.}~\bibnamefont {Nica}}, \bibinfo {author}
  {\bibfnamefont {S.~J.}\ \bibnamefont {Yamamoto}}, \bibinfo {author}
  {\bibfnamefont {P.}~\bibnamefont {Goswami}}, \bibinfo {author} {\bibfnamefont
  {R.}~\bibnamefont {Yu}}, \ and\ \bibinfo {author} {\bibfnamefont
  {S.}~\bibnamefont {Kirchner}},\ }\href@noop {} {\bibfield  {journal}
  {\bibinfo  {journal} {J.~Phys.~Soc.~Jpn.}\ }\textbf {\bibinfo {volume}
  {83}},\ \bibinfo {pages} {061005} (\bibinfo {year} {2014})}\BibitemShut
  {NoStop}%
\bibitem [{\citenamefont {Kirchner}\ \emph {et~al.}(2020)\citenamefont
  {Kirchner}, \citenamefont {Paschen}, \citenamefont {Chen}, \citenamefont
  {Wirth}, \citenamefont {Feng}, \citenamefont {Thompson},\ and\ \citenamefont
  {Si}}]{Kirchner.20}%
  \BibitemOpen
  \bibfield  {author} {\bibinfo {author} {\bibfnamefont {S.}~\bibnamefont
  {Kirchner}}, \bibinfo {author} {\bibfnamefont {S.}~\bibnamefont {Paschen}},
  \bibinfo {author} {\bibfnamefont {Q.}~\bibnamefont {Chen}}, \bibinfo {author}
  {\bibfnamefont {S.}~\bibnamefont {Wirth}}, \bibinfo {author} {\bibfnamefont
  {D.}~\bibnamefont {Feng}}, \bibinfo {author} {\bibfnamefont {J.~D.}\
  \bibnamefont {Thompson}}, \ and\ \bibinfo {author} {\bibfnamefont
  {Q.}~\bibnamefont {Si}},\ }\href {\doibase 10.1103/RevModPhys.92.011002}
  {\bibfield  {journal} {\bibinfo  {journal} {Rev. Mod. Phys.}\ }\textbf
  {\bibinfo {volume} {92}},\ \bibinfo {pages} {011002} (\bibinfo {year}
  {2020})}\BibitemShut {NoStop}%
\bibitem [{\citenamefont {Park}\ \emph {et~al.}(2008)\citenamefont {Park},
  \citenamefont {Sidorov}, \citenamefont {Ronning}, \citenamefont {Zhu},
  \citenamefont {Tokiwa}, \citenamefont {Lee}, \citenamefont {Bauer},
  \citenamefont {Movshovich}, \citenamefont {Sarrao},\ and\ \citenamefont
  {Thompson}}]{Park.08}%
  \BibitemOpen
  \bibfield  {author} {\bibinfo {author} {\bibfnamefont {T.}~\bibnamefont
  {Park}}, \bibinfo {author} {\bibfnamefont {V.~A.}\ \bibnamefont {Sidorov}},
  \bibinfo {author} {\bibfnamefont {F.}~\bibnamefont {Ronning}}, \bibinfo
  {author} {\bibfnamefont {J.-X.}\ \bibnamefont {Zhu}}, \bibinfo {author}
  {\bibfnamefont {Y.}~\bibnamefont {Tokiwa}}, \bibinfo {author} {\bibfnamefont
  {H.}~\bibnamefont {Lee}}, \bibinfo {author} {\bibfnamefont {E.~D.}\
  \bibnamefont {Bauer}}, \bibinfo {author} {\bibfnamefont {R.}~\bibnamefont
  {Movshovich}}, \bibinfo {author} {\bibfnamefont {J.~L.}\ \bibnamefont
  {Sarrao}}, \ and\ \bibinfo {author} {\bibfnamefont {J.~D.}\ \bibnamefont
  {Thompson}},\ }\href@noop {} {\bibfield  {journal} {\bibinfo  {journal}
  {Nature}\ }\textbf {\bibinfo {volume} {456}},\ \bibinfo {pages} {366}
  (\bibinfo {year} {2008})}\BibitemShut {NoStop}%
\bibitem [{\citenamefont {Grube}\ \emph {et~al.}(2017)\citenamefont {Grube},
  \citenamefont {Zaum}, \citenamefont {Stockert}, \citenamefont {Si},\ and\
  \citenamefont {L{\"o}hneysen}}]{Grube.17}%
  \BibitemOpen
  \bibfield  {author} {\bibinfo {author} {\bibfnamefont {K.}~\bibnamefont
  {Grube}}, \bibinfo {author} {\bibfnamefont {S.}~\bibnamefont {Zaum}},
  \bibinfo {author} {\bibfnamefont {O.}~\bibnamefont {Stockert}}, \bibinfo
  {author} {\bibfnamefont {Q.}~\bibnamefont {Si}}, \ and\ \bibinfo {author}
  {\bibfnamefont {H.~v.}\ \bibnamefont {L{\"o}hneysen}},\ }\href@noop {}
  {\bibfield  {journal} {\bibinfo  {journal} {Nature Physics}\ }\textbf
  {\bibinfo {volume} {13}},\ \bibinfo {pages} {742} (\bibinfo {year}
  {2017})}\BibitemShut {NoStop}%
\bibitem [{\citenamefont {Wolf}\ \emph {et~al.}(2011)\citenamefont {Wolf},
  \citenamefont {Tsui}, \citenamefont {Jaiswal-Nagar}, \citenamefont {Tutsch},
  \citenamefont {Honecker}, \citenamefont {Removi{\'c}-Langer}, \citenamefont
  {Hofmann}, \citenamefont {Prokofiev}, \citenamefont {Assmus}, \citenamefont
  {Donath},\ and\ \citenamefont {Lang}}]{Wolf.11}%
  \BibitemOpen
  \bibfield  {author} {\bibinfo {author} {\bibfnamefont {B.}~\bibnamefont
  {Wolf}}, \bibinfo {author} {\bibfnamefont {Y.}~\bibnamefont {Tsui}}, \bibinfo
  {author} {\bibfnamefont {D.}~\bibnamefont {Jaiswal-Nagar}}, \bibinfo {author}
  {\bibfnamefont {U.}~\bibnamefont {Tutsch}}, \bibinfo {author} {\bibfnamefont
  {A.}~\bibnamefont {Honecker}}, \bibinfo {author} {\bibfnamefont
  {K.}~\bibnamefont {Removi{\'c}-Langer}}, \bibinfo {author} {\bibfnamefont
  {G.}~\bibnamefont {Hofmann}}, \bibinfo {author} {\bibfnamefont
  {A.}~\bibnamefont {Prokofiev}}, \bibinfo {author} {\bibfnamefont
  {W.}~\bibnamefont {Assmus}}, \bibinfo {author} {\bibfnamefont
  {G.}~\bibnamefont {Donath}}, \ and\ \bibinfo {author} {\bibfnamefont
  {M.}~\bibnamefont {Lang}},\ }\href {\doibase 10.1073/pnas.1017047108}
  {\bibfield  {journal} {\bibinfo  {journal} {Proceedings of the National
  Academy of Sciences}\ }\textbf {\bibinfo {volume} {108}},\ \bibinfo {pages}
  {6862} (\bibinfo {year} {2011})}\BibitemShut {NoStop}%
\bibitem [{\citenamefont {Nozi{\'{e}}res}\ and\ \citenamefont
  {Blandin}(1980)}]{Nozieres.80}%
  \BibitemOpen
  \bibfield  {author} {\bibinfo {author} {\bibfnamefont {P.}~\bibnamefont
  {Nozi{\'{e}}res}}\ and\ \bibinfo {author} {\bibfnamefont {A.}~\bibnamefont
  {Blandin}},\ }\href@noop {} {\bibfield  {journal} {\bibinfo  {journal}
  {Journal de Physique}\ }\textbf {\bibinfo {volume} {41}},\ \bibinfo {pages}
  {193} (\bibinfo {year} {1980})}\BibitemShut {NoStop}%
\bibitem [{\citenamefont {Tsvelick}(1985)}]{Tsvelik.85}%
  \BibitemOpen
  \bibfield  {author} {\bibinfo {author} {\bibfnamefont {A.~M.}\ \bibnamefont
  {Tsvelick}},\ }\href@noop {} {\bibfield  {journal} {\bibinfo  {journal}
  {Journal of Physics C: Solid State Physics}\ }\textbf {\bibinfo {volume}
  {18}},\ \bibinfo {pages} {159} (\bibinfo {year} {1985})}\BibitemShut
  {NoStop}%
\bibitem [{\citenamefont {Kirchner}(2020)}]{Kirchner.20b}%
  \BibitemOpen
  \bibfield  {author} {\bibinfo {author} {\bibfnamefont {S.}~\bibnamefont
  {Kirchner}},\ }\href {\doibase 10.1002/qute.201900128} {\bibfield  {journal}
  {\bibinfo  {journal} {Adv.\ Quantum Technol.}\ }\textbf {\bibinfo {volume}
  {3}},\ \bibinfo {pages} {1900128} (\bibinfo {year} {2020})}\BibitemShut
  {NoStop}%
\bibitem [{\citenamefont {Withoff}\ and\ \citenamefont
  {Fradkin}(1990)}]{Withoff.90}%
  \BibitemOpen
  \bibfield  {author} {\bibinfo {author} {\bibfnamefont {D.}~\bibnamefont
  {Withoff}}\ and\ \bibinfo {author} {\bibfnamefont {E.}~\bibnamefont
  {Fradkin}},\ }\href {\doibase 10.1103/PhysRevLett.64.1835} {\bibfield
  {journal} {\bibinfo  {journal} {Phys. Rev. Lett.}\ }\textbf {\bibinfo
  {volume} {64}},\ \bibinfo {pages} {1835} (\bibinfo {year}
  {1990})}\BibitemShut {NoStop}%
\bibitem [{\citenamefont {Gonzalez-Buxton}\ and\ \citenamefont
  {Ingersent}(1998)}]{Gonzalez-Buxton.98}%
  \BibitemOpen
  \bibfield  {author} {\bibinfo {author} {\bibfnamefont {C.}~\bibnamefont
  {Gonzalez-Buxton}}\ and\ \bibinfo {author} {\bibfnamefont {K.}~\bibnamefont
  {Ingersent}},\ }\href {\doibase 10.1103/PhysRevB.57.14254} {\bibfield
  {journal} {\bibinfo  {journal} {Phys. Rev. B}\ }\textbf {\bibinfo {volume}
  {57}},\ \bibinfo {pages} {14254} (\bibinfo {year} {1998})}\BibitemShut
  {NoStop}%
\bibitem [{\citenamefont {Ingersent}\ and\ \citenamefont
  {Si}(2002)}]{Ingersent.02}%
  \BibitemOpen
  \bibfield  {author} {\bibinfo {author} {\bibfnamefont {K.}~\bibnamefont
  {Ingersent}}\ and\ \bibinfo {author} {\bibfnamefont {Q.}~\bibnamefont {Si}},\
  }\href {\doibase 10.1103/PhysRevLett.89.076403} {\bibfield  {journal}
  {\bibinfo  {journal} {Phys. Rev. Lett.}\ }\textbf {\bibinfo {volume} {89}},\
  \bibinfo {pages} {076403} (\bibinfo {year} {2002})}\BibitemShut {NoStop}%
\bibitem [{\citenamefont {Fritz}\ \emph {et~al.}(2006)\citenamefont {Fritz},
  \citenamefont {Florens},\ and\ \citenamefont {Vojta}}]{Fritz.06}%
  \BibitemOpen
  \bibfield  {author} {\bibinfo {author} {\bibfnamefont {L.}~\bibnamefont
  {Fritz}}, \bibinfo {author} {\bibfnamefont {S.}~\bibnamefont {Florens}}, \
  and\ \bibinfo {author} {\bibfnamefont {M.}~\bibnamefont {Vojta}},\ }\href
  {\doibase 10.1103/PhysRevB.74.144410} {\bibfield  {journal} {\bibinfo
  {journal} {Phys. Rev. B}\ }\textbf {\bibinfo {volume} {74}},\ \bibinfo
  {pages} {144410} (\bibinfo {year} {2006})}\BibitemShut {NoStop}%
\bibitem [{\citenamefont {Vojta}(2001)}]{Vojta.01}%
  \BibitemOpen
  \bibfield  {author} {\bibinfo {author} {\bibfnamefont {M.}~\bibnamefont
  {Vojta}},\ }\href {\doibase 10.1103/PhysRevLett.87.097202} {\bibfield
  {journal} {\bibinfo  {journal} {Phys. Rev. Lett.}\ }\textbf {\bibinfo
  {volume} {87}},\ \bibinfo {pages} {097202} (\bibinfo {year}
  {2001})}\BibitemShut {NoStop}%
\bibitem [{\citenamefont {Glossop}\ and\ \citenamefont
  {Logan}(2003)}]{Glossop.03}%
  \BibitemOpen
  \bibfield  {author} {\bibinfo {author} {\bibfnamefont {M.~T.}\ \bibnamefont
  {Glossop}}\ and\ \bibinfo {author} {\bibfnamefont {D.~E.}\ \bibnamefont
  {Logan}},\ }\href {\doibase 10.1209/epl/i2003-00306-3} {\bibfield  {journal}
  {\bibinfo  {journal} {Europhysics Letters ({EPL})}\ }\textbf {\bibinfo
  {volume} {61}},\ \bibinfo {pages} {810} (\bibinfo {year} {2003})}\BibitemShut
  {NoStop}%
\bibitem [{\citenamefont {Glossop}\ \emph {et~al.}(2005)\citenamefont
  {Glossop}, \citenamefont {Jones},\ and\ \citenamefont {Logan}}]{Glossop.05}%
  \BibitemOpen
  \bibfield  {author} {\bibinfo {author} {\bibfnamefont {M.~T.}\ \bibnamefont
  {Glossop}}, \bibinfo {author} {\bibfnamefont {G.~E.}\ \bibnamefont {Jones}},
  \ and\ \bibinfo {author} {\bibfnamefont {D.~E.}\ \bibnamefont {Logan}},\
  }\href {\doibase 10.1021/jp0457388} {\bibfield  {journal} {\bibinfo
  {journal} {The Journal of Physical Chemistry B}\ }\textbf {\bibinfo {volume}
  {109}},\ \bibinfo {pages} {6564} (\bibinfo {year} {2005})}\BibitemShut
  {NoStop}%
\bibitem [{\citenamefont {Glossop}\ \emph {et~al.}(2011)\citenamefont
  {Glossop}, \citenamefont {Kirchner}, \citenamefont {Pixley},\ and\
  \citenamefont {Si}}]{Glossop.11}%
  \BibitemOpen
  \bibfield  {author} {\bibinfo {author} {\bibfnamefont {M.~T.}\ \bibnamefont
  {Glossop}}, \bibinfo {author} {\bibfnamefont {S.}~\bibnamefont {Kirchner}},
  \bibinfo {author} {\bibfnamefont {J.~H.}\ \bibnamefont {Pixley}}, \ and\
  \bibinfo {author} {\bibfnamefont {Q.}~\bibnamefont {Si}},\ }\href {\doibase
  10.1103/PhysRevLett.107.076404} {\bibfield  {journal} {\bibinfo  {journal}
  {Phys. Rev. Lett.}\ }\textbf {\bibinfo {volume} {107}},\ \bibinfo {pages}
  {076404} (\bibinfo {year} {2011})}\BibitemShut {NoStop}%
\bibitem [{\citenamefont {Pixley}\ \emph {et~al.}(2012)\citenamefont {Pixley},
  \citenamefont {Kirchner}, \citenamefont {Ingersent},\ and\ \citenamefont
  {Si}}]{Pixley.12}%
  \BibitemOpen
  \bibfield  {author} {\bibinfo {author} {\bibfnamefont {J.~H.}\ \bibnamefont
  {Pixley}}, \bibinfo {author} {\bibfnamefont {S.}~\bibnamefont {Kirchner}},
  \bibinfo {author} {\bibfnamefont {K.}~\bibnamefont {Ingersent}}, \ and\
  \bibinfo {author} {\bibfnamefont {Q.}~\bibnamefont {Si}},\ }\href {\doibase
  10.1103/PhysRevLett.109.086403} {\bibfield  {journal} {\bibinfo  {journal}
  {Phys. Rev. Lett.}\ }\textbf {\bibinfo {volume} {109}},\ \bibinfo {pages}
  {086403} (\bibinfo {year} {2012})}\BibitemShut {NoStop}%
\bibitem [{\citenamefont {Dias~da Silva}\ \emph {et~al.}(2006)\citenamefont
  {Dias~da Silva}, \citenamefont {Sandler}, \citenamefont {Ingersent},\ and\
  \citenamefont {Ulloa}}]{Silva.06}%
  \BibitemOpen
  \bibfield  {author} {\bibinfo {author} {\bibfnamefont {L.~G. G.~V.}\
  \bibnamefont {Dias~da Silva}}, \bibinfo {author} {\bibfnamefont {N.~P.}\
  \bibnamefont {Sandler}}, \bibinfo {author} {\bibfnamefont {K.}~\bibnamefont
  {Ingersent}}, \ and\ \bibinfo {author} {\bibfnamefont {S.~E.}\ \bibnamefont
  {Ulloa}},\ }\href {\doibase 10.1103/PhysRevLett.97.096603} {\bibfield
  {journal} {\bibinfo  {journal} {Phys. Rev. Lett.}\ }\textbf {\bibinfo
  {volume} {97}},\ \bibinfo {pages} {096603} (\bibinfo {year}
  {2006})}\BibitemShut {NoStop}%
\bibitem [{\citenamefont {Zhuravlev}\ \emph {et~al.}(2007)\citenamefont
  {Zhuravlev}, \citenamefont {Zharekeshev}, \citenamefont {Gorelov},
  \citenamefont {Lichtenstein}, \citenamefont {Mucciolo},\ and\ \citenamefont
  {Kettemann}}]{Zhuravlev.07}%
  \BibitemOpen
  \bibfield  {author} {\bibinfo {author} {\bibfnamefont {A.}~\bibnamefont
  {Zhuravlev}}, \bibinfo {author} {\bibfnamefont {I.}~\bibnamefont
  {Zharekeshev}}, \bibinfo {author} {\bibfnamefont {E.}~\bibnamefont
  {Gorelov}}, \bibinfo {author} {\bibfnamefont {A.~I.}\ \bibnamefont
  {Lichtenstein}}, \bibinfo {author} {\bibfnamefont {E.~R.}\ \bibnamefont
  {Mucciolo}}, \ and\ \bibinfo {author} {\bibfnamefont {S.}~\bibnamefont
  {Kettemann}},\ }\href {\doibase 10.1103/PhysRevLett.99.247202} {\bibfield
  {journal} {\bibinfo  {journal} {Phys. Rev. Lett.}\ }\textbf {\bibinfo
  {volume} {99}},\ \bibinfo {pages} {247202} (\bibinfo {year}
  {2007})}\BibitemShut {NoStop}%
\bibitem [{\citenamefont {Si}\ and\ \citenamefont {Smith}(1996)}]{Si.96}%
  \BibitemOpen
  \bibfield  {author} {\bibinfo {author} {\bibfnamefont {Q.}~\bibnamefont
  {Si}}\ and\ \bibinfo {author} {\bibfnamefont {J.~L.}\ \bibnamefont {Smith}},\
  }\href {\doibase 10.1103/PhysRevLett.77.3391} {\bibfield  {journal} {\bibinfo
   {journal} {Phys. Rev. Lett.}\ }\textbf {\bibinfo {volume} {77}},\ \bibinfo
  {pages} {3391} (\bibinfo {year} {1996})}\BibitemShut {NoStop}%
\bibitem [{\citenamefont {Si}\ \emph {et~al.}(1999)\citenamefont {Si},
  \citenamefont {Smith},\ and\ \citenamefont {Ingersent}}]{Si.99}%
  \BibitemOpen
  \bibfield  {author} {\bibinfo {author} {\bibfnamefont {Q.}~\bibnamefont
  {Si}}, \bibinfo {author} {\bibfnamefont {J.~L.}\ \bibnamefont {Smith}}, \
  and\ \bibinfo {author} {\bibfnamefont {K.}~\bibnamefont {Ingersent}},\ }\href
  {\doibase 10.1142/S0217979299002435} {\bibfield  {journal} {\bibinfo
  {journal} {International Journal of Modern Physics B}\ }\textbf {\bibinfo
  {volume} {13}},\ \bibinfo {pages} {2331} (\bibinfo {year}
  {1999})}\BibitemShut {NoStop}%
\bibitem [{\citenamefont {Sengupta}(2000)}]{Sengupta.00}%
  \BibitemOpen
  \bibfield  {author} {\bibinfo {author} {\bibfnamefont {A.~M.}\ \bibnamefont
  {Sengupta}},\ }\href {\doibase 10.1103/PhysRevB.61.4041} {\bibfield
  {journal} {\bibinfo  {journal} {Phys. Rev. B}\ }\textbf {\bibinfo {volume}
  {61}},\ \bibinfo {pages} {4041} (\bibinfo {year} {2000})}\BibitemShut
  {NoStop}%
\bibitem [{\citenamefont {Senthil}\ \emph {et~al.}(2004)\citenamefont
  {Senthil}, \citenamefont {Vojta},\ and\ \citenamefont
  {Sachdev}}]{Senthil.04}%
  \BibitemOpen
  \bibfield  {author} {\bibinfo {author} {\bibfnamefont {T.}~\bibnamefont
  {Senthil}}, \bibinfo {author} {\bibfnamefont {M.}~\bibnamefont {Vojta}}, \
  and\ \bibinfo {author} {\bibfnamefont {S.}~\bibnamefont {Sachdev}},\
  }\href@noop {} {\bibfield  {journal} {\bibinfo  {journal} {Phys. Rev. B}\
  }\textbf {\bibinfo {volume} {69}},\ \bibinfo {pages} {035111} (\bibinfo
  {year} {2004})}\BibitemShut {NoStop}%
\bibitem [{\citenamefont {Paschen}\ \emph {et~al.}(2004)\citenamefont
  {Paschen}, \citenamefont {L{\"u}hmann}, \citenamefont {Wirth}, \citenamefont
  {Gegenwart}, \citenamefont {Trovarelli}, \citenamefont {Geibel},
  \citenamefont {Steglich}, \citenamefont {Coleman},\ and\ \citenamefont
  {Si}}]{Paschen.04}%
  \BibitemOpen
  \bibfield  {author} {\bibinfo {author} {\bibfnamefont {S.}~\bibnamefont
  {Paschen}}, \bibinfo {author} {\bibfnamefont {T.}~\bibnamefont
  {L{\"u}hmann}}, \bibinfo {author} {\bibfnamefont {S.}~\bibnamefont {Wirth}},
  \bibinfo {author} {\bibfnamefont {P.}~\bibnamefont {Gegenwart}}, \bibinfo
  {author} {\bibfnamefont {O.}~\bibnamefont {Trovarelli}}, \bibinfo {author}
  {\bibfnamefont {C.}~\bibnamefont {Geibel}}, \bibinfo {author} {\bibfnamefont
  {F.}~\bibnamefont {Steglich}}, \bibinfo {author} {\bibfnamefont
  {P.}~\bibnamefont {Coleman}}, \ and\ \bibinfo {author} {\bibfnamefont
  {Q.}~\bibnamefont {Si}},\ }\href@noop {} {\bibfield  {journal} {\bibinfo
  {journal} {Nature}\ }\textbf {\bibinfo {volume} {432}},\ \bibinfo {pages}
  {881} (\bibinfo {year} {2004})}\BibitemShut {NoStop}%
\bibitem [{\citenamefont {Friedemann}\ \emph {et~al.}(2010)\citenamefont
  {Friedemann}, \citenamefont {Oeschler}, \citenamefont {Wirth}, \citenamefont
  {Krellner}, \citenamefont {Geibel}, \citenamefont {Steglich}, \citenamefont
  {Paschen}, \citenamefont {Kirchner},\ and\ \citenamefont
  {Si}}]{Friedemann.10}%
  \BibitemOpen
  \bibfield  {author} {\bibinfo {author} {\bibfnamefont {S.}~\bibnamefont
  {Friedemann}}, \bibinfo {author} {\bibfnamefont {N.}~\bibnamefont
  {Oeschler}}, \bibinfo {author} {\bibfnamefont {S.}~\bibnamefont {Wirth}},
  \bibinfo {author} {\bibfnamefont {C.}~\bibnamefont {Krellner}}, \bibinfo
  {author} {\bibfnamefont {C.}~\bibnamefont {Geibel}}, \bibinfo {author}
  {\bibfnamefont {F.}~\bibnamefont {Steglich}}, \bibinfo {author}
  {\bibfnamefont {S.}~\bibnamefont {Paschen}}, \bibinfo {author} {\bibfnamefont
  {S.}~\bibnamefont {Kirchner}}, \ and\ \bibinfo {author} {\bibfnamefont
  {Q.}~\bibnamefont {Si}},\ }\href {\doibase 10.1073/pnas.1009202107}
  {\bibfield  {journal} {\bibinfo  {journal} {Proceedings of the National
  Academy of Sciences}\ }\textbf {\bibinfo {volume} {107}},\ \bibinfo {pages}
  {14547} (\bibinfo {year} {2010})}\BibitemShut {NoStop}%
\bibitem [{\citenamefont {Friedemann}\ \emph {et~al.}(2011)\citenamefont
  {Friedemann}, \citenamefont {Wirth}, \citenamefont {Kirchner}, \citenamefont
  {Si}, \citenamefont {Hartmann}, \citenamefont {Krellner}, \citenamefont
  {Geibel}, \citenamefont {Westerkamp}, \citenamefont {Brando},\ and\
  \citenamefont {Steglich}}]{Friedemann.11}%
  \BibitemOpen
  \bibfield  {author} {\bibinfo {author} {\bibfnamefont {S.}~\bibnamefont
  {Friedemann}}, \bibinfo {author} {\bibfnamefont {S.}~\bibnamefont {Wirth}},
  \bibinfo {author} {\bibfnamefont {S.}~\bibnamefont {Kirchner}}, \bibinfo
  {author} {\bibfnamefont {Q.}~\bibnamefont {Si}}, \bibinfo {author}
  {\bibfnamefont {S.}~\bibnamefont {Hartmann}}, \bibinfo {author}
  {\bibfnamefont {C.}~\bibnamefont {Krellner}}, \bibinfo {author}
  {\bibfnamefont {C.}~\bibnamefont {Geibel}}, \bibinfo {author} {\bibfnamefont
  {T.}~\bibnamefont {Westerkamp}}, \bibinfo {author} {\bibfnamefont
  {M.}~\bibnamefont {Brando}}, \ and\ \bibinfo {author} {\bibfnamefont
  {F.}~\bibnamefont {Steglich}},\ }\href {\doibase 10.1143/JPSJS.80SA.SA002}
  {\bibfield  {journal} {\bibinfo  {journal} {Journal of the Physical Society
  of Japan}\ }\textbf {\bibinfo {volume} {80}},\ \bibinfo {pages} {SA002}
  (\bibinfo {year} {2011})}\BibitemShut {NoStop}%
\bibitem [{\citenamefont {Prochaska}\ \emph {et~al.}(2020)\citenamefont
  {Prochaska}, \citenamefont {Li}, \citenamefont {MacFarland}, \citenamefont
  {Andrews}, \citenamefont {Bonta}, \citenamefont {Bianco}, \citenamefont
  {Yazdi}, \citenamefont {Schrenk}, \citenamefont {Detz}, \citenamefont
  {Limbeck}, \citenamefont {Si}, \citenamefont {Ringe}, \citenamefont
  {Strasser}, \citenamefont {Kono},\ and\ \citenamefont
  {Paschen}}]{Prochaska.20}%
  \BibitemOpen
  \bibfield  {author} {\bibinfo {author} {\bibfnamefont {L.}~\bibnamefont
  {Prochaska}}, \bibinfo {author} {\bibfnamefont {X.}~\bibnamefont {Li}},
  \bibinfo {author} {\bibfnamefont {D.~C.}\ \bibnamefont {MacFarland}},
  \bibinfo {author} {\bibfnamefont {A.~M.}\ \bibnamefont {Andrews}}, \bibinfo
  {author} {\bibfnamefont {M.}~\bibnamefont {Bonta}}, \bibinfo {author}
  {\bibfnamefont {E.~F.}\ \bibnamefont {Bianco}}, \bibinfo {author}
  {\bibfnamefont {S.}~\bibnamefont {Yazdi}}, \bibinfo {author} {\bibfnamefont
  {W.}~\bibnamefont {Schrenk}}, \bibinfo {author} {\bibfnamefont
  {H.}~\bibnamefont {Detz}}, \bibinfo {author} {\bibfnamefont {A.}~\bibnamefont
  {Limbeck}}, \bibinfo {author} {\bibfnamefont {Q.}~\bibnamefont {Si}},
  \bibinfo {author} {\bibfnamefont {E.}~\bibnamefont {Ringe}}, \bibinfo
  {author} {\bibfnamefont {G.}~\bibnamefont {Strasser}}, \bibinfo {author}
  {\bibfnamefont {J.}~\bibnamefont {Kono}}, \ and\ \bibinfo {author}
  {\bibfnamefont {S.}~\bibnamefont {Paschen}},\ }\href {\doibase
  10.1126/science.aag1595} {\bibfield  {journal} {\bibinfo  {journal}
  {Science}\ }\textbf {\bibinfo {volume} {367}},\ \bibinfo {pages} {285}
  (\bibinfo {year} {2020})}\BibitemShut {NoStop}%
\bibitem [{\citenamefont {Millis}(1993)}]{Millis.93}%
  \BibitemOpen
  \bibfield  {author} {\bibinfo {author} {\bibfnamefont {A.~J.}\ \bibnamefont
  {Millis}},\ }\href@noop {} {\bibfield  {journal} {\bibinfo  {journal}
  {Phys.~Rev.~B}\ }\textbf {\bibinfo {volume} {48}},\ \bibinfo {pages} {7183}
  (\bibinfo {year} {1993})}\BibitemShut {NoStop}%
\bibitem [{\citenamefont {Zhu}\ \emph {et~al.}(2004)\citenamefont {Zhu},
  \citenamefont {Kirchner}, \citenamefont {Si},\ and\ \citenamefont
  {Georges}}]{Zhu.04}%
  \BibitemOpen
  \bibfield  {author} {\bibinfo {author} {\bibfnamefont {L.}~\bibnamefont
  {Zhu}}, \bibinfo {author} {\bibfnamefont {S.}~\bibnamefont {Kirchner}},
  \bibinfo {author} {\bibfnamefont {Q.}~\bibnamefont {Si}}, \ and\ \bibinfo
  {author} {\bibfnamefont {A.}~\bibnamefont {Georges}},\ }\href
  {http://prl.aps.org/abstract/PRL/v93/i26/e267201} {\bibfield  {journal}
  {\bibinfo  {journal} {Phys. Rev. Lett.}\ }\textbf {\bibinfo {volume} {93}},\
  \bibinfo {pages} {267201} (\bibinfo {year} {2004})}\BibitemShut {NoStop}%
\bibitem [{\citenamefont {Cai}\ \emph {et~al.}(2020)\citenamefont {Cai},
  \citenamefont {Yu}, \citenamefont {Hu}, \citenamefont {Kirchner},\ and\
  \citenamefont {Si}}]{Cai.20}%
  \BibitemOpen
  \bibfield  {author} {\bibinfo {author} {\bibfnamefont {A.}~\bibnamefont
  {Cai}}, \bibinfo {author} {\bibfnamefont {Z.}~\bibnamefont {Yu}}, \bibinfo
  {author} {\bibfnamefont {H.}~\bibnamefont {Hu}}, \bibinfo {author}
  {\bibfnamefont {S.}~\bibnamefont {Kirchner}}, \ and\ \bibinfo {author}
  {\bibfnamefont {Q.}~\bibnamefont {Si}},\ }\href {\doibase
  10.1103/PhysRevLett.124.027205} {\bibfield  {journal} {\bibinfo  {journal}
  {Phys. Rev. Lett.}\ }\textbf {\bibinfo {volume} {124}},\ \bibinfo {pages}
  {027205} (\bibinfo {year} {2020})}\BibitemShut {NoStop}%
\bibitem [{\citenamefont {Zhu}\ and\ \citenamefont {Si}(2002)}]{Zhu.02}%
  \BibitemOpen
  \bibfield  {author} {\bibinfo {author} {\bibfnamefont {L.}~\bibnamefont
  {Zhu}}\ and\ \bibinfo {author} {\bibfnamefont {Q.}~\bibnamefont {Si}},\
  }\href {\doibase 10.1103/PhysRevB.66.024426} {\bibfield  {journal} {\bibinfo
  {journal} {Phys. Rev. B}\ }\textbf {\bibinfo {volume} {66}},\ \bibinfo
  {pages} {024426} (\bibinfo {year} {2002})}\BibitemShut {NoStop}%
\bibitem [{\citenamefont {Zaránd}\ and\ \citenamefont
  {Demler}(2002)}]{Zarand.02}%
  \BibitemOpen
  \bibfield  {author} {\bibinfo {author} {\bibfnamefont {G.}~\bibnamefont
  {Zaránd}}\ and\ \bibinfo {author} {\bibfnamefont {E.}~\bibnamefont
  {Demler}},\ }\href {\doibase 10.1103/PhysRevB.66.024427} {\bibfield
  {journal} {\bibinfo  {journal} {Phys. Rev. B}\ }\textbf {\bibinfo {volume}
  {66}},\ \bibinfo {pages} {024427} (\bibinfo {year} {2002})}\BibitemShut
  {NoStop}%
\bibitem [{\citenamefont {Bulla}\ \emph {et~al.}(2003)\citenamefont {Bulla},
  \citenamefont {Tong},\ and\ \citenamefont {Vojta}}]{Bulla.03}%
  \BibitemOpen
  \bibfield  {author} {\bibinfo {author} {\bibfnamefont {R.}~\bibnamefont
  {Bulla}}, \bibinfo {author} {\bibfnamefont {N.-H.}\ \bibnamefont {Tong}}, \
  and\ \bibinfo {author} {\bibfnamefont {M.}~\bibnamefont {Vojta}},\ }\href
  {\doibase 10.1103/PhysRevLett.91.170601} {\bibfield  {journal} {\bibinfo
  {journal} {Phys. Rev. Lett.}\ }\textbf {\bibinfo {volume} {91}},\ \bibinfo
  {pages} {170601} (\bibinfo {year} {2003})}\BibitemShut {NoStop}%
\bibitem [{\citenamefont {Glossop}\ and\ \citenamefont
  {Ingersent}(2005)}]{Glossop.05a}%
  \BibitemOpen
  \bibfield  {author} {\bibinfo {author} {\bibfnamefont {M.~T.}\ \bibnamefont
  {Glossop}}\ and\ \bibinfo {author} {\bibfnamefont {K.}~\bibnamefont
  {Ingersent}},\ }\href {\doibase 10.1103/PhysRevLett.95.067202} {\bibfield
  {journal} {\bibinfo  {journal} {Phys. Rev. Lett.}\ }\textbf {\bibinfo
  {volume} {95}},\ \bibinfo {pages} {067202} (\bibinfo {year}
  {2005})}\BibitemShut {NoStop}%
\bibitem [{\citenamefont {Kirchner}\ \emph {et~al.}(2005)\citenamefont
  {Kirchner}, \citenamefont {Zhu}, \citenamefont {Si},\ and\ \citenamefont
  {Natelson}}]{Kirchner.05}%
  \BibitemOpen
  \bibfield  {author} {\bibinfo {author} {\bibfnamefont {S.}~\bibnamefont
  {Kirchner}}, \bibinfo {author} {\bibfnamefont {L.}~\bibnamefont {Zhu}},
  \bibinfo {author} {\bibfnamefont {Q.}~\bibnamefont {Si}}, \ and\ \bibinfo
  {author} {\bibfnamefont {D.}~\bibnamefont {Natelson}},\ }\href {\doibase
  10.1073/pnas.0509519102} {\bibfield  {journal} {\bibinfo  {journal}
  {Proc.~Natl.~Acad.~Sci.~U.S.A.}\ }\textbf {\bibinfo {volume} {102}},\
  \bibinfo {pages} {18824} (\bibinfo {year} {2005})}\BibitemShut {NoStop}%
\bibitem [{\citenamefont {Zamani}\ \emph {et~al.}(2016)\citenamefont {Zamani},
  \citenamefont {Ribeiro},\ and\ \citenamefont {Kirchner}}]{Zamani.16}%
  \BibitemOpen
  \bibfield  {author} {\bibinfo {author} {\bibfnamefont {F.}~\bibnamefont
  {Zamani}}, \bibinfo {author} {\bibfnamefont {P.}~\bibnamefont {Ribeiro}}, \
  and\ \bibinfo {author} {\bibfnamefont {S.}~\bibnamefont {Kirchner}},\ }\href
  {http://stacks.iop.org/1367-2630/18/i=6/a=063024} {\bibfield  {journal}
  {\bibinfo  {journal} {New J. Phys.}\ }\textbf {\bibinfo {volume} {18}},\
  \bibinfo {pages} {063024} (\bibinfo {year} {2016})}\BibitemShut {NoStop}%
\bibitem [{\citenamefont {Vojta}\ and\ \citenamefont
  {Kir\'{c}an}(2003)}]{Vojta.03}%
  \BibitemOpen
  \bibfield  {author} {\bibinfo {author} {\bibfnamefont {M.}~\bibnamefont
  {Vojta}}\ and\ \bibinfo {author} {\bibfnamefont {M.}~\bibnamefont
  {Kir\'{c}an}},\ }\href {\doibase 10.1103/PhysRevLett.90.157203} {\bibfield
  {journal} {\bibinfo  {journal} {Phys. Rev. Lett.}\ }\textbf {\bibinfo
  {volume} {90}},\ \bibinfo {pages} {157203} (\bibinfo {year}
  {2003})}\BibitemShut {NoStop}%
\bibitem [{\citenamefont {Kir\'{c}an}\ and\ \citenamefont
  {Vojta}(2004)}]{Kircan.04}%
  \BibitemOpen
  \bibfield  {author} {\bibinfo {author} {\bibfnamefont {M.}~\bibnamefont
  {Kir\'{c}an}}\ and\ \bibinfo {author} {\bibfnamefont {M.}~\bibnamefont
  {Vojta}},\ }\href {\doibase 10.1103/PhysRevB.69.174421} {\bibfield  {journal}
  {\bibinfo  {journal} {Phys. Rev. B}\ }\textbf {\bibinfo {volume} {69}},\
  \bibinfo {pages} {174421} (\bibinfo {year} {2004})}\BibitemShut {NoStop}%
\bibitem [{\citenamefont {Glossop}\ \emph {et~al.}(2008)\citenamefont
  {Glossop}, \citenamefont {Khoshkhou},\ and\ \citenamefont
  {Ingersent}}]{Glossop.08}%
  \BibitemOpen
  \bibfield  {author} {\bibinfo {author} {\bibfnamefont {M.}~\bibnamefont
  {Glossop}}, \bibinfo {author} {\bibfnamefont {N.}~\bibnamefont {Khoshkhou}},
  \ and\ \bibinfo {author} {\bibfnamefont {K.}~\bibnamefont {Ingersent}},\
  }\href {http://www.sciencedirect.com/science/article/pii/S0921452607011325}
  {\bibfield  {journal} {\bibinfo  {journal} {Physica B: Condensed Matter}\
  }\textbf {\bibinfo {volume} {403}},\ \bibinfo {pages} {1303 } (\bibinfo
  {year} {2008})}\BibitemShut {NoStop}%
\bibitem [{\citenamefont {Pixley}\ \emph {et~al.}(2013)\citenamefont {Pixley},
  \citenamefont {Kirchner}, \citenamefont {Ingersent},\ and\ \citenamefont
  {Si}}]{Pixley.13}%
  \BibitemOpen
  \bibfield  {author} {\bibinfo {author} {\bibfnamefont {J.~H.}\ \bibnamefont
  {Pixley}}, \bibinfo {author} {\bibfnamefont {S.}~\bibnamefont {Kirchner}},
  \bibinfo {author} {\bibfnamefont {K.}~\bibnamefont {Ingersent}}, \ and\
  \bibinfo {author} {\bibfnamefont {Q.}~\bibnamefont {Si}},\ }\href {\doibase
  10.1103/PhysRevB.88.245111} {\bibfield  {journal} {\bibinfo  {journal} {Phys.
  Rev. B}\ }\textbf {\bibinfo {volume} {88}},\ \bibinfo {pages} {245111}
  (\bibinfo {year} {2013})}\BibitemShut {NoStop}%
\bibitem [{\citenamefont {Pixley}\ \emph {et~al.}(2011)\citenamefont {Pixley},
  \citenamefont {Kirchner}, \citenamefont {Glossop},\ and\ \citenamefont
  {Si}}]{Pixley.11}%
  \BibitemOpen
  \bibfield  {author} {\bibinfo {author} {\bibfnamefont {J.~H.}\ \bibnamefont
  {Pixley}}, \bibinfo {author} {\bibfnamefont {S.}~\bibnamefont {Kirchner}},
  \bibinfo {author} {\bibfnamefont {M.~T.}\ \bibnamefont {Glossop}}, \ and\
  \bibinfo {author} {\bibfnamefont {Q.}~\bibnamefont {Si}},\ }\href {\doibase
  10.1088/1742-6596/273/1/012050} {\bibfield  {journal} {\bibinfo  {journal}
  {Journal of Physics: Conference Series}\ }\textbf {\bibinfo {volume} {273}},\
  \bibinfo {pages} {012050} (\bibinfo {year} {2011})}\BibitemShut {NoStop}%
\bibitem [{\citenamefont {Parcollet}\ \emph {et~al.}(1998)\citenamefont
  {Parcollet}, \citenamefont {Georges}, \citenamefont {Kotliar},\ and\
  \citenamefont {Sengupta}}]{Parcollet.98}%
  \BibitemOpen
  \bibfield  {author} {\bibinfo {author} {\bibfnamefont {O.}~\bibnamefont
  {Parcollet}}, \bibinfo {author} {\bibfnamefont {A.}~\bibnamefont {Georges}},
  \bibinfo {author} {\bibfnamefont {G.}~\bibnamefont {Kotliar}}, \ and\
  \bibinfo {author} {\bibfnamefont {A.}~\bibnamefont {Sengupta}},\ }\href
  {\doibase 10.1103/PhysRevB.58.3794} {\bibfield  {journal} {\bibinfo
  {journal} {Phys. Rev. B}\ }\textbf {\bibinfo {volume} {58}},\ \bibinfo
  {pages} {3794} (\bibinfo {year} {1998})}\BibitemShut {NoStop}%
\bibitem [{\citenamefont {Cox}\ and\ \citenamefont
  {Ruckenstein}(1993)}]{Cox.93}%
  \BibitemOpen
  \bibfield  {author} {\bibinfo {author} {\bibfnamefont {D.}~\bibnamefont
  {Cox}}\ and\ \bibinfo {author} {\bibfnamefont {A.}~\bibnamefont
  {Ruckenstein}},\ }\href@noop {} {\bibfield  {journal} {\bibinfo  {journal}
  {Phys.~Rev.~Lett.}\ }\textbf {\bibinfo {volume} {71}},\ \bibinfo {pages}
  {1613} (\bibinfo {year} {1993})}\BibitemShut {NoStop}%
\bibitem [{\citenamefont {Zamani}\ \emph {et~al.}(2013)\citenamefont {Zamani},
  \citenamefont {Chowdhury}, \citenamefont {Ribeiro}, \citenamefont
  {Ingersent},\ and\ \citenamefont {Kirchner}}]{Zamani.13}%
  \BibitemOpen
  \bibfield  {author} {\bibinfo {author} {\bibfnamefont {F.}~\bibnamefont
  {Zamani}}, \bibinfo {author} {\bibfnamefont {T.}~\bibnamefont {Chowdhury}},
  \bibinfo {author} {\bibfnamefont {P.}~\bibnamefont {Ribeiro}}, \bibinfo
  {author} {\bibfnamefont {K.}~\bibnamefont {Ingersent}}, \ and\ \bibinfo
  {author} {\bibfnamefont {S.}~\bibnamefont {Kirchner}},\ }\href
  {https://onlinelibrary.wiley.com/doi/abs/10.1002/pssb.201200928} {\bibfield
  {journal} {\bibinfo  {journal} {physica status solidi (b)}\ }\textbf
  {\bibinfo {volume} {250}},\ \bibinfo {pages} {547} (\bibinfo {year}
  {2013})}\BibitemShut {NoStop}%
\bibitem [{\citenamefont {Affleck}\ and\ \citenamefont
  {Ludwig}(1991)}]{Affleck.91}%
  \BibitemOpen
  \bibfield  {author} {\bibinfo {author} {\bibfnamefont {I.}~\bibnamefont
  {Affleck}}\ and\ \bibinfo {author} {\bibfnamefont {A.~W.~W.}\ \bibnamefont
  {Ludwig}},\ }\href {\doibase 10.1103/PhysRevLett.67.161} {\bibfield
  {journal} {\bibinfo  {journal} {Phys. Rev. Lett.}\ }\textbf {\bibinfo
  {volume} {67}},\ \bibinfo {pages} {161} (\bibinfo {year} {1991})}\BibitemShut
  {NoStop}%
\bibitem [{\citenamefont {Cardy}(1984)}]{Cardy.84}%
  \BibitemOpen
  \bibfield  {author} {\bibinfo {author} {\bibfnamefont {J.~L.}\ \bibnamefont
  {Cardy}},\ }\href {\doibase 10.1016/0550-3213(84)90241-4} {\bibfield
  {journal} {\bibinfo  {journal} {Nuclear Physics B}\ }\textbf {\bibinfo
  {volume} {240}},\ \bibinfo {pages} {514} (\bibinfo {year}
  {1984})}\BibitemShut {NoStop}%
\bibitem [{\citenamefont {Ginsparg}(1989)}]{Ginsparg.89}%
  \BibitemOpen
  \bibfield  {author} {\bibinfo {author} {\bibfnamefont {P.}~\bibnamefont
  {Ginsparg}},\ }in\ \href@noop {} {\emph {\bibinfo {booktitle} {Fields,
  Strings and Critical Phenomena}}},\ \bibinfo {editor} {edited by\ \bibinfo
  {editor} {\bibfnamefont {E.}~\bibnamefont {Br\'ezin}}\ and\ \bibinfo {editor}
  {\bibfnamefont {J.}~\bibnamefont {{Zinn-Justin}}}}\ (\bibinfo  {publisher}
  {Les Houches, Session XLIX},\ \bibinfo {year} {1989})\BibitemShut {NoStop}%
\bibitem [{\citenamefont {Friedan}\ and\ \citenamefont
  {Konechny}(2004)}]{Friedan.04}%
  \BibitemOpen
  \bibfield  {author} {\bibinfo {author} {\bibfnamefont {D.}~\bibnamefont
  {Friedan}}\ and\ \bibinfo {author} {\bibfnamefont {A.}~\bibnamefont
  {Konechny}},\ }\href {\doibase 10.1103/PhysRevLett.93.030402} {\bibfield
  {journal} {\bibinfo  {journal} {Phys. Rev. Lett.}\ }\textbf {\bibinfo
  {volume} {93}},\ \bibinfo {pages} {030402} (\bibinfo {year}
  {2004})}\BibitemShut {NoStop}%
\bibitem [{\citenamefont {Kirchner}\ and\ \citenamefont
  {Si}(2008)}]{Kirchner.08}%
  \BibitemOpen
  \bibfield  {author} {\bibinfo {author} {\bibfnamefont {S.}~\bibnamefont
  {Kirchner}}\ and\ \bibinfo {author} {\bibfnamefont {Q.}~\bibnamefont {Si}},\
  }\href {\doibase 10.1103/PhysRevLett.100.026403} {\bibfield  {journal}
  {\bibinfo  {journal} {Phys. Rev. Lett.}\ }\textbf {\bibinfo {volume} {100}},\
  \bibinfo {pages} {026403} (\bibinfo {year} {2008})}\BibitemShut {NoStop}%
\bibitem [{\citenamefont {Mitchell}\ \emph {et~al.}(2013)\citenamefont
  {Mitchell}, \citenamefont {Vojta}, \citenamefont {Bulla},\ and\ \citenamefont
  {Fritz}}]{Mitchell.13}%
  \BibitemOpen
  \bibfield  {author} {\bibinfo {author} {\bibfnamefont {A.~K.}\ \bibnamefont
  {Mitchell}}, \bibinfo {author} {\bibfnamefont {M.}~\bibnamefont {Vojta}},
  \bibinfo {author} {\bibfnamefont {R.}~\bibnamefont {Bulla}}, \ and\ \bibinfo
  {author} {\bibfnamefont {L.}~\bibnamefont {Fritz}},\ }\href {\doibase
  10.1103/PhysRevB.88.195119} {\bibfield  {journal} {\bibinfo  {journal} {Phys.
  Rev. B}\ }\textbf {\bibinfo {volume} {88}},\ \bibinfo {pages} {195119}
  (\bibinfo {year} {2013})}\BibitemShut {NoStop}%
\bibitem [{\citenamefont {Cheng}\ \emph {et~al.}(2017)\citenamefont {Cheng},
  \citenamefont {Chowdhury}, \citenamefont {Mohammed},\ and\ \citenamefont
  {Ingersent}}]{Cheng.17}%
  \BibitemOpen
  \bibfield  {author} {\bibinfo {author} {\bibfnamefont {M.}~\bibnamefont
  {Cheng}}, \bibinfo {author} {\bibfnamefont {T.}~\bibnamefont {Chowdhury}},
  \bibinfo {author} {\bibfnamefont {A.}~\bibnamefont {Mohammed}}, \ and\
  \bibinfo {author} {\bibfnamefont {K.}~\bibnamefont {Ingersent}},\ }\href
  {\doibase 10.1103/PhysRevB.96.045103} {\bibfield  {journal} {\bibinfo
  {journal} {Phys. Rev. B}\ }\textbf {\bibinfo {volume} {96}},\ \bibinfo
  {pages} {045103} (\bibinfo {year} {2017})}\BibitemShut {NoStop}%
\bibitem [{\citenamefont {Ribeiro}\ \emph {et~al.}(2015)\citenamefont
  {Ribeiro}, \citenamefont {Zamani},\ and\ \citenamefont
  {Kirchner}}]{Ribeiro.15}%
  \BibitemOpen
  \bibfield  {author} {\bibinfo {author} {\bibfnamefont {P.}~\bibnamefont
  {Ribeiro}}, \bibinfo {author} {\bibfnamefont {F.}~\bibnamefont {Zamani}}, \
  and\ \bibinfo {author} {\bibfnamefont {S.}~\bibnamefont {Kirchner}},\ }\href
  {\doibase 10.1103/PhysRevLett.115.220602} {\bibfield  {journal} {\bibinfo
  {journal} {Phys. Rev. Lett.}\ }\textbf {\bibinfo {volume} {115}},\ \bibinfo
  {pages} {220602} (\bibinfo {year} {2015})}\BibitemShut {NoStop}%
\bibitem [{\citenamefont {Baym}\ and\ \citenamefont
  {Kadanoff}(1961)}]{Baym.61}%
  \BibitemOpen
  \bibfield  {author} {\bibinfo {author} {\bibfnamefont {G.}~\bibnamefont
  {Baym}}\ and\ \bibinfo {author} {\bibfnamefont {L.~P.}\ \bibnamefont
  {Kadanoff}},\ }\href@noop {} {\bibfield  {journal} {\bibinfo  {journal}
  {Phys. Rev.}\ }\textbf {\bibinfo {volume} {124}},\ \bibinfo {pages} {287}
  (\bibinfo {year} {1961})}\BibitemShut {NoStop}%
\bibitem [{\citenamefont {Coleman}\ \emph {et~al.}(2005)\citenamefont
  {Coleman}, \citenamefont {Paul},\ and\ \citenamefont {Rech}}]{Coleman.05}%
  \BibitemOpen
  \bibfield  {author} {\bibinfo {author} {\bibfnamefont {P.}~\bibnamefont
  {Coleman}}, \bibinfo {author} {\bibfnamefont {I.}~\bibnamefont {Paul}}, \
  and\ \bibinfo {author} {\bibfnamefont {J.}~\bibnamefont {Rech}},\ }\href
  {\doibase 10.1103/PhysRevB.72.094430} {\bibfield  {journal} {\bibinfo
  {journal} {Phys. Rev. B}\ }\textbf {\bibinfo {volume} {72}},\ \bibinfo
  {pages} {094430} (\bibinfo {year} {2005})}\BibitemShut {NoStop}%
\bibitem [{\citenamefont {Rech}\ \emph {et~al.}(2006)\citenamefont {Rech},
  \citenamefont {Coleman}, \citenamefont {Zarand},\ and\ \citenamefont
  {Parcollet}}]{Rech.06}%
  \BibitemOpen
  \bibfield  {author} {\bibinfo {author} {\bibfnamefont {J.}~\bibnamefont
  {Rech}}, \bibinfo {author} {\bibfnamefont {P.}~\bibnamefont {Coleman}},
  \bibinfo {author} {\bibfnamefont {G.}~\bibnamefont {Zarand}}, \ and\ \bibinfo
  {author} {\bibfnamefont {O.}~\bibnamefont {Parcollet}},\ }\href {\doibase
  10.1103/PhysRevLett.96.016601} {\bibfield  {journal} {\bibinfo  {journal}
  {Phys. Rev. Lett.}\ }\textbf {\bibinfo {volume} {96}},\ \bibinfo {pages}
  {016601} (\bibinfo {year} {2006})}\BibitemShut {NoStop}%
\bibitem [{\citenamefont {Potthoff}(2006)}]{Potthoff.06}%
  \BibitemOpen
  \bibfield  {author} {\bibinfo {author} {\bibfnamefont {M.}~\bibnamefont
  {Potthoff}},\ }\href@noop {} {\bibfield  {journal} {\bibinfo  {journal}
  {Condensed Matter Physics}\ }\textbf {\bibinfo {volume} {9}},\ \bibinfo
  {pages} {557} (\bibinfo {year} {2006})}\BibitemShut {NoStop}%
\bibitem [{\citenamefont {Bulla}\ \emph {et~al.}(1997)\citenamefont {Bulla},
  \citenamefont {Pruschke},\ and\ \citenamefont {Hewson}}]{Bulla.97}%
  \BibitemOpen
  \bibfield  {author} {\bibinfo {author} {\bibfnamefont {R.}~\bibnamefont
  {Bulla}}, \bibinfo {author} {\bibfnamefont {T.}~\bibnamefont {Pruschke}}, \
  and\ \bibinfo {author} {\bibfnamefont {A.~C.}\ \bibnamefont {Hewson}},\
  }\href {\doibase 10.1088/0953-8984/9/47/014} {\bibfield  {journal} {\bibinfo
  {journal} {J.\ Condens.\ Matter Phys.}\ }\textbf {\bibinfo {volume} {9}},\
  \bibinfo {pages} {10463} (\bibinfo {year} {1997})}\BibitemShut {NoStop}%
\bibitem [{\citenamefont {\v{Z}itko}(2009)}]{Zitko.09}%
  \BibitemOpen
  \bibfield  {author} {\bibinfo {author} {\bibfnamefont {R.}~\bibnamefont
  {\v{Z}itko}},\ }\href {\doibase 10.1103/PhysRevB.80.125125} {\bibfield
  {journal} {\bibinfo  {journal} {Phys. Rev. B}\ }\textbf {\bibinfo {volume}
  {80}},\ \bibinfo {pages} {125125} (\bibinfo {year} {2009})}\BibitemShut
  {NoStop}%
\bibitem [{\citenamefont {Maldacena}\ and\ \citenamefont
  {Stanford}(2016)}]{Maldacena.16}%
  \BibitemOpen
  \bibfield  {author} {\bibinfo {author} {\bibfnamefont {J.}~\bibnamefont
  {Maldacena}}\ and\ \bibinfo {author} {\bibfnamefont {D.}~\bibnamefont
  {Stanford}},\ }\href {\doibase 10.1103/PhysRevD.94.106002} {\bibfield
  {journal} {\bibinfo  {journal} {Phys. Rev. D}\ }\textbf {\bibinfo {volume}
  {94}},\ \bibinfo {pages} {106002} (\bibinfo {year} {2016})}\BibitemShut
  {NoStop}%
\bibitem [{\citenamefont {Florens}\ and\ \citenamefont
  {Rosch}(2004)}]{Florens.04}%
  \BibitemOpen
  \bibfield  {author} {\bibinfo {author} {\bibfnamefont {S.}~\bibnamefont
  {Florens}}\ and\ \bibinfo {author} {\bibfnamefont {A.}~\bibnamefont
  {Rosch}},\ }\href {\doibase 10.1103/PhysRevLett.92.216601} {\bibfield
  {journal} {\bibinfo  {journal} {Phys. Rev. Lett.}\ }\textbf {\bibinfo
  {volume} {92}},\ \bibinfo {pages} {216601} (\bibinfo {year}
  {2004})}\BibitemShut {NoStop}%
\bibitem [{\citenamefont {Zinn-Justin}(2020)}]{JeanZinnJustin}%
  \BibitemOpen
  \bibfield  {author} {\bibinfo {author} {\bibfnamefont {J.}~\bibnamefont
  {Zinn-Justin}},\ }\href@noop {} {}\bibinfo {howpublished} {private
  communication} (\bibinfo {year} {2020})\BibitemShut {NoStop}%
\bibitem [{\citenamefont {Dai}\ \emph {et~al.}(2008)\citenamefont {Dai},
  \citenamefont {Si},\ and\ \citenamefont {Bolech}}]{Dai.08}%
  \BibitemOpen
  \bibfield  {author} {\bibinfo {author} {\bibfnamefont {J.}~\bibnamefont
  {Dai}}, \bibinfo {author} {\bibfnamefont {Q.}~\bibnamefont {Si}}, \ and\
  \bibinfo {author} {\bibfnamefont {C.~J.}\ \bibnamefont {Bolech}},\
  }\href@noop {} {\enquote {\bibinfo {title} {The {B}ose-{F}ermi {K}ondo model
  with a singular dissipative spectrum: Exact solutions and their
  implications},}\ } (\bibinfo {year} {2008}),\ \bibinfo {note}
  {arXiv:0712.3280v2}\BibitemShut {NoStop}%
\bibitem [{\citenamefont {Casini}\ \emph {et~al.}(2016)\citenamefont {Casini},
  \citenamefont {Landea},\ and\ \citenamefont {Torroba}}]{Casini.16}%
  \BibitemOpen
  \bibfield  {author} {\bibinfo {author} {\bibfnamefont {H.}~\bibnamefont
  {Casini}}, \bibinfo {author} {\bibfnamefont {I.~S.}\ \bibnamefont {Landea}},
  \ and\ \bibinfo {author} {\bibfnamefont {G.}~\bibnamefont {Torroba}},\ }\href
  {\doibase 10.1007/JHEP10(2016)140} {\bibfield  {journal} {\bibinfo  {journal}
  {Journal of High Energy Physics}\ }\textbf {\bibinfo {volume} {2016}},\
  \bibinfo {pages} {140} (\bibinfo {year} {2016})}\BibitemShut {NoStop}%
\bibitem [{\citenamefont {Pixley}\ \emph {et~al.}(2015)\citenamefont {Pixley},
  \citenamefont {Chowdhury}, \citenamefont {Miecnikowski}, \citenamefont
  {Stephens}, \citenamefont {Wagner},\ and\ \citenamefont
  {Ingersent}}]{Pixley.15}%
  \BibitemOpen
  \bibfield  {author} {\bibinfo {author} {\bibfnamefont {J.~H.}\ \bibnamefont
  {Pixley}}, \bibinfo {author} {\bibfnamefont {T.}~\bibnamefont {Chowdhury}},
  \bibinfo {author} {\bibfnamefont {M.~T.}\ \bibnamefont {Miecnikowski}},
  \bibinfo {author} {\bibfnamefont {J.}~\bibnamefont {Stephens}}, \bibinfo
  {author} {\bibfnamefont {C.}~\bibnamefont {Wagner}}, \ and\ \bibinfo {author}
  {\bibfnamefont {K.}~\bibnamefont {Ingersent}},\ }\href {\doibase
  10.1103/PhysRevB.91.245122} {\bibfield  {journal} {\bibinfo  {journal} {Phys.
  Rev. B}\ }\textbf {\bibinfo {volume} {91}},\ \bibinfo {pages} {245122}
  (\bibinfo {year} {2015})}\BibitemShut {NoStop}%
\bibitem [{\citenamefont {Lebanon}\ \emph {et~al.}(2006)\citenamefont
  {Lebanon}, \citenamefont {Rech}, \citenamefont {Coleman},\ and\ \citenamefont
  {Parcollet}}]{lebanon2006}%
  \BibitemOpen
  \bibfield  {author} {\bibinfo {author} {\bibfnamefont {E.}~\bibnamefont
  {Lebanon}}, \bibinfo {author} {\bibfnamefont {J.}~\bibnamefont {Rech}},
  \bibinfo {author} {\bibfnamefont {P.}~\bibnamefont {Coleman}}, \ and\
  \bibinfo {author} {\bibfnamefont {O.}~\bibnamefont {Parcollet}},\ }\href
  {\doibase 10.1103/PhysRevLett.97.106604} {\bibfield  {journal} {\bibinfo
  {journal} {Phys. Rev. Lett.}\ }\textbf {\bibinfo {volume} {97}},\ \bibinfo
  {pages} {106604} (\bibinfo {year} {2006})}\BibitemShut {NoStop}%
\end{thebibliography}%
\end{document}